\documentclass[twocolumn,trackchanges]{aastex62}
\usepackage{hyperref}
\usepackage{url}

\graphicspath{{./}{figures/}}

\received{January 1, 2018}
\revised{January 7, 2018}
\accepted{\today}
\submitjournal{ApJ}

\shorttitle{Time-delay of MgII Emission Response for HE 0435-4312}
\shortauthors{Zaja\v{c}ek et al.}

\begin{document}

\title{Time Delay of MgII Emission Response for the Luminous Quasar HE 0435-4312: Towards Application of High-Accretor Radius-Luminosity Relation in Cosmology}

\correspondingauthor{Michal Zaja\v{c}ek}
\email{zajacek@cft.edu.pl}

\author[0000-0001-6450-1187]{Michal Zaja\v{c}ek}
\affiliation{Center for Theoretical Physics, Polish Academy of Sciences, Al. Lotnik\'ow 32/46, 02-668 Warsaw, Poland}

\author[0000-0001-5848-4333]{Bo\.{z}ena Czerny}
\affiliation{Center for Theoretical Physics, Polish Academy of Sciences, Al. Lotnik\'ow 32/46, 02-668 Warsaw, Poland}

\author[0000-0002-7843-7689]{Mary Loli Martinez--Aldama}
\affiliation{Center for Theoretical Physics, Polish Academy of Sciences, Al. Lotnik\'ow 32/46, 02-668 Warsaw, Poland}

\author[0000-0002-0297-3346]{Mateusz Ra\l{}owski}
\affiliation{Astronomical Observatory of the Jagiellonian University, Orla 171, 30-244 Cracow, Poland}

\author[0000-0002-6105-6492]{Aleksandra Olejak}
\affiliation{Nicolaus Copernicus Astronomical Center, Polish Academy of Sciences, ul. Bartycka 18, 00-716 Warsaw, Poland}

\author{Robert Przy\l{}uski}
\affiliation{Center for Theoretical Physics, Polish Academy of Sciences, Al. Lotnik\'ow 32/46, 02-668 Warsaw, Poland}

\author[0000-0002-5854-7426]{Swayamtrupta Panda}
\affiliation{Center for Theoretical Physics, Polish Academy of Sciences, Al. Lotnik\'ow 32/46, 02-668 Warsaw, Poland}
\affiliation{Nicolaus Copernicus Astronomical Center, Polish Academy of Sciences, ul. Bartycka 18, 00-716 Warsaw, Poland}

\author[0000-0002-2005-9136]{Krzysztof Hryniewicz}
\affiliation{National Centre for Nuclear Research, Pasteura 7, 02-093 Warsaw, Poland}

\author[0000-0003-2656-6726]{Marzena \'Sniegowska}
\affiliation{Center for Theoretical Physics, Polish Academy of Sciences, Al. Lotnik\'ow 32/46, 02-668 Warsaw, Poland}
\affiliation{Nicolaus Copernicus Astronomical Center, Polish Academy of Sciences, ul. Bartycka 18, 00-716 Warsaw, Poland}

\author[0000-0002-7604-9594]{Mohammad-Hassan Naddaf}
\affiliation{Center for Theoretical Physics, Polish Academy of Sciences, Al. Lotnik\'ow 32/46, 02-668 Warsaw, Poland}

\author{Raj Prince}
\affiliation{Center for Theoretical Physics, Polish Academy of Sciences, Al. Lotnik\'ow 32/46, 02-668 Warsaw, Poland}

\author{Wojtek Pych}
\affiliation{Nicolaus Copernicus Astronomical Center, Polish Academy of Sciences, ul. Bartycka 18, 00-716 Warsaw, Poland}

\author[0000-0002-9443-4138]{Grzegorz Pietrzy\'nski}
\affiliation{Nicolaus Copernicus Astronomical Center, Polish Academy of Sciences, ul. Bartycka 18, 00-716 Warsaw, Poland}

\author[0000-0001-9704-690X]{Catalina Sobrino Figaredo}
\affiliation{Astronomisches Institut - Ruhr Universitaet Bochum, Germany}

\author[0000-0002-5918-8656]{Martin Haas}
\affiliation{Astronomisches Institut - Ruhr Universitaet Bochum, Germany}

\author{Justyna \'Sredzi\'nska}
\affiliation{Nicolaus Copernicus Astronomical Center, Polish Academy of Sciences, ul. Bartycka 18, 00-716 Warsaw, Poland}

\author{Magdalena Krupa}
\affiliation{Astronomical Observatory of the Jagiellonian University, Orla 171, 30-244 Cracow, Poland}

\author{Agnieszka Kurcz}
\affiliation{Astronomical Observatory of the Jagiellonian University, Orla 171, 30-244 Cracow, Poland}
 
\author[0000-0001-5207-5619]{Andrzej Udalski}
\affiliation{Astronomical Observatory, University of Warsaw, Al. Ujazdowskie 4,  00-478 Warsaw, Poland}

\author[0000-0002-5760-0459]{Vladim\'ir Karas}
\affiliation{Astronomical Institute of the Czech Academy of Sciences, Bo\v{c}ní II 1401, CZ-14100 Prague, Czech Republic}

\author[0000-0003-4745-3923]{Marek Sarna}
\affiliation{Nicolaus Copernicus Astronomical Center, Polish Academy of Sciences, ul. Bartycka 18, 00-716 Warsaw, Poland}

\author{Hannah L. Worters}
\affiliation{South African Astronomical Observatory, PO Box 9, Observatory, 7935 Cape Town, South Africa}

\author[0000-0003-3904-6754]{Ramotholo R. Sefako}
\affiliation{South African Astronomical Observatory, PO Box 9, Observatory, 7935 Cape Town, South Africa}

\author{Anja Genade}
\affiliation{South African Astronomical Observatory, PO Box 9, Observatory, 7935 Cape Town, South Africa}



\begin{abstract}

Using the six years of the spectroscopic monitoring of the luminous quasar HE 0435-4312 ($z=1.2231$) with the Southern African Large Telescope (SALT), in combination with the photometric data (CATALINA, OGLE, SALTICAM, and BMT), we determined the rest-frame time-delay of $296^{+13}_{-14}$ days between the MgII broad-line emission and the ionizing continuum using seven different time-delay inference methods. Artefact time-delay peaks and aliases were mitigated using the bootstrap method, prior weighting probability function as well as by analyzing unevenly sampled mock light curves. The MgII emission is considerably variable with the fractional variability of $\sim 5.4\%$, which is comparable to the continuum variability ($\sim 4.8\%$). Because of its high luminosity ($L_{3000}=10^{46.4}\,{\rm erg\,s^{-1}}$), the source is beneficial for a further reduction of the scatter along the MgII-based radius-luminosity relation and its extended versions, especially when the high-accreting subsample that has an RMS scatter of $\sim 0.2$ dex is considered. This opens up a possibility to use the high-accretor MgII-based radius-luminosity relation for constraining cosmological parameters. With the current sample of 27 reverberation-mapped sources, the best-fit cosmological parameters $(\Omega_{\rm m}, \Omega_{\Lambda})=(0.19; 0.62)$ are consistent with the standard cosmological model within 1$\sigma$ confidence level.

\end{abstract}



\section{Introduction} \label{sec:intro}

Broad emission lines with line widths of several $1000\,{\rm km\,s^{-1}}$ are one of the main characteristic features of the optical and UV spectra of active galactic nuclei \citep[AGN;][]{1943ApJ....97...28S,1959ApJ...130...38W,1963Natur.197.1040S}, specifically of type I where the broad line region (BLR) is not obscured by the dusty molecular torus \citep{1993ARA&A..31..473A,1995PASP..107..803U}. However, the scattered polarized light can reveal broad lines even for obscured type II AGN \citep[type II NGC 1068 as the first case,][]{1985ApJ...297..621A}, which implies the universal presence of the BLR for accreting supermassive black holes (SMBH). The low-luminous systems, such as the Galactic Center \citep{2010RvMP...82.3121G,2017FoPh...47..553E,2018MNRAS.480.4408Z} or M87 \citep{2003ApJ...584..164S}, do not reveal the presence of broad lines. But even for some sources with lower accretion rates, broad lines can be present \citep{2019MNRAS.488L...1B} and the exact accretion limit for their appearance was analyzed to some extent by e.g. \citet{2009ApJ...701L..91E}, who estimated the bolometric luminosity limit of $5 \times 10^{39} (M_{\bullet}/10^7\,M_{\odot})^{2/3}\,{\rm erg\,s^{-1}}$, where $M_{\bullet}$ is the black hole mass in units of $10^7$ Solar masses ($M_{\odot}$). However, several crucial questions remain yet unanswered. Mainly the transition from the geometrically thin disk accretion flows to geometrically thick advection-domination accretion flows at lower acretion rates is still unclear and is likely related to the boundary conditions, the ability of the flow to cool, the feeding mechanism (warm stellar winds or an inflow of cold gas from larger scales), the associated initial angular momentum and the resulting circularization radius. In addition, not only does the formation of the BLR but also its properties seem to depend on the accretion-rate, which motivates the study of higher-accreting sources \citep{2015ApJ...806...22D,2018ApJ...856....6D}, such as in particular HE 0413-4031 \citep{2020ApJ...896..146Z}.  

The BLR has mostly been studied indirectly via the reverberation mapping, i.e. by inferring the time-delay between the ionizing UV continuum emission of an accretion disc and the broad-line emission \citep{1982ApJ...255..419B,2004AN....325..248P,2009NewAR..53..140G,2019arXiv190800742C,2020OAst...29....1P} using typically the interpolated cross-correlation function \citep{1998PASP..110..660P,peterson2004,2018ascl.soft05032S} or other methods \citep[see e.g.][for an overview and an application of seven methods in total.]{2019arXiv190703910Z,2020ApJ...896..146Z}. The high correlation between the continuum and the line-emission fluxes implies that the line emission is mostly a reprocessed thermal emission of an accretion disc. From the rest-frame time-delay $\tau_{\rm BLR}$, it is straightforward to estimate the size of the BLR, $R_{\rm BLR}=c\tau_{\rm BLR}$, and in combination with the single-epoch line full width at half maximum (FWHM) or the line dispersion $\sigma$, which serve as proxies for the BLR virial velocity, one can infer the central black hole mass, $M_{\bullet}=fc\tau_{\rm BLR}\mathrm{FWHM}^2 G^{-1}$. The factor $f$ is known as the virial factor and takes into account the geometrical and kinematical characteristics of the BLR. Although $f$ is typically of the order of unity, it introduces a factor of $\sim 2-3$ uncertainty in the black hole mass. By comparing the black hole masses inferred from accretion-disc spectra with the single-epoch spectroscopy masses, \citet{mejia2018} found that the value of $f$ is approximately inversely proportional to the broad-line FWHM\footnote{More precisely, using the general dependency between $f$ and the line FWHM in the form $f=(\text{FWHM}_{\rm obs}(\text{line})/\text{FWHM}_{\rm obs}^{0})^{\beta}$, with $\text{FWHM}_{\rm obs}^0$ and $\beta$ being the searched parameters, \citet{mejia2018} got $\text{FWHM}_{\rm obs}^{0}=3200 \pm 800\,{\rm km\,s^{-1}}$, $\beta=-1.21 \pm 0.24$ for MgII and $\text{FWHM}_{\rm obs}^{0}=4550 \pm 1000\,{\rm km\,s^{-1}}$, $\beta=-1.17 \pm 0.11$ for H$\beta$ measurements.}, which provides a way to better estimate $f$ for individual sources.

The optical reverberation mapping studies using H$\beta$ line showed that there is a simple power-law relation between the size of the BLR and the monochromatic luminosity of AGN at 5100\AA\, \citep{kaspi2000, 2004ApJ...613..682P}, so-called radius-luminosity (RL) relation. After a proper removal of the host galaxy starlight \citep{bentz2013}, the slope of the power law is close to $0.5$, which is expected from simple photoionization arguments. The importance of the radius-luminosity relation lies in its application to infer black hole masses from single-epoch spectroscopy, where FWHM or the line dispersion $\sigma$ serves as a proxy for the velocity of virialized BLR clouds and the source monochromatic luminosity serves as a proxy for the rest-frame time delay, and hence the BLR radius, via the RL relation.

Another, more recent application of the RL relation is the possibility to utilize it for obtaining absolute monochromatic luminosities. From the measured flux densities, one can calculate luminosity distances and use them for constraining cosmological parameters \citep{haas2011,watson2011,czerny2013,martinez_aldama2019,2019FrASS...6...75P,PTF100}. The problem for cosmological applications is that the RL relation exhibits a scatter, which has increased with the accumulation of more sources, especially including those with higher accretion rates \citep[super-Eddington accreting massive black holes--SEAMBHs sample;][]{2015ApJ...806...22D,2018ApJ...856....6D}. 

The scatter is present for both lower-redshift H$\beta$ sources \citep{grier2017} and also for higher-redshift MgII sources that follow an analogous RL relation \citep{czerny2019,2020ApJ...896..146Z,2020ApJ...901...55H}. The scatter was attributed to the accretion-rate intensity, with the basic trend that the largest departure from the nominal RL relation is exhibited by the highly accreting sources \citep{2018ApJ...856....6D}. The correction to the time-delay was proposed based on the Eddington ratio and the dimensionless accretion rate \citep{martinez_aldama2019}, however, these two quantities depend on the time-delay via the black hole mass. To break down the interdependency, \citet{2020ApJ...899...73F} proposed to make use of independent, observationally inferred quantities related to the optical/UV spectral energy distribution. The relative FeII strength correlated with the accretion-rate intensity is especially efficient in reducing the scatter for H$\beta$ sources to only $0.19$ dex \citep{2019ApJ...886...42D,2020MNRAS.491.5881Y}. The same effect is observed for the extended radius-luminosity relations for the MgII reverberation-mapped sources \citep[68 sources; ][]{Mart_nez_Aldama_2020}. \citet{Mart_nez_Aldama_2020} divide the sources into low and high accretors, where the high accretors show a much smaller scatter of only $\sim 0.2$ dex. The extended RL relation including the relative FeII strength further reduces the scatter down to $0.17$ dex for the high-accreting subsample.

Massive reverberation monitoring campaigns are currently performed by several groups, for instance the Australian Dark Energy Survey \citep[OzDES,][]{2019MNRAS.487.3650H}, the Sloan Digital Sky Survey Reverberation Mapping project \citep[SDSS-RM, ][]{2015ApJS..216....4S,2019ApJS..241...34S}, the Dark Energy Survey \citep[DES, ][]{2018ApJS..239...18A,2018SPIE10704E..0DD,2020MNRAS.493.5773Y}, the super-Eddington accreting massive black holes \citep[SEAMBHs, ][]{2015ApJ...806...22D,2018ApJ...856....6D}, but an individual source monitoring can also contribute significantly, especially if its luminosity is at the extreme values of the radius-luminosity relation. This is because the most luminous quasars have very low sky densities, so they are not suitable for reverberation mapping multi-object spectroscopy (MOS-RM) programs, and instead require an individual object monitoring. This is why programs like the one presented here are necessary. Luminous sources are expected to be beneficial in terms of increasing the RL correlation coefficient and it can also lead to the reduction in the scatter. In the current paper, we present results of the time delay of MgII line for the last of three intermediate redshift very luminous quasars monitored for several years with the Southern African Large Telescope (SALT). The source HE 0435-4312 ($z=1.2231$) hosts a supermassive black hole of $2.2\times 10^9\,M_{\odot}$ that is highly accreting with the Eddington ratio of $0.58$ according to the SED best-fit of \citet{sredzinska2017}. The peculiarity of the source is a smooth shift of the MgII line peak first towards longer wavelengths, while currently the shift proceeds towards shorter wavelengths. This line shift could hint at the presence of a supermassive black hole binary \citep{sredzinska2017}.

In this paper, we measure the time delay of HE~0435-4312 using seven methods. Subsequently, we study the position of the source in the RL relation, including its extended versions, and how the source affects the correlation coefficient as well as the scatter. Finally, we look at the potential applicability of the MgII high-accreting subsample for cosmological studies.

The paper is structured as follows. In Section~\ref{sec:obs}, we describe observations including both spectroscopy and photometry. In Section~\ref{sec_results_spectroscopy}, we analyze the mean and the rms spectrum, spectral fits of individual observations, and the variability of the continuum and the line-emission light curves. The core of the paper is Section~\ref{sec:result}, where we summarize the mean rest-frame time delay between the MgII emission and the continuum as inferred from seven different methods. The position of the source in the RL relation and its extended versions is also analyzed in this section. Subsequently, in Section~\ref{sec:discussion}, we discuss the aspect of MgII emission variability as well as the application of the high-accreting MgII subsample in cosmology. Finally, we present our conclusions in Section~\ref{sec:conclusions}.


\section{Observations} \label{sec:obs}

Here we present the optical photometric and spectroscopic observations of the quasar HE 0435-4312 ($z=1.232$, $V=17.1\,{\rm mag}$) with J2000 coordinates RA$=04{\rm h}37\,{\rm m}11.8{\rm s}$, Dec$=-43{\rm d}06{\rm m}04{\rm s}$ according to the NED database\footnote{NASA/IPAC Extragalactic Database: \url{http://ned.ipac.caltech.edu/}.}. Due to its large optical flux density, it was found during the Hamburg ESO quasar survey \citep{wisotzki2000}. Previously, \citet{sredzinska2017} reported ten spectroscopic observations performed by SALT Robert Stobie Spectrograph (RSS) over the course of three years (from Dec. 23/24, 2012 to Dec.7/8, 2015). The main result of their analysis was the detection of the fast displacement of MgII line with respect to the quasar rest frame by $104 \pm 14\,{\rm km\,s^{-1}\,yr^{-1}}$. In this paper the number of spectroscopic observations increased to 25, which together with 81 photometric measurements allows for the analysis of the time-delay response of MgII line with respect to the variable continuum. Previously we detected a time-delay for two other luminous quasars: $562^{+116}_{-68}$ days for CTS C30.10 \citep{czerny2019} and $303^{+29}_{-33}$ days for HE 0413-4031 \citep{2020ApJ...896..146Z}, both in the rest frames of the corresponding sources. These intermediate-redshift sources are of high importance for constraining the MgII-based radius-luminosity relation. Especially sources with either low or high luminosities are needed to constrain the slope of the RL relation and evaluate the scatter along it \citep{Mart_nez_Aldama_2020}.

\subsection{Photometry}
\label{sec:photometry}

The photometric data were combined from a few dedicated monitoring projects, described in more detail in \citet{2020ApJ...896..146Z}. The source has been monitored in the V band as part of the OGLE-IV survey \citep{2015AcA....65....1U} done
with the 1.3 m Warsaw telescope at the Las Campanas Observatory, Chile. The exposure times were typically around 240 sec, and the photometric errors were small, of order of 0.005 magnitude (see Table~\ref{tab:phot1}). In the later epochs the quasar was observed, again in V band, with the $40\,{\rm cm}$ Bochum Monitoring Telescope (BMT)\footnote{BMT is a part of the Universitaetssternwarte Bochum located near Cerro Armazones in Chile. For more information, see \citet{2013AN....334.1115R}.}. These data showed a systematic offset of 0.2 magnitudes towards larger magnitudes with respect to the overlapping OGLE data, which we corrected for by the shift of all of the BMT points. SALT spectroscopic observations were also supplemented, whenever possible, with the SALTICAM imaging in g band. We have analysed all of these data, however the two data sets (20 August 2013 and 27 January 2019) showed significant discrepancy with the other measurements. The first of the two sets of observations was done during full moon, and with the moon-target separation relatively large, spectroscopic observations were not affected, but the photometric observations were still affected. During the second set of observations the night was dark but seeing was about 2.5" during the photometry, dropping down to 2" during the spectroscopic exposures. We were unable to correct the data for these effects, and we did not include these data points in further considerations. Because of the collection of the data in $g$-band, we allowed for the grey shift of all the SALTICAM data, and the shift was determined using epochs when they coincided with the more precise OGLE set collected in the V-band. Finally, we supplemented our photometry with the lightcurve from Catalina Sky Survey \footnote{\url{http://nunuku.caltech.edu/cgi-bin/getcssconedb_release_img.cgi}} which is not of a very high quality (with the uncertainties of $0.02-0.03$ mag) but nicely covers the early period since 2005 till 2013. We have binned this data to reduce the scatter. Table ~\ref{tab:phot1} contains only the data points which were used in time delay measurements. All the data points are presented in the upper panel of Figure~\ref{fig:zestaw1}. 

Recently, the quasar was monitored in the V-band with a median sampling of 14 days using Lesedi, a 1-m telescope at the South African Astronomical Observatory (SAAO), with the Sutherland High Speed Optical Camera (SHOC).
SHOC has a 5.7 $\times$ 5.7 square arcmin field-of-view (FoV).
Each observation consists of 9 dithered 60\,s exposures.
They are corrected for bias and flatfields (using dusk or dawn sky flats).
Astrometry is obtained using the SCAMP tool \footnote{https://www.astromatic.net/software/scamp}.
The resulting median-combined image has a 7.5 $\times$ 7.5 square arcmin FoV centered on the quasar.
The light curves were created using 5 calibration stars located on the same image as the quasar. The preliminary results are consistent with the last photometric point from SALT/SALTICAM.

\subsection{Spectroscopy}
\label{sec:spectr}

Spectroscopic observations of HE 0435-4312 were performed with 11-m telescope SALT, with Robert Stobie Spectrograph (RSS; \citealt{burgh2003,kobulnicky2003,smith2006}
) in a long-slit mode, with the slit width of 2". We used RSS PG1300 grating, with the grating tilt angle of $26.75$ deg. Order blocking was done with the blue PC04600 filter. Two exposure times were always made, each of about 820 s. The same setup has been used throughout the whole monitoring period, from 23 Dec 2012 till 20 Aug 2020. Observations were always performed in the service mode.

The raw data reduction was perfomed by the SALT observatory staff, and the final reduction was performed by us with the use of the IRAF package. All the details were given in \citet{sredzinska2017}, where the results from the first three years of this campaign were presented. We followed exactly the same procedure for all 25 observations to minimize the possibility of unwanted systematic differences.

In order to get the flux calibration, we performed a weighted spline interpolation of the first order (with inverse measurement errors as weights) between the photometric and spectroscopic observations epochs, thus an apparent $V$ magnitude was assigned to each spectrum. Taking as a reference the composite spectrum and the continuum with a slope of $\alpha_\lambda$=-1.56 from \citet{vandenberk2001}, we just normalized each spectrum to the $V$ magnitudes (5500 \AA).

\subsection{Spectroscopic data fitting}
\label{sec:spec_fit}

The reduced and calibrated spectra were fitted in the 2700 \AA - 2900 \AA~ range in the rest frame. The basic model components were as in \citet{sredzinska2017}. The data were represented by (i) continuum in the form of a power law with arbitrary slope and normalization, (ii) the FeII pseudo-continuum, and (iii) two kinematic components representing MgII line, each of the kinematic components was represented by two doublet components. The line is unresolved, and the doublet ratio could not be well constrained, so it was fixed at 1:1 ratio \citep[see][for the discussion]{sredzinska2017}. For the kinetic shapes we used Lorentzian profiles since they provided somewhat better representation of the data in $\chi^2$ terms than the Gaussian ones.  All the parameters were fitted simultaneously. In order to determine the redshift and the most appropriate FeII template, we studied in detail observation 23 which also covered the region around 3000 \AA~ in the rest frame (see Appendix~\ref{sect:redshift}). Thus for the adopted redshift $z = 1.2231$, the FeII template was very slightly modified in comparison with \citet{sredzinska2017}, and pseudo-continuum was kept at  FWHM of 3100 km s$^{-1}$. Thus there were eight free parameters of the model: power law normalization and slope, normalization of the FeII pseudo-continuum, the width and normalizations of the two kinematic components of MgII, and the position of the second component, with the other one set at the quasar rest frame, together with FeII. 


\section{Results: spectroscopy}
\label{sec_results_spectroscopy}

%

\subsection{Determination of the mean and the rms spectra}
\label{sec_mean_spectrum}

\begin{figure}
    \centering
    \includegraphics[width=\columnwidth]{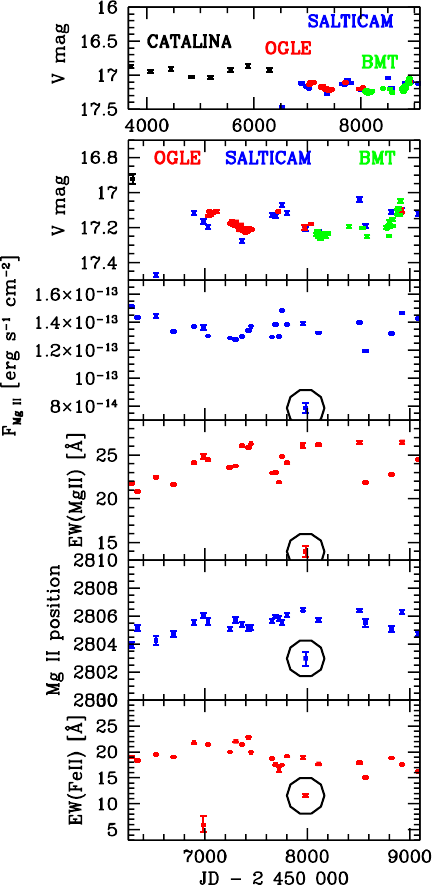}
    \caption{Light curves and the time evolution of emission line properties measured in the rest frame. Points marked with a black circle (obs. 19) were removed from the analysis.}
    \label{fig:zestaw1}
\end{figure}

We determined the mean and the rms spectrum for HE 0435-4312 as we did before for quasar HE 0413-4031 \citep{2020ApJ...896..146Z}. Due to the particularly low signal to noise ratio ($\sim 7.5$) shown by the spectrum no. 19, which is clearly an outlier in the light curve (Fig.~\ref{fig:zestaw1}), we do not consider it for the estimation of the mean and rms spectra. We followed the methodology for constructing the mean and the rms spectra as outlined in  \citet{2004ApJ...613..682P}. In particular, we formed the mean spectrum (without spectrum no. 19) using
\begin{equation}
    \overline{F(\lambda)}=\frac{1}{N}\sum_{i=1}^{N} F_i(\lambda)\,,
    \label{eq_mean_spectrum}
\end{equation}
where $F_i(\lambda)$ is the $i$th spectrum out of total $N$ spectra in the measured database.
Subsequently, the rms spectrum, initially taking into account all spectral components (MgII line$+$continuum$+$FeII pseudocontinuum), was constructed using 
\begin{equation}
    S(\lambda)=\left\{\frac{1}{N-1}\sum_{i=1}^{N}\left[F_i(\lambda)-\overline{F(\lambda)} \right]^2 \right\}^{1/2}\,.
    \label{eq_rms_spectrum}
\end{equation}

\begin{figure}
    \centering
    \includegraphics[width=\columnwidth]{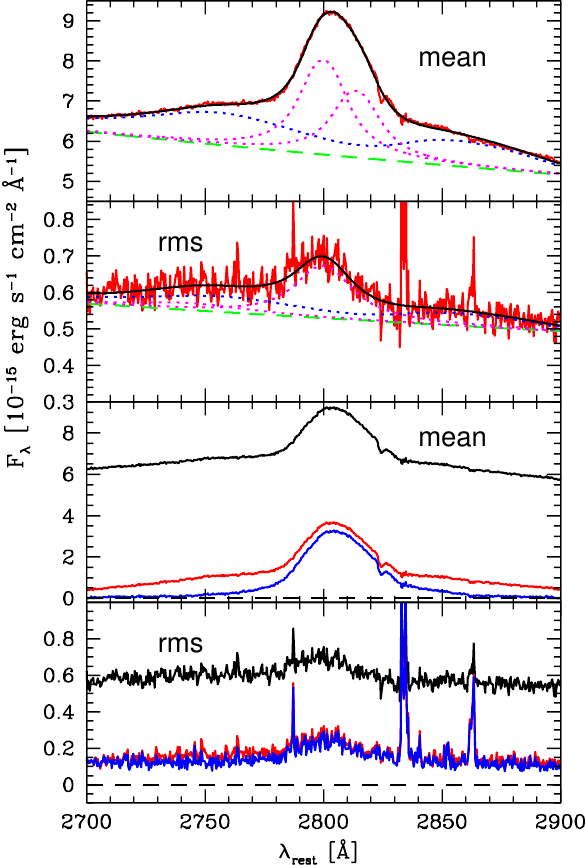}
    \caption{{\bf Upper panel:} Mean spectrum (red curve) obtained without observation 19. We also show its decomposition into two Lorentzian components of MgII line (dotted magenta), FeII pseudo-continuum (dotted blue line), and a power law (green dashed line). {\bf Upper middle panel:} RMS spectrum and its decomposition. The spectral component notation is the same as for the mean spectrum in the upper panel. {\bf Lower middle panel:} The mean spectrum in absolute calibration. The solid black line denotes the mean spectrum including all the components (line, continuum, and FeII pseudocontinuum). The red solid line stands for the mean spectrum constructed after the removal of the continuum contribution from each spectrum. The blue solid line stands for the line-only mean spectrum, after the additional removal of FeII pseudocontinuum. {\bf Lower panel:} A properly calibrated rms spectrum, with the black line denoting the total rms (line$+$continuum$+$FeII pseudocontinuum). The rms spectrum after the subtraction of the continuum is denoted by the red line, while the line-only rms is depicted by the blue line.}
    \label{fig:zestaw2}
\end{figure}

The constructed mean and the rms spectrum is shown in Fig.~\ref{fig:zestaw2} (black solid lines) in the upper and the upper middle panel, respectively.
The quasar is not strongly variable, so the normalization of the rms is very low, and the spectrum is noisy, with visible effects of occasional imperfect sky subtraction. However, the overall quality of the rms spectrum is still suitable for the analysis. In both the mean and the rms spectra, the MgII line modelling required two kinematic components, since line asymmetry is clearly visible. The result is shown in Figure~\ref{fig:zestaw2} with spectral components depicted with different colored lines described in the figure caption. For the fitting, we finally used the redshift as determined by \citet{sredzinska2017}, but with a slightly modified FeII template based on the d11 template of \citet{bruhweiler2008}. We also compared and analysed other FeII templates based on the updated CLOUDY model \citep{ferland2017} as well as the six-transition model by \citet{kovacevic15} and \citet{2019MNRAS.484.3180P}. For a detailed discussion of different FeII templates, in total eight set-ups with different redshifts as well as Lorentzian or Gaussian line component profiles, see Appendix~\ref{sect:redshift}. 

The overall shapes of the mean and the rms spectra are similar (see Figure~\ref{fig:zestaw2}). The FWHM of MgII line in the mean spectrum is $3695^{+21}_{-21}\,{\rm km\,s^{-1}}$, the line in the rms spectrum might be slightly broader, $3886^{+143}_{-341}\, {\rm km\,s^{-1}}$, but consistent within the error margins. A much larger difference is seen in the line dispersion: in the mean spectrum it is much smaller than in the rms spectrum ($2707^{+10}_{-6}$ km s$^{-1}$ vs. $3623^{+76}_{-412}$ km s$^{-1}$). The FWHM and $\sigma$ is larger in the rms spectrum most likely due to its noisy nature, although there is an indication of a trend of both MgII FWHM and $\sigma$ in the rms spectrum being larger than in the mean spectrum based on the analysis of the SDSS-RM sample \citep{2019ApJ...882....4W}. However, the FWHM to $\sigma$ ratio for both the mean and the rms spectra is far from the value 2.35 expected for a Gaussian shape. The source can be classified as A-type source according to \citet{sulentic2000} classification. This is consistent with the Eddington ratio $0.58$ determined by \citet{sredzinska2017} for this object. There is also an interesting change in the line position between the mean and the rms spectrum, as determined from the first moment of the distribution: 2805 \AA~ vs. 2792 \AA. 

Following the spectral studies by \citet{2015ApJS..217...26B} and \citet{2019ApJ...882....4W}, we show in Fig.~\ref{fig:zestaw2} (middle lower and lower panels) the mean and the rms spectrum constructed taking into account all the spectral components (black lines; MgII line$+$continuum$+$FeII pseudocontinuum), the mean and the rms spectrum that have the continuum power-law subtracted from each spectrum (red lines), and finally the spectra with FeII pseudocontinuum subtracted from individual spectra, which represents the true, line-only mean and the rms spectrum (blue lines). For the rms spectra, we do not detect any significant difference in the line width, which is expected to be smaller for the total-flux rms than for the line-only rms spectrum \citep{2015ApJS..217...26B,2019ApJ...882....4W}. This can be attributed to the overall noisy nature of our rms spectra that are constructed from only 24 individual spectra. According to \citet{2015ApJS..217...26B} and \citet{2019ApJ...882....4W}, the difference in the line width becomes smaller for a sufficiently long duration of the campaign. However, in our case, the spectroscopic monitoring was only $\sim 4.25$ times longer than the emission-line time lag in the observer's frame (see Section~\ref{sec:result}). For such a short duration, the distribution of the line-width ratios between the total-flux and the line-only rms spectra is broad - between 0.5 and 1.0, with the peak close to 1.0, see Appendix~C in \citet{2015ApJS..217...26B}.

\subsection{Spectral fits of individual observations}

The fits to individual spectroscopic data sets were done in the same way as for the mean spectrum. In all 25 data sets, two kinematic components of the MgII line were needed to represent well the line shape. The normalization of FeII pseudocontinuum was allowed to vary for each individual spectrum, while the FeII width was fixed to FWHM$=3100\,{\rm km\,s^{-1}}$. The MgII component kinematically related to the FeII is slightly narrower, having on average FWHM of $2128 \pm 28$ km s$^{-1}$, while the second shifted component is somewhat broader, with FWHM of $2262 \pm 90$ km s$^{-1}$. However, we stress here that the broad band modelling by \citet{sredzinska2017} did not support any identification of these components with separate regions, so the two components are just a mathematical representation of the slightly asymmetric line shape.

The parameters for observation 19 were considerably different, but as we already mentioned in Section~\ref{sec_mean_spectrum}, this observation was of a particularly low quality despite the fact that actually three exposures were made this time. Cirrus clouds were present during the whole night, and even more clouds started coming during the quasar second exposure, so the third exposure was done. Nevertheless, all the parameters were determined with the errors a few times larger than for the remaining observations.

The average value of the equivalent width of the MgII line is $23.6 \pm 0.5$ \AA, if observation 19 was included. If observation 19 was not taken into account, the mean value increased to $24.1 \pm 0.6$ \AA, and the value was similar to the value determined from the mean spectrum: $23.9$ \AA. Such values are perfectly in agreement with the properties of the Bright Quasar Sample studied by \citet{forster2001}, where the mean EW of MgII broad component was found to be $27.4^{+8.5}_{-6.3}$ \AA. We do not see any traces of the narrow component neither in the mean spectrum, nor in the individual data sets.

The average value of the EW of FeII, $18.2 \pm 0.7$ \AA\, is also similar to the value in the mean spectrum, $18.9$ \AA. The relative error is larger than for MgII line since FeII pseudo-continuum is more strongly coupled to the power-law continuum during the fitting procedure.

The dispersion in the measurements between observations partially represents the statistical errors, but partially reflect the intrinsic evolution of the source in time. This evolution is studied in the next section.

\subsection{Light curves: variability and trends}

The quasar HE 0435-4312 is not strongly variable in terms of the continuum emission. The whole photometric lightcurve, including the CATALINA data, covers 14 years, and the fractional variability amplitude for the continuum is  8.9 \% in flux. During the period covered by SALT data (7 years) it is $\sim 4.9 \%$. Fortunately, the MgII line flux shows significant variability, comparable to the continuum, at the level of 5.4 \% during this period. We do not observe a suppressed MgII variability in this source, unlike seen in much larger samples \citep{goad1999,woo2008,zhu2017}. \citet{2020MNRAS.493.5773Y} also detect the response of MgII emission to the continuum for 16 extreme variability quasars, but with a smaller variability amplitude, $\Delta \log{L(\text{MgII})}=(0.39 \pm 0.07)\Delta \log{L(3000\AA)}$. A low MgII variability was modeled to be the result of a relatively large Eddington ratio \citep{Guo2020} but HE 0435-4312 is also a source with a rather large Eddington ratio of 0.58 \citep{sredzinska2017}. Overall, the fractional variability of MgII line for our source is comparable to the variability of $\sim 10\%$ on 100-day timescales as inferred for the SDSS-RM ensemble study \citep{2015ApJ...811...42S}.

The variations in quasar parameters are overall smooth. The quality of the observation 6 was not very high, as discussed in \citet{sredzinska2017}, but it still allowed to obtain the MgII line parameters properly. However, observation 19 created considerable problems. The registered number of photons was much lower (by a factor of a few) in comparison with typical observations, even the background was rather low, and clouds were apparently present in the sky. We did the data fitting and the derived values of the model parameters form clear outliers when compared with the trends (see Figure~\ref{fig:zestaw1}). Therefore, in the remaining analysis and the time delay determination, we did not use this observation.

In \citet{sredzinska2017} the change of the line position was discussed in much detail, since during the first three years the MgII line seems to move systematically towards longer wavelength with a surprising speed of $104 \pm 14\,{\rm km\,s^{-1}\,yr^{-1}}$ with respect to the quasar rest frame. However, now this trend has seemingly stopped, and in the recent observations it seems to be reversed. Such emission line behaviour is frequently considered  as a signature of a binary black hole \citep[e.g. ][for a review]{popovic2012}. However, to claim such a phenomenon, it would require extensive tests which are beyond the current paper aimed at the MgII line time delay measurement.

\section{Results: Time-delay determination} \label{sec:result}

To determine the most probable time-delay between the continuum and MgII line-emission, we applied several methods as previously in \citet{czerny2019} and \citet{2020ApJ...896..146Z}, namely:
\begin{itemize}
    \item Interpolated Cross-Correlation Function (ICCF),
    \item Discrete Correlation Function (DCF),
    \item $z$-transformed Discrete Correlation Function (zDCF),
    \item the JAVELIN package,
    \item Measures of data regularity/randomness -- von Neumann and Bartels estimators,
    \item $\chi^2$ method.
\end{itemize}
These seven methods are described in detail in Appendix~\ref{sec_time-delay}, including their strengths over other methods. It is beneficial to compare more methods since our light curves are irregularly sampled and the continuum light curve is heterogeneous, i.e. coming from four different instruments (CATALINA, OGLE, SALTICAM, and BMT). After the exclusion of low-quality data points and outliers, we finally obtained 79 continuum measurements with the mean cadence of $69.0$ days (maximum 597.6 days, minimum 0.75 days) and 24 MgII light-curve data points with the mean cadence of $121.6$ days (maximum 398.9 days, minimum 25.9 days). 

For our set of light curves, there were several candidate time delays present for different methods. A significant time delay between 600 and 700 days in the observer's frame was present for all the methods and we summarize the obtained values for this peak in Table~\ref{tab_time_delay_all_methods} for d11$_{\rm mod}$ FeII template and the redshift of $z=1.2231$. The time-delay is not affected significantly by a different FeII template, in particular the template of \citet{kovacevic15} and \citet{2019MNRAS.484.3180P} (hereafter denoted as KDP15) with a slightly different best-fit redshift of $z=1.2330$. The ICCF peak for KDP15 is at 692 days for the observer's frame, see Fig.~\ref{fig_ICCF_KDP15} in Appendix~\ref{sec_time-delay}.

\begin{table}[h!]
    \centering
     \caption{Overview of the best time delays for different methods. The time delay is expressed in light days in the observer's frame. The last two rows contain the observer's frame mean time delay as well as the mean rest-frame time delay for the redshift of $z=1.2231$.}
     \resizebox{\columnwidth}{!}{%
    \begin{tabular}{c|c}
    \hline
    \hline
    Method & Time delay in the observer's frame [days]\\
    \hline
    ICCF & $672^{+49}_{-37}$  \\
    DCF     & $656^{+18}_{-73}$ \\
    zDCF     &  $646^{+63}_{-57}$\\
    JAVELIN  &  $645^{+55}_{-41}$ \\
    von Neumann estimator & $635^{+32}_{-66}$\\
    Bartels estimator  &  $644^{+27}_{-45}$\\
    $\chi^2$ method  & $706^{+70}_{-61}$ \\
    \hline 
    Observer's frame mean & $658^{+29}_{-31}$  \\
    Rest-frame mean $(z=1.2231)$ & $296^{+13}_{-14}$ \\
    \hline
    \end{tabular}}
    \label{tab_time_delay_all_methods}
\end{table}

The significance of the peak between 600 and 700 days was evaluated using the bootstrap method for several time-delay techniques, i.e. by randomly selected light-curve subsets. In addition, for the JAVELIN method, we applied the alias mitigation using down-weighting by the number of overlapping data points, see Subsection~\ref{subsec_javelin} in the Appendix. With this technique \citep[see also][]{grier2017}, secondary peaks for the time delay longer than 700 days could effectively be suppressed. For the assessment of other time-delay artefact peaks, we generated mock light curves using the Timmer-Koenig algorithm \citep{1995A&A...300..707T} using the same light curve cadence as the observed one, see Appendix~\ref{appendix_alias_mitigation} for a detailed discussion. From the constructed time-delay probability distributions for all the seven methods, we could identify clear artefacts due to the sampling for time-delays $\lesssim 200$ days as well as for $\gtrsim 800$ days in the observer's frame. The recovery of the true time delay appears to be challenging for the given cadence and the duration of the observations, but the combination of more methods is beneficial to identify the best candidate for the true time delay.

\subsection{Final time-delay for MgII line}

\begin{figure}
    \centering
    \includegraphics[width=\columnwidth]{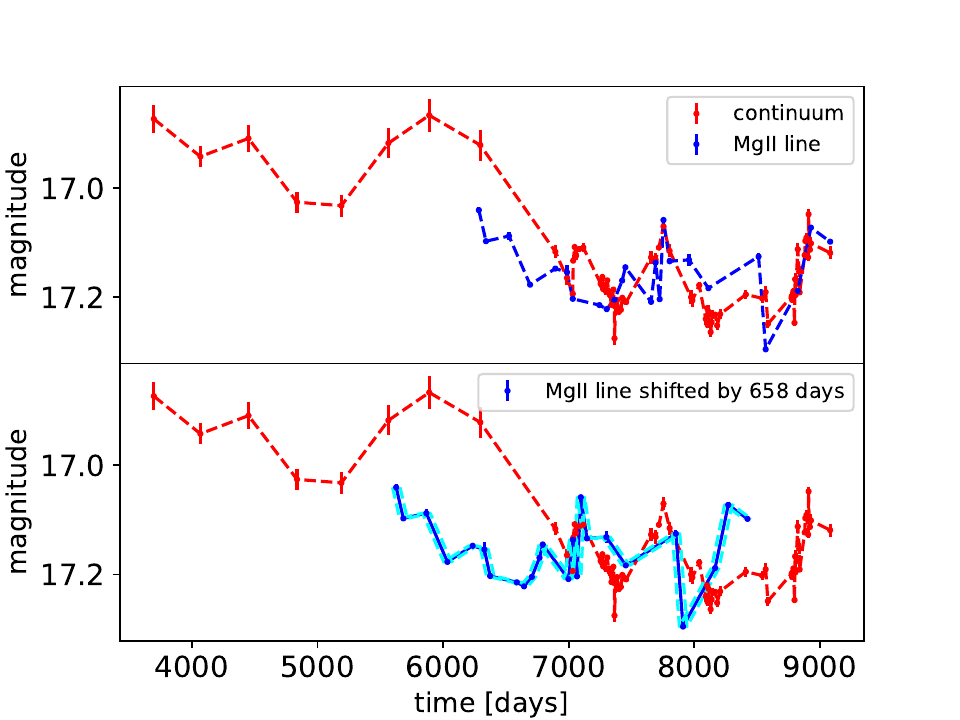}
    \caption{Comparison of the continuum and the shifted MgII light curves. {\bf Top panel:} The continuum and the original MgII light curve. {\bf Bottom panel:} The continuum and the MgII light curve shifted by the mean time-delay of $658$ days in the observer's frame.}
    \label{fig_cont_shifted_line}
\end{figure}

Combining all the seven methods listed in Table~\ref{tab_time_delay_all_methods}, the mean value in the observer's frame is $\overline{\tau}_{\rm obs}=658^{+29}_{-31}$ days. We visually compare the continuum light curve and the original as well as the time-shifted MgII line light curve in Fig.~\ref{fig_cont_shifted_line}. For an easier comparison, the MgII line is shifted towards larger magnitudes by the difference in the mean values of both light curves ($1.74$ mag). The correlation between the continuum and the shifted MgII light curve, although present, is not visually improved with respect to the zero  time lag. This is not unexpected since our source does not exhibit such a large variability amplitude as low Eddington-ratio sources. In addition, even when the line-emission light curve is shifted by the fiducial time-delay, it can intrinsically exhibit periods when the line emission is decorrelated with respect to the continuum emission, which is referred to as the emission-line or the BLR holiday \citep[studied in more detail for NGC 5548,][]{2019ApJ...882L..30D}. This justifies the need for using several robust statistical methods to assess the best time-delay. We also tried to subtract a linear trend from both light curves, but since for both of them the slope is consistent with zero within fitting uncertainties, it did not yield an improvement. Even for a noticeable linear trend, as for instance for HE 0413-4031 \citep{2020ApJ...896..146Z}, detrending actually led to a decrease in the correlation coefficient at the fiducial time delay. Hence, the linear trend subtraction should be tested on a larger set of sources to assess statistical relevance in terms of time-delay determination.

Subsequently, we obtain the mean rest-frame value of $\overline{\tau}_{\rm rest}=\overline{\tau}_{\rm obs}/(1+z)=296^{+13}_{-14}$ days for the redshift of $z=1.2231$. The light-travel distance of the MgII emission zone can then be estimated as $R_{\rm MgII}\sim c\overline{\tau}_{\rm rest}=7.7^{+0.3}_{-0.4}\times 10^{17}\,{\rm cm}=0.249^{+0.011}_{-0.012}\,{\rm pc}$. The rest-frame time delay is within uncertainties comparable to the time delay of the previously analyzed highly-accreting quasar HE 0413-4031 \citep[$z=1.38$,][]{2020ApJ...896..146Z}. 

\subsection{Black hole mass estimate}

Taking into account the rest-frame time-delay of $\overline{\tau}_{\rm rest}=296^{+13}_{-14}$ days and the MgII FWHM in the mean spectrum of FWHM$_{\rm MgII}=3695^{+21}_{-21}\,{\rm km\,s^{-1}}$, we can estimate the central black hole mass under the assumption that MgII emission clouds are virialized. Using the anticorrelation between the virial factor and the line FWHM according to \citet{mejia2018},
\begin{equation}
    f_{\rm MR}=\left(\frac{\mathrm{FWHM}_{\mathrm{MgII}}}{3200\pm 800\,{\rm km\,s^{-1}}} \right)^{-1.21 \pm 0.29}\,,
    \label{eq_fmr}
\end{equation}
we obtain $f_{\rm MR}\sim 0.84$. The virial black hole mass can then be calculated using the MgII FWHM in the mean spectrum and the measured time delay as $M_{\rm vir}^{\rm FWHM}(f_{\rm MR})=(6.6^{+0.3}_{-0.3})\times 10^8\,M_{\odot}$. Adopting the virial factor according to \citet{2015ApJ...801...38W}, $f_{\rm Woo}=1.12$ (based on FWHM of the H$\beta$ line), we obtain the virial black hole mass $M_{\rm vir}^{\rm FWHM}(f_{\rm Woo})=(8.8^{+0.4}_{-0.4})\times 10^8\,M_{\odot}$. Here we adopted the FWHM from the mean spectrum since it is better constrained than the rms FWHM. The mean values are, however, consistent within uncertainties, which is in agreement with the general correlation of MgII line widths measured in the mean and the rms spectra using the SDSS-RM sample \citep{2019ApJ...882....4W}.

Using instead the MgII line dispersion in the mean spectrum, $\sigma=2707^{+10}_{-6}\,{\rm km\, s^{-1}}$, and the associated virial factor $f_{\sigma}=4.47$ (based on the H$\beta$ line dispersion) according to \citet{2015ApJ...801...38W}, the black hole mass is estimated as $M_{\rm vir}^{\sigma}(f_{\sigma})=(1.89^{+0.08}_{-0.09})\times 10^9\,M_{\odot}$. This value is consistent with the value inferred from the broad-band SED fitting using the model of a thin accretion disc according to \citet{sredzinska2017}, where they obtained $M_{\rm SED}=2.2\times 10^{9}\,M_{\odot}$. Hence for our source, using the line dispersion inferred from the mean spectrum (the rms spectrum is too noisy to reliably measure $\sigma$) appears to be beneficial for constraining the virial SMBH mass. Using FWHM, yields the virial mass below the one inferred from the broad-band fitting.

Next, we estimate the Eddington ratio. Using our continuum light curve, the mean $V$-band magnitude is $17.15\pm 0.09$ mag. With the redshift of $z=1.2231$ and the mean Galactic foreground extinction in the $V$-band of $0.045$ mag according to NED\footnote{\url{https://ned.ipac.caltech.edu/}}, we determine the luminosity at 3000\AA\,, $\log{(L_{3000}[{\rm erg\,s^{-1}}])}=46.359^{+0.038}_{-0.034}$, for which we apply the conversions of \citet{2015AcA....65..251K}. To estimate the bolometric luminosity, the corresponding bolometric correction can be obtained via the simple power-law scaling by \citet{2019MNRAS.488.5185N}, $\kappa_{\rm bol}=25\times[L_{3000}/10^{42}\,{\rm erg\,s^{-1}}]^{-0.2}\sim 3.36$, which yields $L_{\rm bol}=3.36L_{3000}\sim 7.68\times 10^{46}\,{\rm erg\,s^{-1}}$. The Eddington limit can be estimated for $M_{\bullet}\simeq 2\times 10^{9}\,M_{\odot}$, which was obtained via the SED fitting as well as the virial mass using the line dispersion, $L_{\rm Edd}\simeq 2.5\times 10^{47}\,(M_{\bullet}/2\times 10^9\,M_{\odot})\,{\rm erg\,s^{-1}}$. Finally, the Eddington ratio is $\eta=L_{\rm bol}/L_{\rm Edd}\sim 0.31$, which is about a factor of two smaller than the Eddington ratio of $0.58$ obtained by \citet{sredzinska2017} using the SED fitting. Still, the source is highly accreting, with $\eta$ comparable to HE~0413-4031 \citep[$\eta=0.4$,][]{2020ApJ...896..146Z}.

The high-accreting sources exhibit the largest scatter with respect to the standard RL relation with a trend of shorter time delays by a factor of a few than expected based on their monochromatic luminosity. This was studied for the H$\beta$ reverberation mapped sources \citep{2015ApJ...806...22D,2016ApJ...825..126D,Mart_nez_Aldama_2019}, and confirmed for the higher redshift MgII reverberation mapped sources as well \citep{2020ApJ...896..146Z,2020ApJ...901...55H,Mart_nez_Aldama_2020}, which suggests a common mechanism for time-delay shortening driven by the accretion rate. The relation between the rest-frame time-delay of H$\beta$ line and the linear combination of the monochromatic luminosity at 5100\AA\, and the relative strength of FeII line (flux ratio between FeII and H$\beta$ lines) yields a low scatter of only $\sigma_{\rm rms}=0.196$ dex \citep{2019ApJ...886...42D}, which suggests that the relative FeII strength and hence the accretion rate can account for most of the scatter.

In addition, highly accreting MgII reverberation-mapped sources exhibit a generally lower scatter along the multidimensional RL relations \citep{Mart_nez_Aldama_2020}. This motivates us to further study how high-luminosity and highly accreting sources such as HE 0435-4312 affect the scatter along the radius-luminosity relation alone as well as in combination with independent observables, such as the MgII line FWHM, the FeII strength $R_{\rm FeII}$, and the fractional variability $F_{\rm var}$ of the continuum, where the latter two parameters are correlated with the accretion rate.  

\subsection{Position in the radius-luminosity plane}

Given the high luminosity of HE 0435-4312, it is expected to be beneficial for constraining the MgII-based radius-luminosity relation. We add HE 0435-4312 to the original sample of 68 MgII reverberation-mapped sources studied in \citet{Mart_nez_Aldama_2020}. To characterize the accretion rate of our source and in order to compare it with other sources, we make use of the dimensionless accretion rate $\dot{\mathcal{M}}$ expressed specifically for 3000\AA\, as \citep{2014ApJ...793..108W,Mart_nez_Aldama_2020},
\begin{equation}
    \dot{\mathcal{M}}=26.2\left(\frac{L_{44}}{\cos{\theta}}\right)^{3/2}m_7^{-2}\,,
    \label{eq_mdot_3000}
\end{equation}
where $L_{44}$ is the luminosity at 3000\AA\, expressed in units of $10^{44}\,{\rm erg\,s^{-1}}$ and $m_7$ is the central black hole mass expressed in $10^7\,M_{\odot}$. The angle $\theta$ is the inclination angle with respect to the accretion disc and we set $\cos{\theta}=0.75$, which represents the mean inclination angle for type I AGN according to the studies of the dusty torus covering factor \citep[see e.g.,][]{lawrenceandelvis2010,ichikawa2015}.

\begin{figure*}[h!]
    \centering
    \includegraphics[width=\columnwidth]{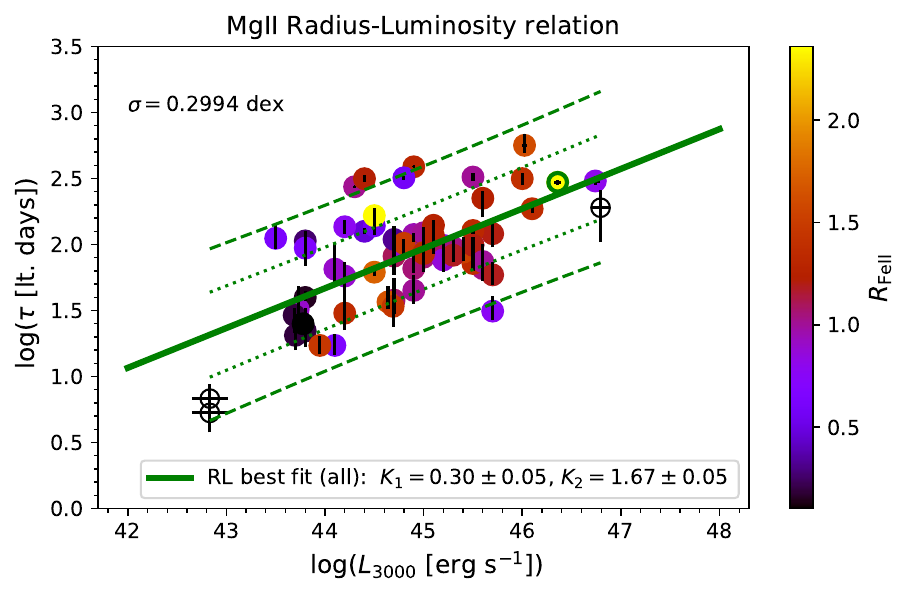}
    \includegraphics[width=\columnwidth]{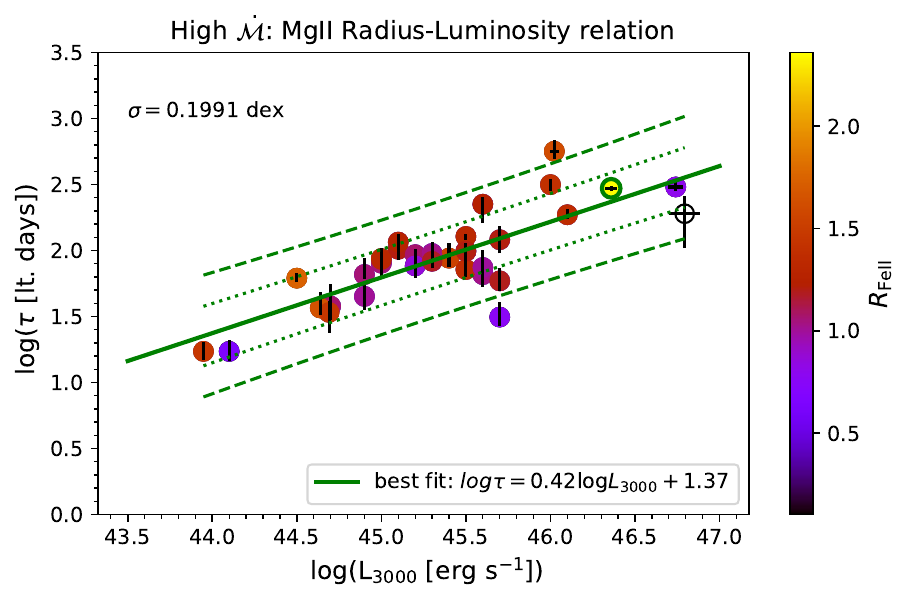}
    \caption{MgII-based radius-luminosity (RL) relation. {\bf Left panel:} MgII-based RL relation including all 68 MgII reverberation-mapped sources studied in \citet{Mart_nez_Aldama_2020} and the new source HE 0435-4312 denoted by a green circle. The relative UV FeII strength $R_{\rm FeII}$ is colour-coded for each source according to the colour axis on the right (except for NGC 4151 and CTS252, for which $R_{\rm FeII}$ was not available, and these sources are denoted by blank circles). For HE 0435-4312, we obtained $R_{\rm FeII}=\text{EW(FeII)}/\text{EW(MgII)}\sim 2.36$. {\bf Right panel:} The same as in the left panel, but for highly-accreting sources with $\log{\dot{\mathcal{M}}}>0.2167$, where the division of high accretors was taken from \citet{Mart_nez_Aldama_2020} according to the median value of their full sample. The scatter along the RL relation is noticeably lower for high accretors than for the full sample. In both panels, dashed and dotted lines denote the 68$\%$ and 95$\%$ confidence intervals, respectively.}
    \label{fig_RL_high}
\end{figure*}

Using Eq.~\ref{eq_mdot_3000} and the black hole mass estimate based on the mean-spectrum line dispersion and the broad-band SED fitting, we obtain $\dot{\mathcal{M}}=4.0^{+0.7}_{-0.6}$ or $\log{\dot{\mathcal{M}}}=0.61^{+0.07}_{-0.06}$ for HE 0435-4312, which puts it into the high-accretor category according to \citet{Mart_nez_Aldama_2020}, where all the MgII reverberation-mapped sources with $\log{\dot{\mathcal{M}}}>0.2167$ belong (median value of their MgII sample). Using the smaller SMBH mass based on MgII FWHM in the mean spectrum yields a larger $\dot{\mathcal{M}}$ by a factor of a few, $\log{\dot{\mathcal{M}}}=1.51^{+0.07}_{-0.06}$ (based on $f_{\rm MR}$) and $\log{\dot{\mathcal{M}}}=1.26^{+0.07}_{-0.06}$ (based on $f_{\rm Woo}$). In the further analysis, we adopt $\dot{\mathcal{M}}$ value based on the mean-spectrum line dispersion because of the consistency of $M^{\sigma}_{\rm vir}(f_{\sigma})$ with the SMBH mass inferred from the broad-band fitting of \citet{sredzinska2017}. In Fig.~\ref{fig_RL_high} (left panel), we plot the RL relation for all 69 sources including HE 0435-4312 (green circle). In the right panel of  Fig.~\ref{fig_RL_high}, we restrict the RL relation only to highly-accreting sources, which results in a significantly reduced RMS scatter of $\sigma_{\rm rms}=0.1991$ dex versus $\sigma_{\rm rms}=0.2994$ dex for the full sample. Adding HE 0435-4312 results in the small, but detectable reduction of scatter for both the full sample ($0.2994$ versus $0.3014$) as well as for the highly-accreting subsample ($0.1991$ versus $0.2012$) with respect to the original MgII sample of 68 sources \citep{Mart_nez_Aldama_2020}.

\begin{figure*}[h!]
    \centering
    \includegraphics[width=\columnwidth]{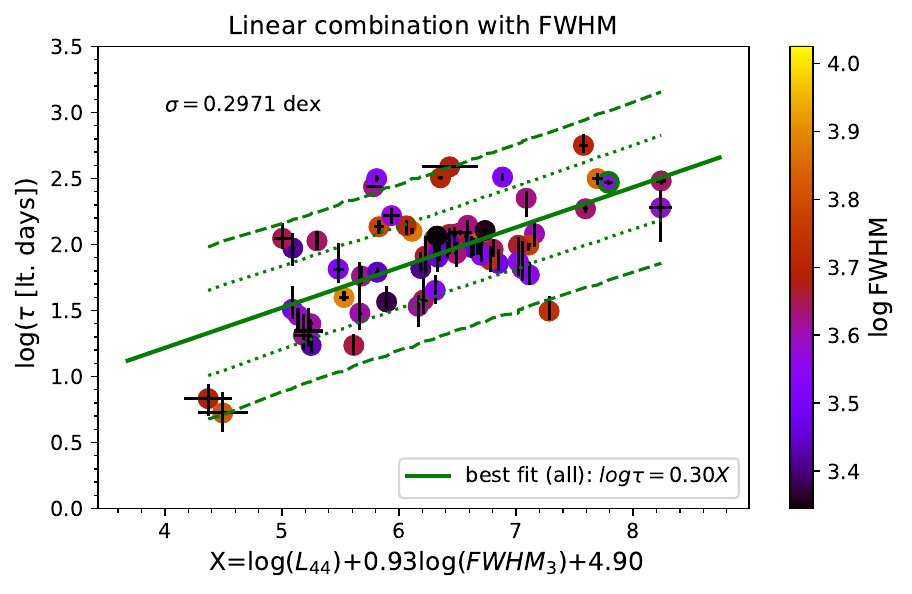}
    \includegraphics[width=\columnwidth]{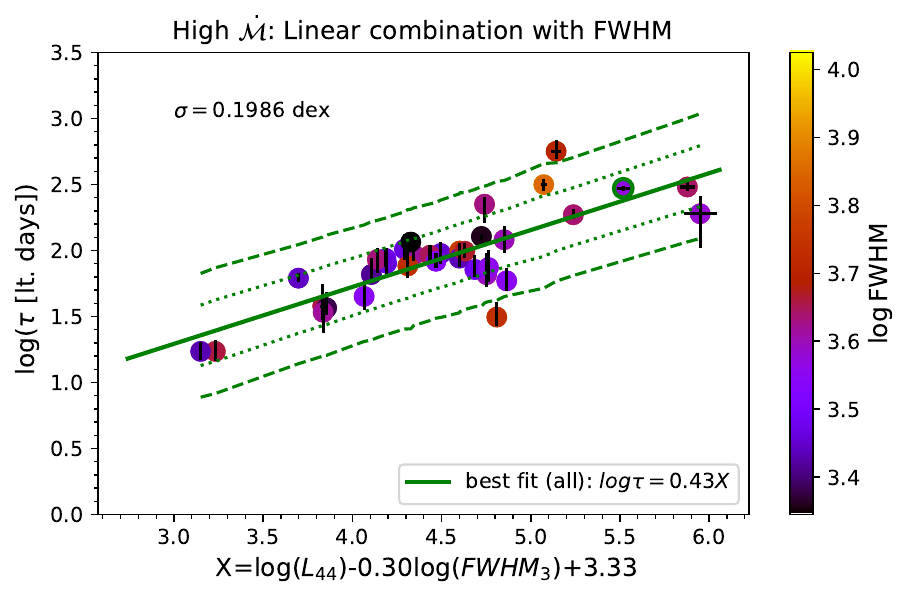}
    \caption{Rest-frame time delay expressed as a function of $\log{(L_{44})}$ and $\log{(\text{FWHM}_3)}$. {\bf Left panel:} The linear combination studied for all 68 MgII reverberation-mapped sources studied in \citet{Mart_nez_Aldama_2020} and a new source HE 0435-4312 denoted by a green circle. {\bf Right panel:} As in Figure~\ref{fig_RL_high}, the linear combination is restricted to highly-accreting sources with the same division according to $\dot{\mathcal{M}}$. In both panels, each source is colour-coded according to the corresponding MgII FWHM.}
    \label{fig_RL_FWHM_high}
\end{figure*}

\begin{figure*}[h!]
    \centering
    \includegraphics[width=\columnwidth]{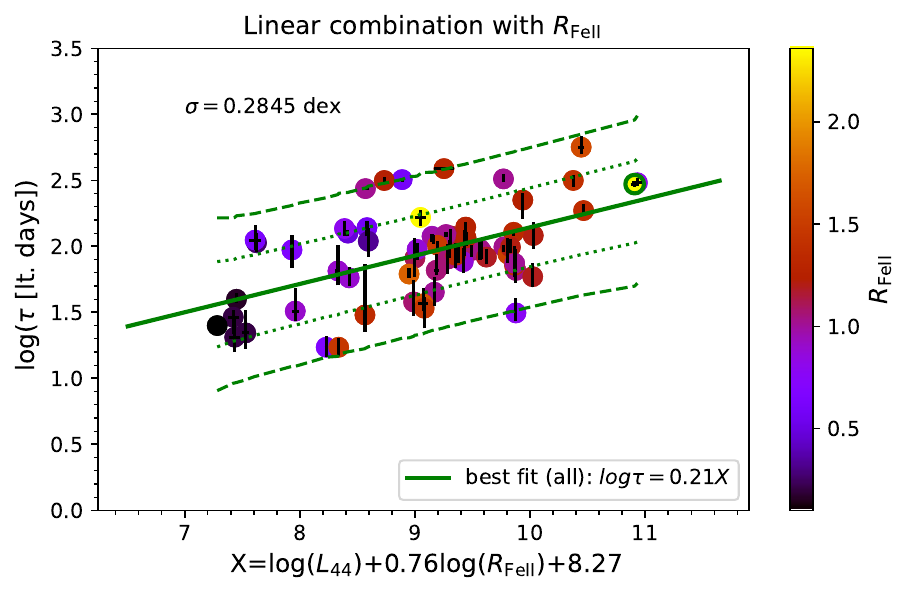}
    \includegraphics[width=\columnwidth]{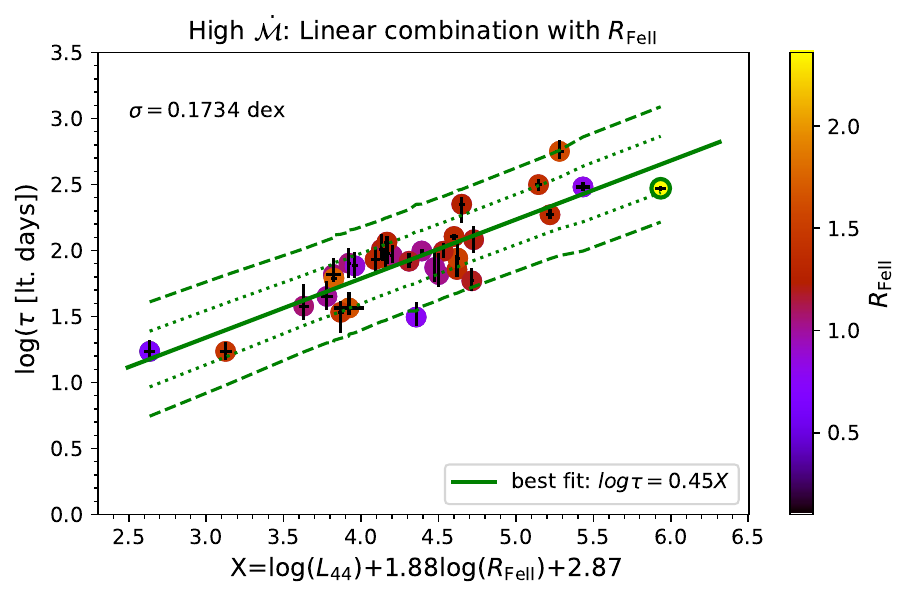}
    \caption{Rest-frame time delay expressed as a function of $\log{(L_{44})}$ and $\log{(R_{\rm FeII})}$. {\bf Left panel:} The linear combination studied for all 68 MgII reverberation-mapped sources studied in \citet{Mart_nez_Aldama_2020} and a new source HE 0435-4312 denoted by a green circle. {\bf Right panel:} As in Figure~\ref{fig_RL_high}, the linear combination is restricted to highly-accreting sources with the same division according to $\dot{\mathcal{M}}$. In both panels, each source is colour-coded according to the corresponding relative UV FeII strength $R_{\rm FeII}$.}
    \label{fig_RL_RFeII_high}
\end{figure*}

\begin{figure*}[h!]
    \centering
    \includegraphics[width=\columnwidth]{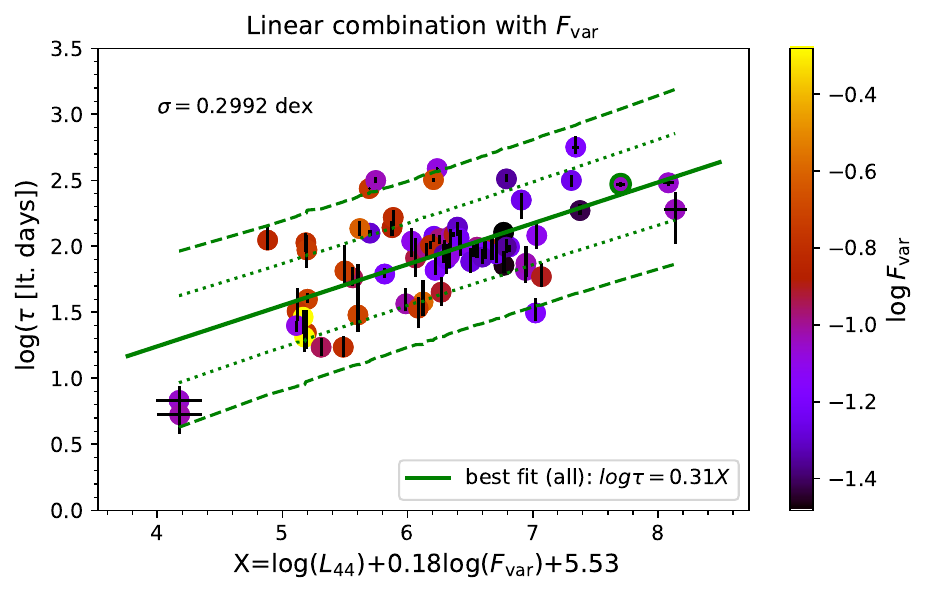}
    \includegraphics[width=\columnwidth]{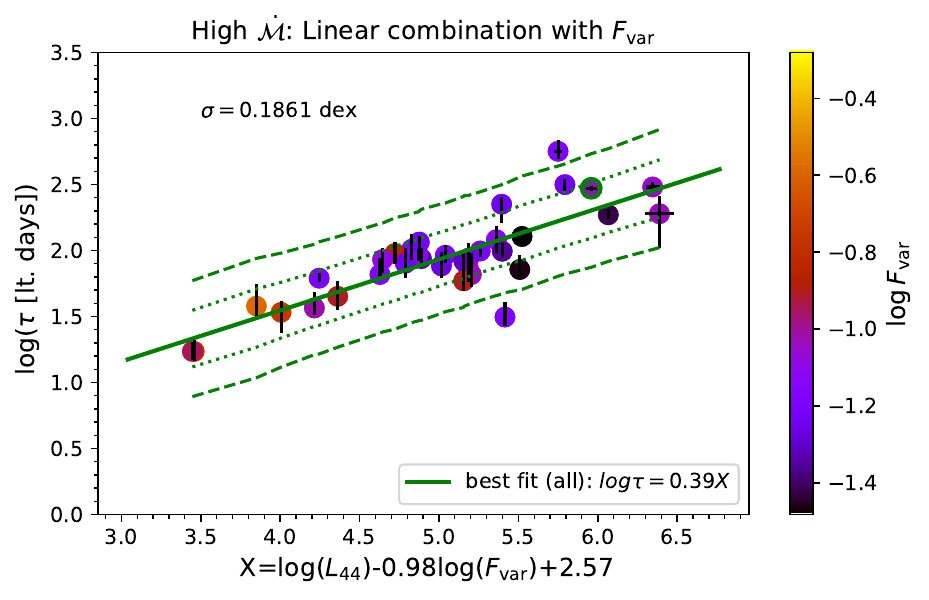}
    \caption{Rest-frame time delay expressed as a function of $\log{(L_{44})}$ and $\log{(F_{\rm var})}$. {\bf Left panel:} The linear combination studied for all 68 MgII reverberation-mapped sources studied in \citet{Mart_nez_Aldama_2020} and a new source HE 0435-4312 denoted by a green circle. {\bf Right panel:} As in Figure~\ref{fig_RL_high}, the linear combination is restricted to highly-accreting sources with the same division according to $\dot{\mathcal{M}}$. In both panels, each source is colour-coded according to the corresponding continuum fractional variability $F_{\rm var}$.}
    \label{fig_RL_RFvar_high}
\end{figure*}

In the extended RL relations that include other independent observables in the linear combination of logarithms, namely MgII line FWHM, the relative FeII strength with respect to MgII line $R_{\rm FeII}$, and the continuum fractional variability $F_{\rm var}$, HE 0435-4312 lies within 1-$\sigma$ of the mean relation for both the full sample and the high-accretion subsample, see Fig.~\ref{fig_RL_FWHM_high} for the combination with FWHM, Fig.~\ref{fig_RL_RFeII_high} that includes $R_{\rm FeII}$, and Fig.~\ref{fig_RL_RFvar_high} that utilizes the continuum $F_{\rm var}$. When one restricts the analysis to the high-accretion subsample, the RMS scatter drops below $0.2$ dex in all combinations, with the smallest scatter exhibited by the RL relation including $R_{\rm FeII}$ ($\sigma_{\rm rms}=0.1734$ dex). In Table~\ref{tab_multivariate_regression_HE0435}, we summarize the best-fit parameters as well as the RMS scatter for all the RL relations for the high-accretion subsample including HE 0435-4312. Our high-luminosity source is beneficial for the reduction of RMS scatter, except for the combination including $R_{\rm FeII}$, where the RMS scatter marginally increases in comparison with the best-fit result of \citet{Mart_nez_Aldama_2020}. For all RL relations, adding the new quasar leads to the increase in the Pearson's correlation coefficient. For the comparison of the RMS scatter and the correlation coefficients with and without the inclusion of HE 0435-4312, see the last four columns of Table~\ref{tab_multivariate_regression_HE0435}.

\begin{table*}[]
  \caption{Parameters for the standard RL relation as well as multidimensional RL relations including independent observables, namely FWHM, $R_{\rm FeII}$, and $F_{\rm var}$, for the high-accretion subsample (35 sources including  HE 0435-4312). The last four columns include RMS scatter and Pearson's correlation coefficient with and without HE 0435-4312 for comparison.}
  \hspace{-3cm}
     \resizebox{1.2\textwidth}{!}{
    \begin{tabular}{c|c|c|c|c|c|c|c}
    \hline
    \hline
    $\log{\tau_{\rm obs}}=$  & $K_1$ & $K_2$ & $K_3$ & $\sigma_{\rm rms}$ [dex] & $r$ & $\sigma_{\rm rms}$ [dex] (without) & $r$ (without)\\
    \hline
  $K_1\log{L_{44}}+K_2$ & $0.422 \pm 0.055$  & $1.374 \pm 0.082$ & - &  $0.1991$ & $0.80$  &  $0.2012$ & $0.78$  \\
   $K_1\log{L_{44}}+K_2\log{{\rm FWHM_3}}+K_3$ & $0.43 \pm 0.06$  & $-0.13 \pm 0.31$  & $1.44 \pm 0.17$  & $0.1986$ &  $0.80$ & $0.2007$ &  $0.78$\\
     $K_1\log{L_{44}}+K_2\log{R_{\rm FeII}}+K_3$ & $0.45\pm 0.05$ & $0.84 \pm 0.29$ & $1.28 \pm 0.08$ & $0.1734$ & $0.85$ & $0.1718$ & $0.84$\\
  $K_1\log{L_{44}}+K_2\log{F_{\rm var}}+K_3$ & $0.39 \pm 0.06$ & $-0.38 \pm 0.18$ & $0.99 \pm 0.19$ & $0.1861$ & $0.83$ & $0.1863$ & $0.82$\\  \hline
     \hline
    \end{tabular}}
    \label{tab_multivariate_regression_HE0435}
\end{table*}

The dimensionless accretion rate $\dot{\mathcal{M}}$ of MgII sources defined by Eq.~\ref{eq_mdot_3000} is intrinsically correlated with the rest-frame time-delay via the black hole mass ($\dot{\mathcal{M}}\propto \tau^{-2}$, while the Eddington ratio $\eta\propto \tau^{-1}$) as discussed by \citet{Mart_nez_Aldama_2020}. Therefore we do not use $\dot{\mathcal{M}}$ in extended RL relations, as the correlation would be artificially enhanced. We merely use $\dot{\mathcal{M}}$ for the division of the sample into the high- and low-accretors. For the extended RL relations, we prefer the independent observables as MgII FWHM, UV $R_{\rm FeII}$, and $F_{\rm var}$. Characterizing the accretion-rate intensity by $\dot{\mathcal{M}}$ is justified by its correlation with the relative UV FeII strength $R_{\rm FeII}$, which in turn is related to the accretion rate \citep{2011ApJ...736...86D,Mart_nez_Aldama_2020}, which was analogously shown for optical FeII strength \citep{2014Natur.513..210S,2019ApJ...886...42D}. For the whole sample of MgII sources, including HE 0435-4312, for which $R_{\rm FeII}$ can be estimated (in total 66 sources), the Spearman correlation coefficient between  $\dot{\mathcal{M}}$ and $R_{\rm FeII}$ is positive with $\rho=0.440$ (with the $p$-value of $2.19\times 10^{-4}$). For the same sample, the relative FeII strength is in turn positively correlated with the Eddington ratio $\eta=L_{\rm bol}/L_{\rm Edd}$ with $\rho=0.447$ ($p=1.71 \times 10^{-4}$). This justifies the division into low- and high-accretors according to $\dot{\mathcal{M}}$.

On the other hand, $\dot{\mathcal{M}}$ is anticorrelated with respect to the continuum variability $F_{\rm var}$ with $\rho=-0.480$ ($p=2.97 \times 10^{-5}$, 69 sources). Furthermore, the parameter $\dot{\mathcal{M}}$ is anticorrelated with MgII FWHM, although the (anti)correlation is weaker in comparison with $R_{\rm FeII}$ and $F_{\rm var}$ with $\rho=-0.386$ ($p=1.05 \times 10^{-3}$, 69 sources). The anticorrelation between MgII FWHM and $\dot{\mathcal{M}}$, and hence also $R_{\rm FeII}$, is expected from optical studies of Eigenvector 1 of the quasar main sequence \citep{1992ApJS...80..109B,sulentic2000}, which can be traced in the UV plane as well \citep{2020ApJ...900...64S}. These (anti)correlations between $\dot{\mathcal{M}}$ and other quantities across classical and extended RL relations are also apparent in the colour-coded plots in Figs.~\ref{fig_RL_high}, \ref{fig_RL_FWHM_high}, \ref{fig_RL_RFeII_high}, and \ref{fig_RL_RFvar_high}. The scatter reduction for high-$\dot{\mathcal{M}}$ subsample could therefore arise due to a lower variability and the stabilisation of the luminosity and black hole mass ratio for the sources accreting close to the Eddington limit as discussed in detail in \citet{Mart_nez_Aldama_2020}; see also \citet{2010ApJ...716L..31A} and \citet{2014MNRAS.442.1211M}.

\begin{figure*}
    \centering
    \includegraphics[width=\textwidth]{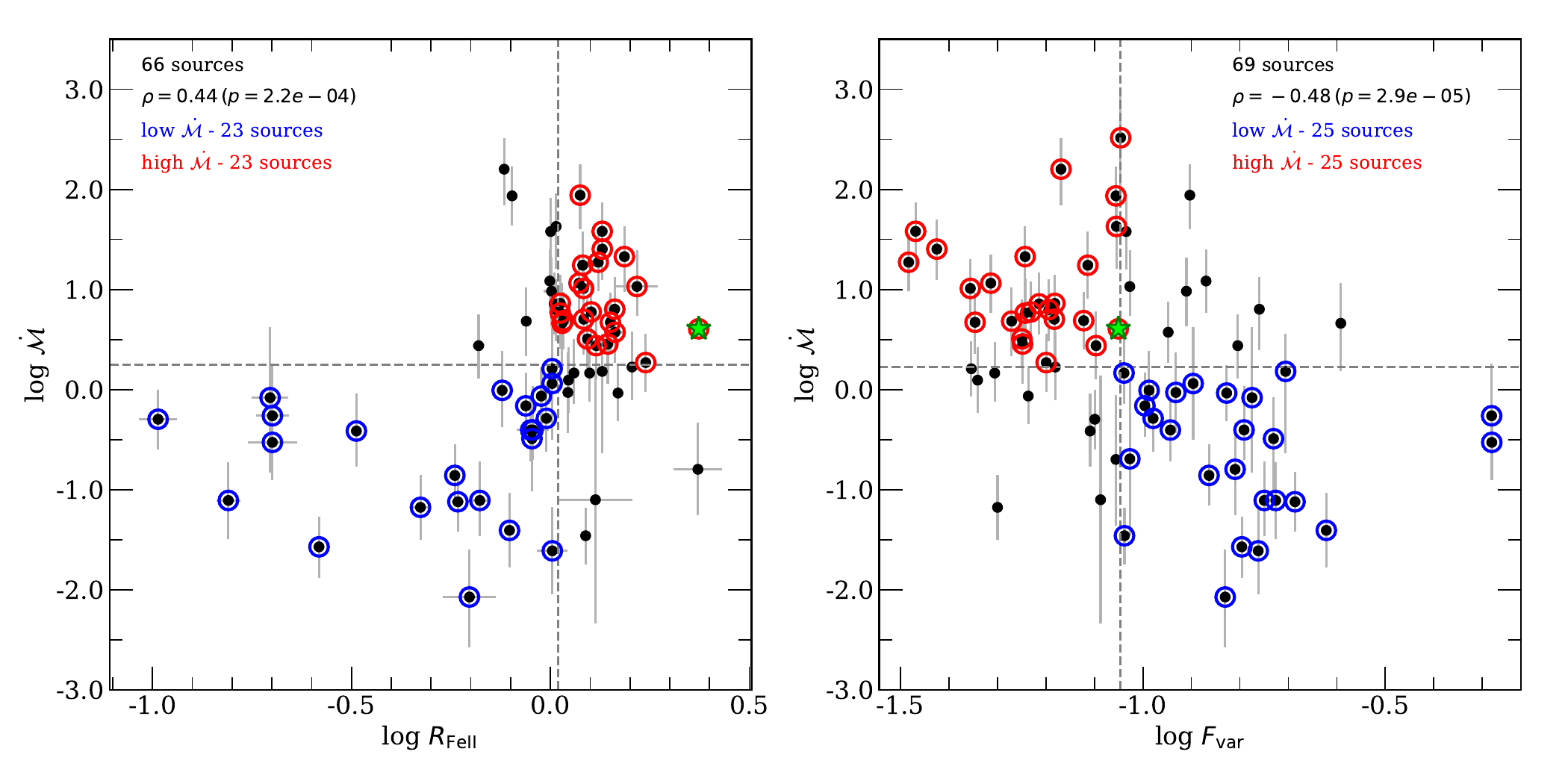}
    \caption{Relation between the dimensionless accretion rate $\dot{\mathcal{M}}$ and the relative UV FeII strength $R_{\rm FeII}$ and the continuum fractional variability $F_{\rm var}$. {\bf Left panel:} The positive correlation between $\dot{\mathcal{M}}$ and $R_{\rm FeII}$ (Spearman correlation coefficient $\rho=0.44$; $p=2.2\times 10^{-4}$). The dashed lines mark the median values of each parameter. When the sample is divided according to the median of $R_{\rm FeII}$, $\sim 70\%$ of high-/low-$\dot{\mathcal{M}}$ sources are present. The star symbol marks HE 0435-4312. {\bf Right panel:} The anticorrelation between $\dot{\mathcal{M}}$ and $F_{\rm var}$ (Spearman correlation coefficient $\rho=-0.48$; $p=2.9\times 10^{-5}$). The dashed lines mark the median values of each parameter. When the MgII sample is divided according to the median value of $F_{\rm var}$, $\sim 72\%$ of high-/low-$\dot{\mathcal{M}}$ sources are included. The star symbol depicts HE 0435-4312.}
    \label{fig_mdot_correlation}
\end{figure*}

In order to clarify our idea of the sample division quantitatively, we divided the sample analyzed by \citet{Mart_nez_Aldama_2020}, with the studied source HE 0435-4312 included, considering the $R_{\rm FeII}$ and $F_{\rm var}$ median values ($\log{R_{\rm FeII}}=0.02$, $\log{F_{\rm var}}=-1.05$) instead of $\dot{\mathcal{M}}$ ($\log{\dot{\mathcal{M}}}=0.23$). Since the correlation between $\dot{\mathcal{M}}$ and $R_{\rm FeII}$ is positive, we expect that the high-FeII emitters show the highest $\dot{\mathcal{M}}$ values, i.e. above the corresponding median value. On the other hand, the correlation between $F_{\rm var}$ and $\dot{\mathcal{M}}$ is negative, hence highly accreting AGNs show a lower variability or $F_{\rm var}$ values with respect to the median value of each parameter. In the case of the $\dot{\mathcal{M}}$-$R_{\rm FeII}$ correlation, we find that $\sim 70\%$ of the sample satisfies the criteria with respect to the median values of $R_{\rm FeII}$ and $\dot{\mathcal{M}}$.  Meanwhile, for the $\dot{\mathcal{M}}$-$F_{\rm var}$ correlation, $\sim 72\%$ of our sources are within the median values considered. This is graphically shown in Fig.~\ref{fig_mdot_correlation}, where we depict the correlation $\dot{\mathcal{M}}$--$R_{\rm FeII}$ (left panel) and the anticorrelation $\dot{\mathcal{M}}$--$F_{\rm var}$ (right panel) for the studied sample of MgII sources. Therefore, we can conclude that the division into low and high accretors based on $\dot{\mathcal{M}}$ is analogous to a division based on the independent parameters $R_{\rm FeII}$ and $F_{\rm var}$. To illustrate this, in Fig.~\ref{fig_RL_high} we use $R_{\rm FeII}$ for colour-coding individual sources in the RL relation for the whole sample (left panel) as well as the high-accretor subsample (right panel).

\section{Discussion} \label{sec:discussion}

By combining the SALT spectroscopy and the photometric data from more instruments, we were able to determine the rest-frame time-delay of $296^{+13}_{-14}$ days between MgII broad line and the underlying continuum at 3000\,\AA. Fitting MgII line with the underlying continuum and FeII pseudocontinuum is rather complex, but we showed in Appendix~\ref{sect:redshift} that the time-delay distribution for a different FeII template is comparable and does not affect the main time-delay analysis presented in this paper.

The rest-frame time-delay is essentially the same as for another luminous and highly-accreting quasar HE~0413-4031 analysed in \citet{2020ApJ...896..146Z}. While the RL relation for the collected sample of 69 MgII reverberation-mapped sources has a relatively large scatter of $\sim 0.3$ dex, it is significantly reduced to $0.2$ dex when we consider only high-accretors, where also HE~0435-4312 with the Eddington ratio of $\sim 0.31-0.58$ belongs (the corresponding dimensionless accretion rate is $\dot{\mathcal{M}}=4.0^{+0.7}_{-0.6}$). The further reduction of the scatter is achieved in the 3D RL relations that include independent observables (FWHM, $R_{\rm FeII}$, and $F_{\rm var}$). Especially, the linear combination including the FeII relative strength leads to the smallest scatter of $0.17$ dex. Because of the correlation between $R_{\rm FeII}$ and $\dot{\mathcal{M}}$ \citep{Mart_nez_Aldama_2020}, this indicates that the scatter along the RL relation is driven by the accretion rate, as we already showed in \citet{2020ApJ...896..146Z} for a sample of only 11 MgII sources.   

In the following subsection, we further discuss some aspects of the MgII variability, focusing on the interpretation of a relatively large variability of MgII line for our source. Then, using the sample of 35 high accretors that exhibit a small scatter along the RL relation, we show how it is possible to apply the radius-luminosity relation to constrain cosmological parameters.

\subsection{Variability of MgII emission}

The fractional variability of the MgII line of HE 0435-4312 is $F_{\rm var}^{\rm line}\simeq 5.4\%$ (excluding observation 19). The continuum fractional variability is larger, $F_{\rm var}^{\rm cont}\simeq 8.8\%$ during $14.74$ years. However, when the continuum light curve is separated into the first 7 years (CATALINA), $F_{\rm var}^{\rm cont1}\simeq 4.8\%$, and the last 6 years (SALT monitoring), $F_{\rm var}^{\rm cont2}\simeq 4.8\%$, then both values are comparable to the fractional variability of MgII emission. This implies that MgII-emitting clouds reprocess the continuum emission very well for our source. In other words, both the triggering continuum and a line echo have similar amplitudes, suggesting that sharp echoes are present and that the MgII emitting region lies on (or close to) an iso-delay surface, which is an important geometrical condition. In addition, the variability time scale of the continuum must be long enough, i.e. longer than the light-travel time through the locally optimally-emitting cloud (LOC) model of the BLR. It seems that for our source, both conditions are fulfilled. 

This is in contrast with the study of \citet{Guo2020} who used the CLOUDY code and the LOC model to study the response of MgII emission to the variable continuum. They found that at the Eddington ratio of $\sim 0.4$, the MgII emission saturates and does not further increase with the rise in the continuum luminosity. Observationally, \citet{2020MNRAS.493.5773Y} found for extreme-variability quasars that MgII emission responds to the continuum but with a smaller amplitude, $\Delta \log L(MgII)=(0.39 \pm 0.07) \Delta \log L(3000\AA)$. Although our source has the high Eddington ratio comparable to that studied in \citet{Guo2020}, $\eta\sim 0.3-0.6$, its MgII emission responds very efficiently to the variable continuum. For the previous highly-accreting luminous quasar HE~0413-4031 that we studied \citep{2020ApJ...896..146Z}, we also showed that MgII line can respond strongly to the continuum increase, $\Delta \log L(MgII)=(0.82 \pm 0.26) \Delta \log L(3000\AA)$, as shown by the intrinsic Baldwin effect. 

The possible interpretation of the discrepancy between the saturation of MgII emission at larger Eddington ratios modelled nominally by \citet{Guo2020} and our two highly-accreting sources, HE~0435-4312 and HE~0413-4031, which exhibit a strong response of the MgII emission to the continuum, is the order of magnitude difference in the black hole mass as well as the studied luminosity at 3000\AA. In \citet{Guo2020}, they used $M_{\bullet}=10^8\,M_{\odot}$ and $L_{3000}=10^{44}-10^{45}\,{\rm erg\,s^{-1}}$ as well as $(R_{\rm out},\Gamma)=(10^{17.5},-2)$ for the outer radius and the slope of the LOC model, respectively. For our source, all of the characteristic parameters are increased, $M_{\bullet}\simeq 2\times 10^9\,M_{\odot}$, $L_{3000}\simeq 10^{46.36}\,{\rm erg\,s^{-1}}$, and $R_{\rm MgII}=10^{17.9}\,{\rm cm}$. Mainly the larger outer radius implies that not all of the MgII-emitting gas is fully ionized at these scales and can exhibit ``partial breathing'' as is also shown by \citet{Guo2020} for the case with $R_{\rm out}=10^{18}\,{\rm cm}$, when the MgII line luminosity continues to rise with the increase in the continuum luminosity.

\subsection{Constraining cosmological parameters using MgII high-accretion subsample}
\label{subsec_cosmology}

\begin{figure*}
    \centering
    \includegraphics[width=\columnwidth]{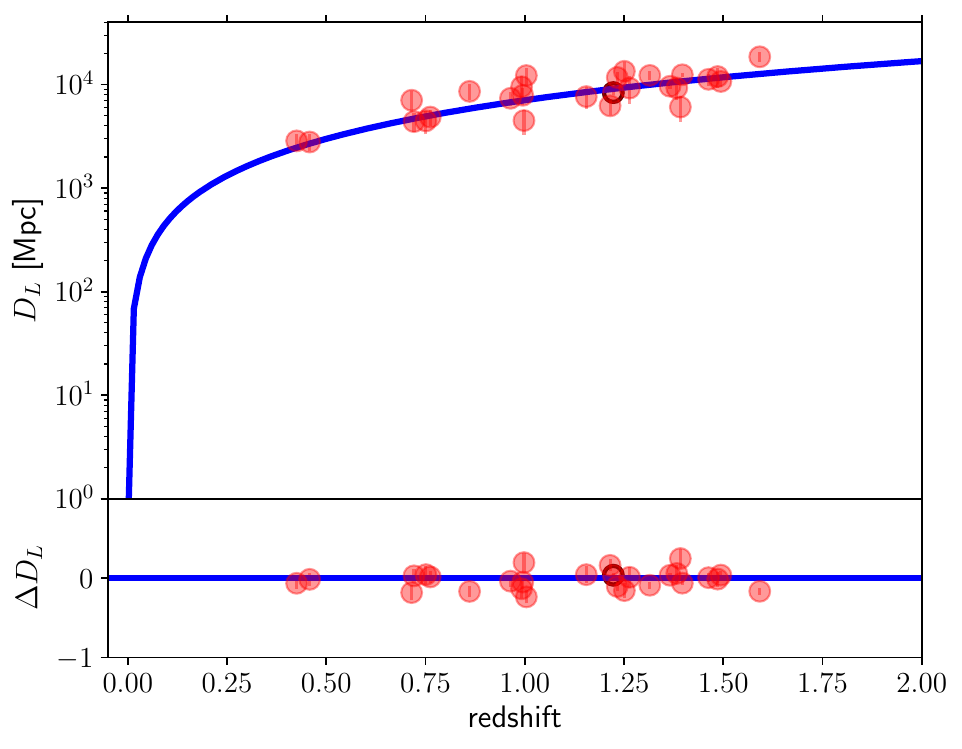}
    \includegraphics[width=\columnwidth]{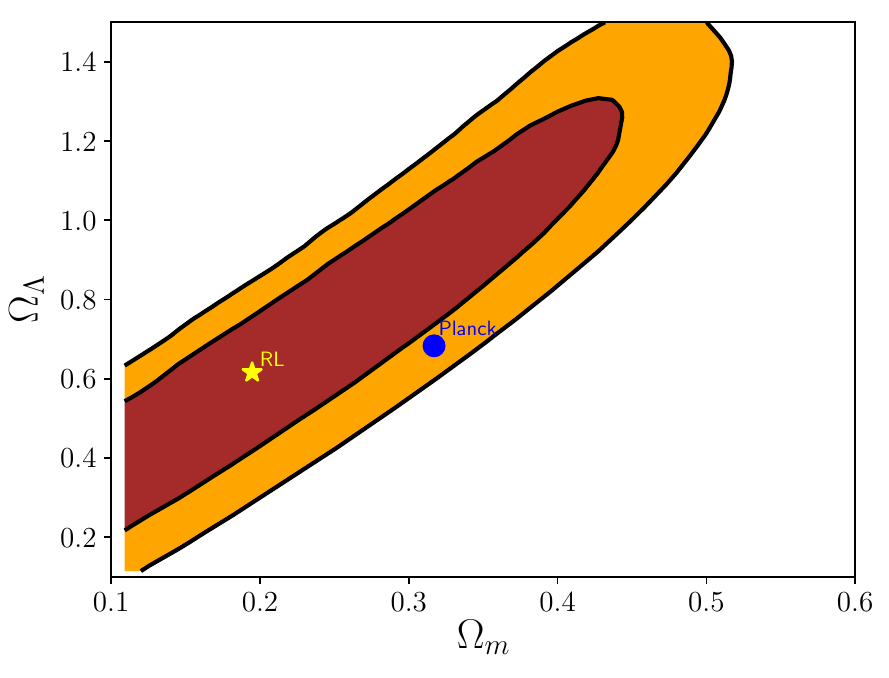}
    \caption{Constraining cosmological parameters using the highly-accreting subsample of MgII reverberation-mapped sources, including HE~0435-4312. {\bf Left panel:} The quasar Hubble diagram using 27 sources (red points), after removal of 8 outliers, with the best-fit standard $\Lambda$CDM model (black solid line; $\Omega_{\rm m} = 0.252^{+0.051}_{-0.044}$ with the fixed $H_0 = 67.36\,{\rm km\,s^{-1}\,Mpc^{-1}}$ according to \citeauthor{2020A&A...641A...6P}, \citeyear{2020A&A...641A...6P}). The bottom panel shows the residuals $\Delta D_{\rm L}=\log(D_{\rm L}^{\rm exp}/D_{\rm L})$, where $D_{\rm L}^{\rm exp}$ is the expected luminosity distance and $D_{\rm L}$ is the measured one. HE~0435-4312 is denoted with a black circle. {\bf Right panel:} The contour plot showing the confidence levels at 68\% (orange) and 95\% (brown) for the general $\Lambda$CDM model based on $\chi^2$ fitting. The yellow star depicts the best-fit $(\Omega_{\rm m}, \Omega_{\Lambda})=(0.19,0.62)$ and the blue filled circle denotes the cosmological constraints as provided by \citet{2020A&A...641A...6P}.}
    \label{fig_cosmology_mgII}
\end{figure*}

Since the RL relation for a high-accretion MgII subsample showed a relatively small scatter of $\sim 0.2$ dex, we attempt to use it for the cosmology purposes. We adopt the general approach outlined in \citet{Mart_nez_Aldama_2019}, that is we assume a perfect relation between the absolute luminosity and the measured time delay, and having absolute luminosity and the observed flux we can determine the luminosity distance for each source. We do not use here the best-fit relation given in Fig.~\ref{fig_RL_high}, right panel, since this relation was calibrated for a specific assumed cosmology. Therefore we assume a relation in a general form
\begin{equation}
    \log L_{3000} = \alpha \log \tau - \beta + 44,
    \label{eq_L3000_tau}
\end{equation}
and we treat $\alpha$ and $\beta$ as free parameters. We consider first the case of a flat cosmology, and we assume the value of the Hubble constant $H_0 = 67.36\,{\rm km\,s^{-1}\,Mpc^{-1}}$ adopted from \citet{2020A&A...641A...6P}, and we minimize the fits to the predicted luminosity distance by varying $\Omega_{\rm m}$, $\alpha$, and $\beta$. The preliminary fit was highly unsatisfactory, and we used a sigma-clipping method to remove the outliers (8 sources). The final sample was well fitted with the standard $\Lambda$CDM model, for $\Omega_{\rm m} = 0.252^{+0.051}_{-0.044}$ (1-$\sigma$ error). This result is fully consistent with the \citet{2020A&A...641A...6P} data with $\Omega_{\rm m}^{\rm Planck}= 0.3153 \pm 0.0073$. The obtained values for the remaining two parameters were $\alpha = 1.8$ and $\beta = 2.1$, which would correspond to a slightly different RL relation, $\log \tau = 0.56 \log L_{3000} +1.18$. The results are illustrated in Fig.~\ref{fig_cosmology_mgII}, left panel. We also attempted to constrain both $\Omega_{\Lambda}$ and $\Omega_{\rm m}$, waiving the assumption of a flat space. The parameters $\alpha$ and $\beta$ in Eq.~\ref{eq_L3000_tau} were kept fixed to the values inferred from the flat-cosmology fit, i.e. $\alpha=1.8$ and $\beta=2.1$. The result is given in the right panel of Fig.~\ref{fig_cosmology_mgII}. The best fit in this case implies a slightly lower $\Omega_{\rm m}$, but the contour error is large and covers the best fit from \citet{2020A&A...641A...6P} within 1$\sigma$ confidence level. The data do not yet tightly constrain the cosmological parameters, but the MgII high-accretion subsample does not imply any departures from the standard model, and the data quality is much higher, and constraints better in comparison with the larger sample based on H$\beta$ line and discussed in \citet{Mart_nez_Aldama_2019}, and somewhat better than from the mixed sample discussed in \citet{PTF100}.

\section{Conclusions} \label{sec:conclusions}

Our main conclusions of the reverberation-mapping analysis of the source HE 0435-4312 ($z=1.2231$, $\log{(L_{3000}[{\rm erg\,s^{-1}}])}=46.359^{+0.038}_{-0.034}$)  can be summarized as follows:
\begin{itemize}
    \item Using seven different methods, we determined the mean rest-frame time-delay of $296^{+13}_{-14}$ days, with all the methods giving the consistent time-delay peak within uncertainties that were determined using the bootstrap or the maximum-likelihood method. By the combination of the bootstrap method, the weighting by the number of overlapping data points, and the analysis of mock light curves, we could classify the other prominent time-delay peaks as artefacts due to the particular sampling and a limited observational duration. 
    \item The fractional variability of MgII emission ($\sim 5.4\%$) is comparable to the continuum variability, hence for our source, the MgII line flux has not reached the saturation level despite the high Eddington ratio of $\sim 0.3-0.6$. This is most likely due to the large black hole mass of $\sim 2 \times 10^9\,M_{\odot}$ and the large extent of the MgII-emitting region, $R_{\rm MgII}=10^{17.9}\,{\rm cm}$.
    \item Given the large luminosity of HE 0435-4312, it is beneficial for decreasing the scatter and increasing the correlation coefficient for the MgII-based radius-luminosity relation. The scatter is $\sim 0.2$ dex for the high-accreting subsample of MgII sources, where HE 0435-4312 belongs. A further reduction in the scatter is achieved using linear combinations with independent observables (FWHM, relative FeII strength, and fractional variability), which indicates that the scatter along the radius-luminosity relation is mainly driven by the accretion rate.
    \item A low scatter of the radius-luminosity relation for the high-accreting subsample motivates us to apply these sources for constraining cosmological parameters. Given the current number of sources (27, after removal of 8 outliers) and the data quality, the best-fit values of the cosmological parameters $(\Omega_{\rm m}, \Omega_{\Lambda})=(0.19; 0.62)$ are consistent with the standard cosmological model within 1$\sigma$ confidence level. 
\end{itemize}

\software{IRAF \citep{1986SPIE..627..733T,1993ASPC...52..173T},  JAVELIN \citep{2011ApJ...735...80Z,2013ApJ...765..106Z,2016ApJ...819..122Z}, PyCCF \citep{2018ascl.soft05032S}, vnrm.py \citep{2017ApJ...844..146C}, zdcf\_v2.f90 \citep{1997ASSL..218..163A}, plike\_v4.f90 \citep{1997ASSL..218..163A}, delay\_chi2.f \citep{2013A&A...556A..97C}, sklearn \citep{scikit-learn}, statsmodels \citep{seabold2010statsmodels}, numpy \citep{numpy}, matplotlib \citep{hunter07}}

\acknowledgments

The  project  is  based  on  observations made with the SALT under programs 2012-2-POL-003, 2013-1-POL-RSA-002,  2013-2-POL-RSA-001,  2014-1-POL-RSA-001,  2014-2-SCI-004,  2015-1-SCI-006,  2015-2-SCI-017, 2016-1-SCI-011, 2016-2-SCI-024, 2017-1-SCI-009, 2017-2-SCI-033, 2018-1-MLT-004 (PI: B. Czerny). 
The authors acknowledge the financial support by the National Science Centre, Poland, grant No.~2017/26/A/ST9/00756 (Maestro 9), and by the Ministry of Science and Higher Education (MNiSW) grant DIR/WK/2018/12. GP acknowledges the grant MNiSW DIR/WK/2018/09. KH acknowledges support by the Polish National Science Centre grant
2015/18/E/ST9/00580. MZ was supported by the NAWA under the agreement PPN/WYM/2019/1/00064 to
perform a three-month exchange stay at the Charles University and the Astronomical Institute of the Czech Academy of Sciences in Prague. MH and CSF are supported by Deutsche Forschungsgemeinschaft grants CH71/33 and CH71/34. The OGLE project has received funding from the National Science
Centre, Poland, grant MAESTRO 2014/14/A/ST9/00121. The Polish participation in SALT is funded by grant No. MNiSW DIR/WK/2016/07. MZ, VK, and BC acknowledge PPN/BCZ/2019/1/00069 (M\v{S}MT 8J20PL037) for the support of the Polish-Czech mobility programme. This paper uses observations made at the South African Astronomical Observatory (SAAO).

%




\appendix

\setcounter{table}{0}
\renewcommand\thetable{\Alph{section}.\arabic{table}}

\section{Photometric and spectroscopic data}
\label{sec_photometry_spectroscopy}

\begin{center}
\begin{longtable}[c]{c|c|c|c}
\caption{Table of continuum magnitudes with uncertainties. The epoch is given in Julian Dates ($-2450 000$). The last column denotes three different instruments used to obtain the photometry data: 1. CATALINA, 2. SALTICAM, 3.  OGLE, 4. BMT. The BMT photometry points were shifted by $0.2$ mag toward lower magnitudes to optimize the match with the OGLE photometry.} 
\label{tab:phot1} \\
    \hline
    \hline
{JD} & {magnitude ($V$-band)}  & {Error} & {Instrument} \\
-2 450 000 & [mag] & [mag] & No. \\
     \hline
   3696.98828   &    16.8733330   &    2.51584481E-02  & 1 \\
   4067.76562   &    16.9422932   &    1.88624337E-02  & 1 \\
   4450.48828   &    16.9090614   &    2.42404137E-02  & 1 \\
   4837.18359   &    17.0258694   &    1.94998924E-02  & 1 \\
   5190.14453   &    17.0322189   &    2.00808570E-02  & 1 \\
   5564.26562   &    16.9177780   &    2.75434572E-02  & 1 \\
   5889.63672   &    16.8666668   &    2.98607871E-02  & 1 \\
   6296.05078   &    16.9212513   &    2.82704569E-02  & 1 \\
   6893.60449   &    17.1165180   &    1.20000001E-02  & 2 \\
   6986.31592   &    17.1647282   &    1.20000001E-02  & 2 \\
   7032.45020   &    17.1936970   &    1.20000001E-02  & 2 \\
   7035.63428   &    17.1329994   &    4.00000019E-03  & 3 \\
   7047.64209   &    17.1079998   &    4.00000019E-03  & 3 \\
   7058.61475   &    17.1240005   &    8.00000038E-03  & 3 \\
   7083.54346   &    17.1119995   &    4.99999989E-03  & 3 \\
   7116.50684   &    17.1089993   &    8.00000038E-03  & 3 \\
   7253.89209   &    17.1749992   &    4.00000019E-03  & 3 \\
   7261.88330   &    17.1779995   &    4.00000019E-03  & 3 \\
   7267.91504   &    17.1630001   &    4.99999989E-03  & 3 \\
   7273.84717   &    17.1849995   &    4.99999989E-03  & 3 \\
   7283.84912   &    17.1779995   &    4.00000019E-03  & 3 \\
   7295.84277   &    17.1889992   &    6.00000005E-03  & 3 \\
   7306.78125   &    17.1690006   &    4.00000019E-03  & 3 \\
   7317.74023   &    17.1889992   &    4.99999989E-03  & 3 \\
   7327.77490   &    17.1959991   &    6.00000005E-03  & 3 \\ 
   7340.70654   &    17.2140007   &    4.00000019E-03  & 3\\ 
   7355.69482   &    17.1860008   &    4.99999989E-03  & 3 \\
   7363.66650   &    17.2159996   &    4.00000019E-03  & 3 \\
   7364.53760   &    17.2753944   &    1.20000001E-02  & 2 \\
   7374.70947   &    17.2080002   &    4.00000019E-03  & 3 \\
   7385.55762   &    17.2049999   &    4.00000019E-03  & 3 \\
   7398.61768   &    17.2269993   &    4.00000019E-03  & 3 \\
   7415.58545   &    17.2229996   &    4.00000019E-03  & 3 \\
   7426.56641   &    17.2010002   &    4.00000019E-03  & 3 \\
   7436.52539   &    17.2049999   &    4.99999989E-03  & 3 \\
   7447.52783   &    17.2080002   &    4.00000019E-03  & 3 \\
   7457.52246   &    17.2089996   &    4.00000019E-03  & 3 \\
   7655.49414   &    17.1270580   &    1.20000001E-02  & 2 \\
   7692.38037   &    17.1313496   &    1.20000001E-02  & 2 \\
   7717.70557   &    17.1089993   &    4.00000019E-03  & 3 \\
   7754.46582   &    17.0703278   &    1.20000001E-02  & 2 \\
   7803.32959   &    17.1150951   &    1.20000001E-02  & 2 \\
   7973.91357   &    17.2000008   &    1.30000003E-02  & 3 \\
   7984.59424   &    17.2056389   &    1.20000001E-02  & 2 \\
   8038.86182   &    17.1779995   &    6.00000005E-03  & 3 \\
   8091.25000   &    17.2399998   &    9.99999978E-03  & 4 \\
   8102.25000   &    17.2199993   &    4.00000019E-03  & 4 \\
   8104.25000   &    17.2430000   &    8.99999961E-03  & 4 \\ 
   8105.00000   &    17.2430000   &    4.99999989E-03  & 4\\
   8107.25000   &    17.2509995   &    6.00000005E-03  & 4 \\
   8112.48926   &    17.2262497   &    1.20000001E-02  & 2 \\
   8130.25000   &    17.2639999   &    8.99999961E-03  & 4 \\
   8138.00000   &    17.2459984   &    7.00000022E-03  & 4 \\
   8147.00000   &    17.2319984   &    8.00000038E-03  & 4 \\
   8166.00000   &    17.2319984   &    2.00000009E-03  & 4 \\
   8182.00000   &    17.2519989   &    8.00000038E-03  & 4 \\
   8206.00000   &    17.2309990   &    8.99999961E-03  & 4 \\ 
   8410.25000   &    17.1949997   &    8.99999961E-03  & 4 \\
   8540.00000   &    17.2019997   &    8.00000038E-03  & 4 \\
   8568.24609   &    17.1906471   &    1.20000001E-02  & 2 \\
   8585.50000   &    17.2489986   &    8.00000038E-03  & 4 \\
   8775.75000   &    17.1980000   &    7.00000022E-03  & 4 \\
   8778.75000   &    17.2049999   &    4.00000019E-03  & 4 \\
   8788.75000   &    17.1889992   &    4.99999989E-03  & 4 \\
   8797.75000   &    17.2469997   &    4.99999989E-03  & 4 \\
   8802.50000   &    17.1669998   &    1.09999999E-02  & 4 \\
   8805.50000   &    17.1959991   &    3.00000003E-03  & 4 \\
   8823.55176   &    17.1121445   &    1.20000001E-02  & 2 \\
   8834.75000   &    17.1529999   &    9.99999978E-03  & 4 \\
   8838.75000   &    17.1910000   &    4.00000019E-03  & 4 \\
   8878.75000   &    17.1229992   &    3.00000003E-03  & 4 \\
   8883.50000   &    17.0970001   &    2.00000009E-03  & 4 \\
   8892.50000   &    17.0909996   &    8.99999961E-03  & 4 \\
   8906.50000   &    17.1269989   &    6.00000005E-03  & 4 \\
   8909.50000   &    17.0479984   &    8.99999961E-03  & 4 \\
   8913.50293   &    17.0949993   &    4.99999989E-03  & 3 \\
   8920.50293   &    17.1130009   &    4.00000019E-03  & 3 \\
   8928.26728   &    17.1011489   &    0.012        &    2 \\
   9081.58692   &    17.1188643   &    0.012        &    2 \\
   \hline
\end{longtable}
\end{center}

\section{Redshift and FeII template}
\label{sect:redshift}

The absolute precise redshift determination for quasar HE 0435-4312 is challenging, although quasar spectrum is not affected by absorption.
The original redshift for quasar $z = 1.232 \pm 0.001$ was reported by \citet{wisotzki2000} from the position of MgII line. \citet{marziani2009} measured the redshift at the basis of H$\beta$ line from VLT/ISAAC IR spectra, deriving $z = 1.2321 \pm 0.0014$. Narrow [OIII] line was also visible in the spectrum, but is was strongly affected by atmospheric absorption, so it could not serve as a reliable redshift mark. \citet{sredzinska2017} determined the redshift as $z = 1.2231$ using the data from SALT collected during the first three years of the SALT monitoring of this source. In this paper, the two kinematic components were used to fit the MgII line, one component at the systemic redshift, together with FeII template, and the shift of the second was a model parameter. What is more, \citet{sredzinska2017} showed that the MgII line position systematically changes with time. Thus the redshift was mostly determined by the location of the strong FeII emission at 2750 \AA. 

In the presently available set of observations, one set, obtained on 5 December 2019 (observation 23), was obtained with a slightly shifted setup, so it covered the spectral range almost up to 3000 \AA~ in the rest frame (in all other observations, this region overlaps with a CCD gap). Observed quasar spectra have usually a clear gap just above 2900 \AA, which could serve as additional help in constraining the redshift, as well as the optimum FeII template.

\begin{figure}
    \centering
  \includegraphics[width=0.95\columnwidth]{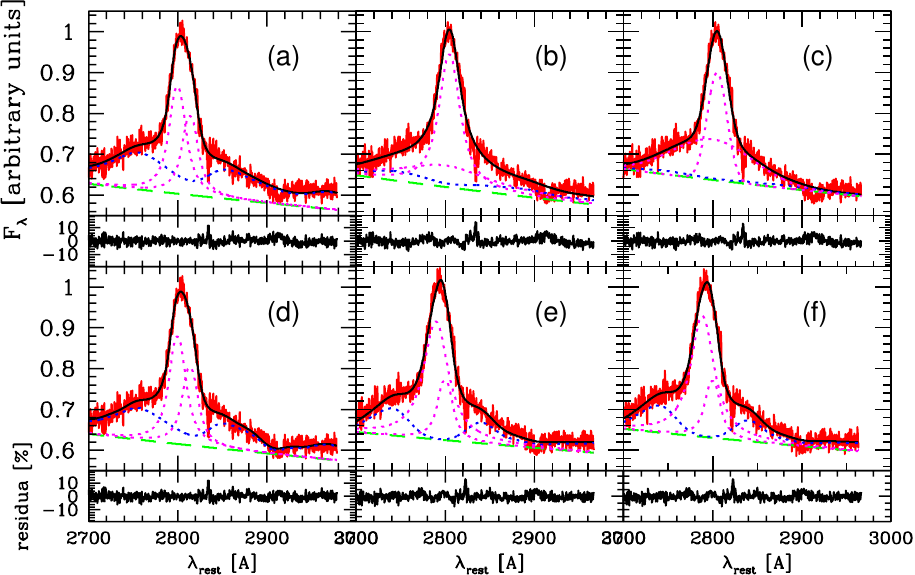}  
   \caption{Selected models of observation 23 from Table~\ref{tab_chi2_z_Fe}. }
    \label{fig:z_and_templlates}
\end{figure}

The basic template d11-m20-20.5-735 (marked later as d11) of \citet{bruhweiler2008} (with parameters: plasma density $10^{11}$ cm$^{-3}$, turbulence velocity 20 km s$^{-1}$, and log of ionization parameter in cm$^{-2}$ s$^{-1}$ of 20.5) favored by \citet{sredzinska2017} does not fit well the region of 2900 \AA~ dip (see Figure~\ref{fig:z_and_templlates}). We also checked other theoretical templates provided by \citet{bruhweiler2008}, but all of them predicted considerable emissivity at that spectral region. Since the CLOUDY code has been modified over the years, we calculated new FeII templates using the newest version of the code \citep{ferland2017}, but with the same input as in the template d11, including the spectral shape of the incident continuum. New results did not solve the problem, and actually gave much worse fit (see Table~\ref{tab_chi2_z_Fe}), if the two Lorentzian model was assumed for the MgII line, as in \citet{sredzinska2017}. If, instead, we used two Gaussians for the line fitting, the resulting $\chi^2$ became much better (see Table~\ref{tab_chi2_z_Fe}) but still higher than the older template, and the emission above 2900 \AA~ was still overproduced. We thus experimented with simple removal of a few transitions from the original d11 template. We removed transitions at $\lambda$2896.32, 2901.96, 2907.61, and 2913.28 \AA, creating $d11_{{\rm mod}}$ template. This clearly allowed to achieve a satisfactory representation of the data, and favored the same redshift as in \citet{sredzinska2017}.
 
 Next, we tried the KDP15 FeII templates \citep{kovacevic15,2019MNRAS.484.3180P} which have less transitions but an additional flexibility of arbitrarily changing the normalization of each of the six transitions. This template provided a good fit for another of the quasars monitored with SALT \citep{2020ApJ...896..146Z}. The clear disadvantage is that fitting these templates is much more time consuming, since then the spectral model has 13 free parameters instead of 8, when the FeII template has just one normalization as a free parameter. Nevertheless, we fitted all parameters at the same time, as in our standard approach.
 
\begin{figure*}
\begin{tabular}{cc}
        \includegraphics[width=0.45\columnwidth]{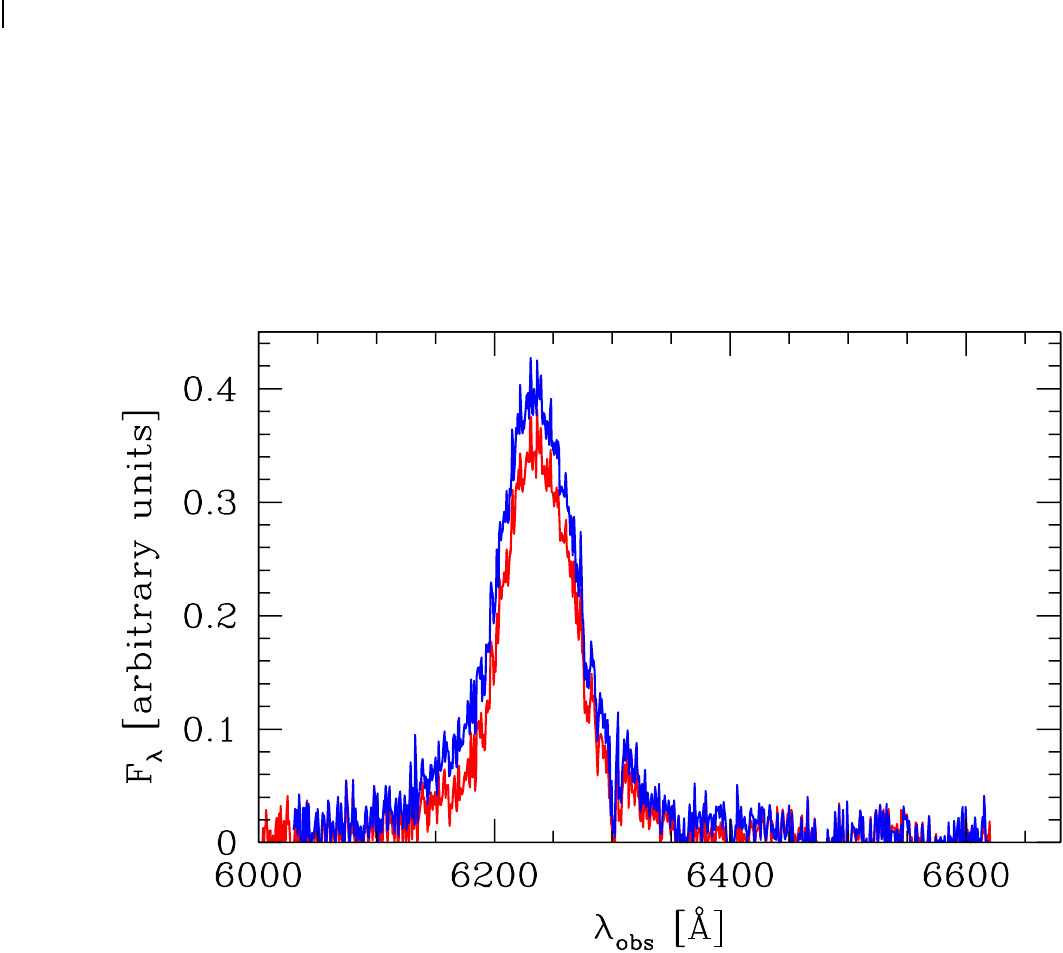}
        \includegraphics[width=0.5\columnwidth]{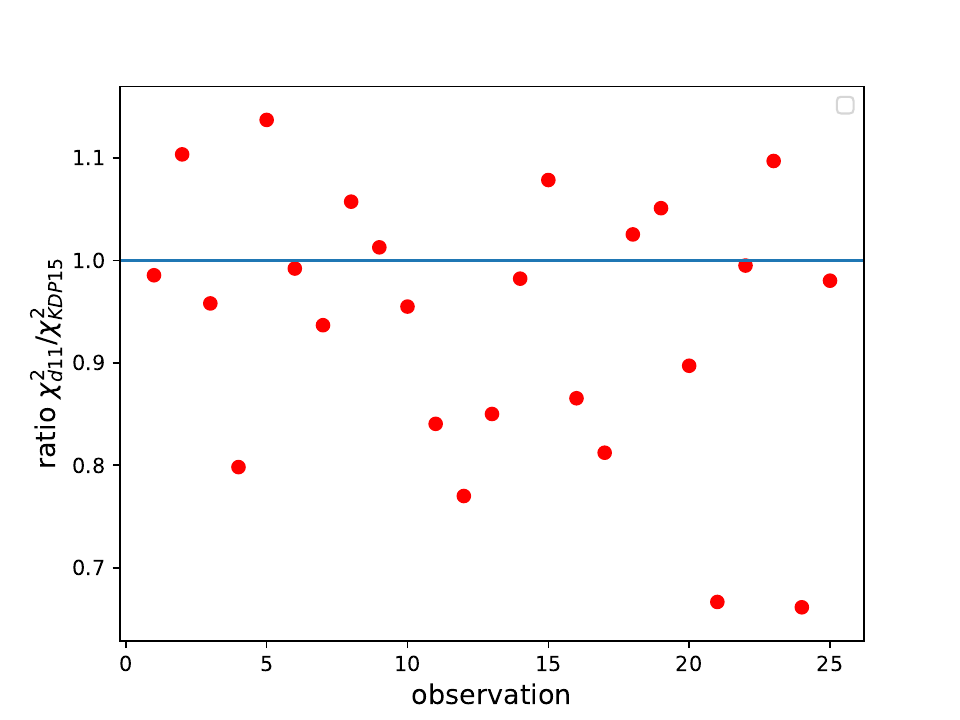}
\end{tabular}        
    \caption{Comparison between the fits of the observed MgII and FeII complex performed using $d11_{\rm mod}$ and KDP15 models. {\bf Left panel:} MgII line shape in the observed frame after subtraction of FeII and power law with models  $d11_{\rm mod}$ from panel (d) of Figure~\ref{fig:z_and_templlates} (red line) and KDP15 from panel (f) (blue line). The normalization is the same as in Fig.~\ref{fig:z_and_templlates}, so the slight difference in the line normalization between the two models is visible. {\bf Right panel:} The ratio of $\chi^2$ values between the fits performed using $d11_{\rm mod}$ model ($z=1.2231$) and the KDP15 model ($z=1.2330)$ for available observational epochs. For 17 epochs out of 25, $\chi^2_{\rm d11_{\rm mod}}<\chi^2_{\rm KDP15}$.}
    \label{fig:MgII_comparison}
\end{figure*}

 The KDP15 template gave rather poor fit for the low values of redshift favored by \citet{sredzinska2017}, but the model gave much lower $\chi^2$ values for the redshift range suggested by \citet{marziani2009} (see Table~\ref{tab_chi2_z_Fe}). The best fit for this template was achieved for the redshift $z =1.2330$, even slightly higher than in \citet{marziani2009}. However, formal $\chi^2$ is still higher than for the $d11_{{\rm mod}}$ template. Since the two redshifts and the FeII templates used for modelling in these two cases are so different, we checked how this difference in the spectrum decomposition actually affects the MgII line. For that purpose, we subtracted the fitted FeII template and continuum from the original spectrum. The result is displayed in Figure~\ref{fig:MgII_comparison} (left panel) in the observed frame. As we see, the difference in the actual MgII shape is not large, despite large differences in the FeII templates. FWHM of the line is only slightly broader for KDP15, 3632 km s$^{-1}$ vs. 3507 km s$^{-1}$, as determined from the total line shape. Line dispersion $\sigma$ measured as the second moment differs more (3060 km s$^{-1}$ vs. 3423 km s$^{-1}$) but in both cases FWHM/$\sigma$ ratio is much smaller than expected for a Gaussian profile.
 
 Since fitting full 6 transition KDP15 template is time-consuming, and we do not expect considerable changes in FeII shape in the SALT data, we constructed a single parameter new template based on KDP15, taking the relative values of the 6 components from the best fit of observation 23 in the extended wavelength range, and using the (best) FeII broadening of 4000 km s$^{-1}$. With this new template KDP15, we refitted all SALT observations in the range 2700 - 2900 \AA~. For 17 observations  $\chi^2_{d11_{\rm mod}}<\chi^2_{\rm KDP15}$, for 8 KDP15 fitted the data better, but generally not significantly, see Fig~\ref{fig:MgII_comparison} (right panel). We thus perform most of the time delay computations with the $d11_{\rm mod}$ template, and we illustrate the role of the template in the time-delay determination for the ICCF method, see Subsection~\ref{subsec_ICCF}, where we show that the KDP15 template is associated with the comparable cross-correlation function and similar peak and centroid distributions as those for $d11_{\rm mod}$, see Fig.~\ref{fig_ICCF_KDP15}.

\begin{table*}[h!]
    \centering
    \caption{Summary of the redshift and FeII templates used to fit observation 23.}
    \begin{tabular}{l|r|r|r|r|r}
    \hline
    \hline
    MgII model & FeII template & redshift & FWHM (FeII) & $\chi^2$& panel~in~Fig.~\ref{fig:z_and_templlates}\\
    \hline
    LL & d11     & 1.2231   &  3100 & 1377.59436 & a\\
    LL & d11BV08-C17 & 1.2231 & 3100 & 1818.21631 & b\\
    GG & d11BV08-C17 & 1.2231 & 3100 & 1505.38 & c \\
    GG & d12-MaFer-C17 & 1.2321 & 2800 & 1517.13757 \\
    LL & $d11_{mod}$ & 1.2231 & 3100 & 1296.09912 & d\\
    LL & KDP15 & 1.2321   &  3600 & 1408.91748 & \\
    LL & KDP15 & 1.2321   &  4000 & 1400.57812 & e \\
    LL & KDP15 & 1.2330   &  4000 & 1370.64978 & f \\
     \hline
    \end{tabular}
    \label{tab_chi2_z_Fe}
\end{table*}

\section{Overview of time-delay determination methods}
\label{sec_time-delay}
 
\subsection{Interpolated cross-correlation function (ICCF)}
\label{subsec_ICCF} 

We first analyzed the continuum and MgII line-emission light curves using the interpolated cross-correlation function (ICCF), which belongs to the standard and well-tested methods for assessing the time-delay between the continuum and the line emission flux density \citep{1998PASP..110..660P}. Light curves are typically unevenly sampled, while the ICCF works with the continuum-line emission pairs with a certain time-step of $\Delta t=t_{i+1}-t_{i}$. The cross-correlation function for 2 light curves, $x_i$ and $y_i$, with the same time-step of $\Delta t$ achieved by the interpolation, evaluated for the time-shift of $\tau_{k}=k\Delta t$ ($k=1$, \ldots, $N-1$), is defined as
\begin{equation}
    CCF(\tau_{k})=\frac{(1/N)\sum_{i=1}^{N-k} (x_i-\overline{x})(y_{i+k}-\overline{y})}{[(1/N)\sum_{i=1}^N (x_i-\overline{x})^2]^{1/2}[(1/N)\sum_{i=1}^N (y_i-\overline{y})^2]^{1/2}}\,.
    \label{eq_CCF}
\end{equation}

The same time-step $\Delta t$ can be achieved by interpolating the continuum light curve with respect to the line-emission light curve and vice versa (asymmetric ICCF). Typically, both interpolations are averaged to obtain the symmetric ICCF.

For the time-delay analysis, we use the python implementation of ICCF in code PYCCF by \citet{2018ascl.soft05032S}, which is based on the algorithm by \citet{1998PASP..110..660P}. The code allows to perform both the asymmetric as well as the symmetric interpolation. Based on the Monte Carlo techniques of the random subset selection (RSS) and flux randomization (FR), one can obtain the centroid and the peak distributions and their corresponding uncertainties.

\begin{figure*}
    \centering
    \includegraphics[width=\textwidth]{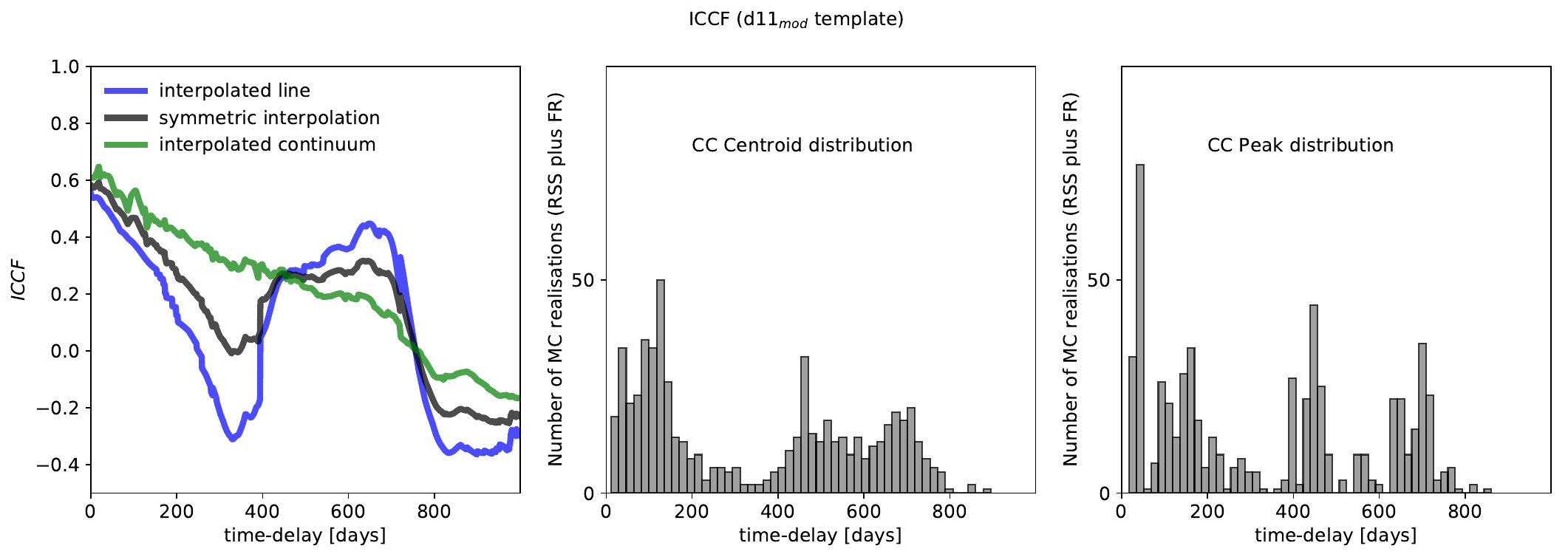}
    \caption{The interpolated cross-correlation function (ICCF) as a function of the time-delay expressed in days in the observer's frame. The calculation is performed for the d11$_{\rm mod}$ template and the redshift of $z=1.2231$. \textbf{Left panel:} The ICCF as a function of the time-delay expressed in days in the observer's frame for three types of interpolation: interpolated continuum light curve (green solid line), interpolated line light curve (blue solid line), and the symmetric interpolation (black solid line). \textbf{Middle panel:} The distribution of the cross-correlation centroids for the symmetric interpolation. \textbf{Right panel:} The distribution of the cross-correlation peaks for the symmetric interpolation.}
    \label{fig_ICCF}
\end{figure*}

We studied the time-delay between the continuum light curve consisting of 81 measurements, with 8 Catalina-survey averaged detections, 16 SALTICAM measurement, 27 BMT data, and 30 OGLE data. Two SALTICAM measurements were excluded based on the poor quality. The emission-line light curve consists of 25 SALT measurements, where the 19th measurement has a poor quality due to weather conditions and is excluded from the further analysis. Hence, we have 79 continuum points with the mean cadence of $69.0$ days and 24 MgII flux density measurements with the mean cadence of $121.6$ days. We set the interpolation interval to one day. For the redshift of $z=1.2231$ and d11$_{\rm mod}$ template, we display the ICCF as a function of time-delay in the observer's frame in Fig.~\ref{fig_ICCF} (left panel) for both asymmetric and the symmetric interpolation. In the middle and the right panel, we show the centroid and the peak distributions for the symmetric interpolation based on 3000 Monte Carlo realizations of random subset selection and flux randomization. The centroid and the peak are generally not well defined. The peak value of the correlation function for the symmetric interpolation is $0.32$ for the time-delay of 635 days, which is less than for our previous quasars CTS C30.10 \citep[peak CCF of $\sim 0.65$,][]{czerny2019,2019arXiv190703910Z} and HE 0413-4031 \citep[peak CCF of $0.8$,][]{2020ApJ...896..146Z}. In the next step, we focus on the surroundings of this peak and we analyze the CCF centroid and peak distributions in the interval between 500 and 1000 days. The results for all interpolations are summarized in Table~\ref{tab_ICCF}. For the symmetric interpolation, we obtain the centroid time-delay of $\tau_{\rm cent}=663^{+66}_{-40}$ days and the peak time-delay of $\tau_{\rm peak}=672^{+49}_{-37}$ days.

\begin{figure*}
    \centering
    \includegraphics[width=\textwidth]{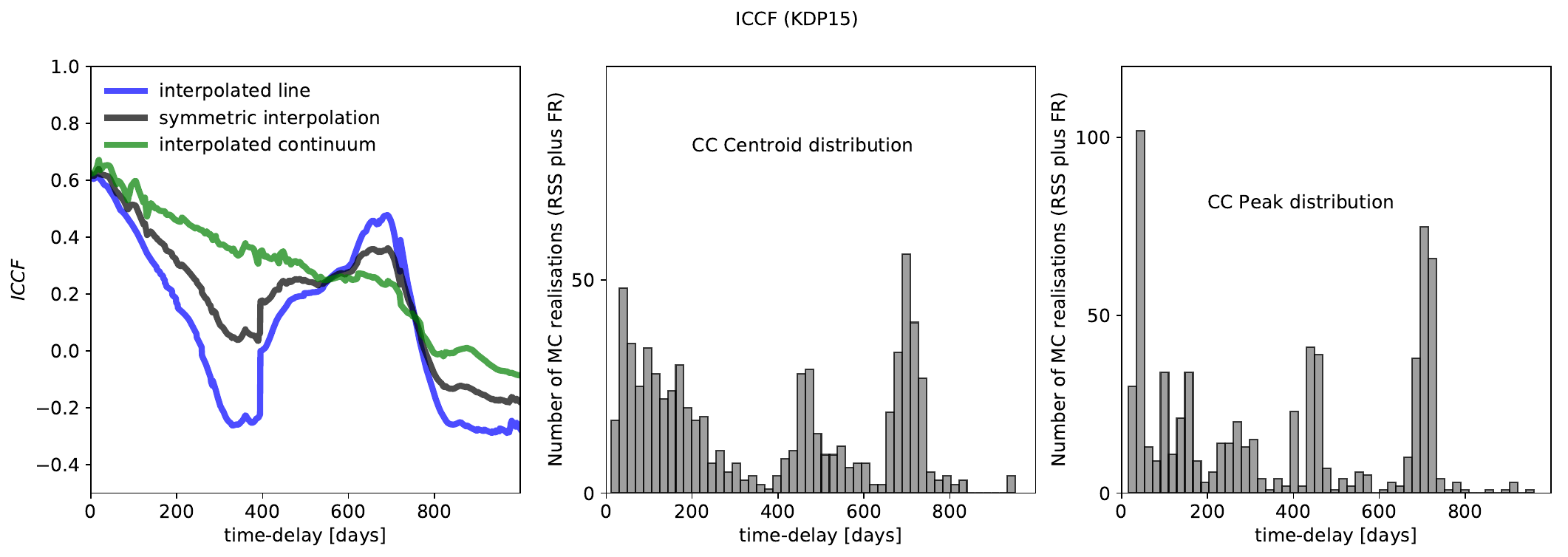}
    \caption{The same ICCF analysis is in Fig.~\ref{fig_ICCF} but for the KDP15 template (case \textit{f} in Table~\ref{tab_chi2_z_Fe}) and the redshift of $z=1.2330$.}
    \label{fig_ICCF_KDP15}
\end{figure*}

We also performed the time-delay analysis using the ICCF for the MgII light curve derived using the KDP15 template (case \textit{f} in Table~\ref{tab_chi2_z_Fe}). For the symmetric ICCF, the CCF peak value is $692$ days with CCF=$0.36$, which is comparable to the analysis performed for the d11$_{\rm mod}$ template. The CCF peak and centroid distributions also contain the same subpeaks, with the peak close to 700 days being the most prominent, see Fig.~\ref{fig_ICCF_KDP15}.

\begin{table*}[h!]
    \centering
    \caption{Summary of the time-delay determination for the quasar HE 0435-4312 using the interpolated cross-correlation function (ICCF). We list the centroids as well as the peaks for the interpolated continuum light curve, the interpolated line emission light curve, and for the symmetric interpolation.Time delays are expressed in days in the observer's frame of reference.}
    \begin{tabular}{c|c}
    \hline
    \hline
    Interpolation method  & Time delay ($z=1.2231$)\\
    \hline
     Interpolated continuum -- centroid [days]   & $635^{+31}_{-34}$ \\
     Interpolated continuum -- peak [days] & $630^{+42}_{-46}$\\
     Interpolated line -- centroid [days]  & $673^{+60}_{-50}$ \\
     Interpolated line -- peak [days]  & $683^{+38}_{-48}$ \\
     Symmetric -- centroid [days]  &  $663^{+66}_{-40}$ \\
     Symmetric -- peak [days]   & $672^{+49}_{-37}$  \\
     \hline
    \end{tabular}
    \label{tab_ICCF}
\end{table*}

\subsection{Discrete Correlation Function (DCF)}
\label{subsec_DCF}

For unevenly sampled and sparse light curves, the ICCF can distort the best time-delay due to adding additional data points to one or both light curves. In that case, the discrete correlation function (DCF) introduced by \citet{1988ApJ...333..646E} is better suited to determine the best time lag. 

In general, the DCF value is determined for any pair of two light curves $(x_i,y_j)$ whose time difference $\Delta t_{ij}$ falls into the time-delay bin of size $\delta \tau$, $\tau-\delta \tau/2<\Delta t_{ij}<\delta \tau$, $\tau+\delta \tau/2$, where $\tau$ is a given time delay. First, one calculates the unbinned DCF,

\begin{equation}
    \mathrm{UDCF}_{ij}=\frac{(x_i-\overline{x})(y_j-\overline{y})}{\sqrt{s_{x}-\sigma_x^2}\sqrt{s_{y}-\sigma_y^2}}\,,
    \label{eq_udcf}
\end{equation}
where $\overline{x}$ and $\overline{y}$ are light curve means in a given time-delay bin, $s_x$ and $s_y$ are corresponding light curve variances, and $\sigma_x$ and $\sigma_y$ are the mean measurement errors of the light curve points in the given time-delay interval.

\begin{figure*}
    \centering
    \includegraphics[width=0.49\columnwidth]{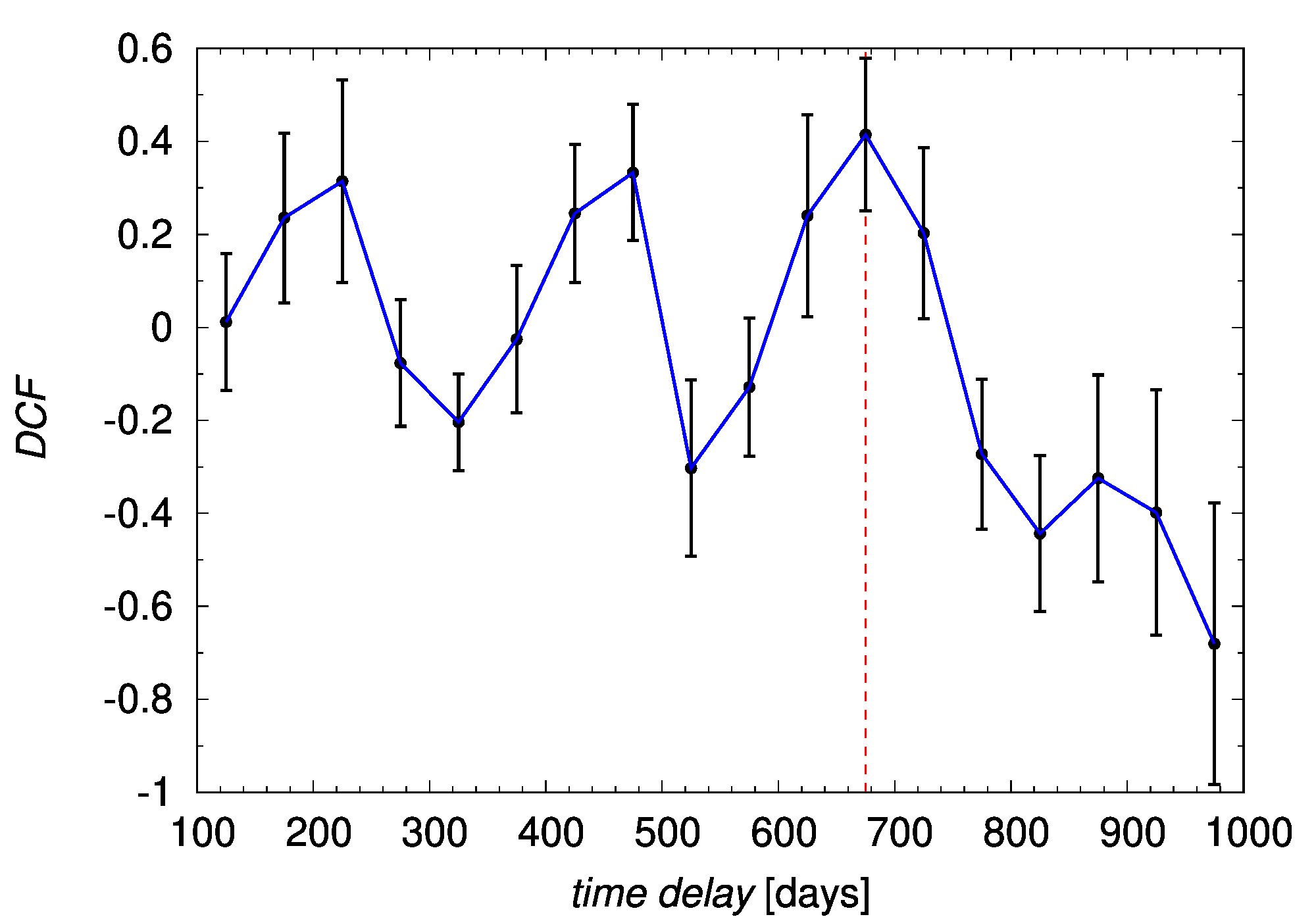}
     \includegraphics[width=0.49\columnwidth]{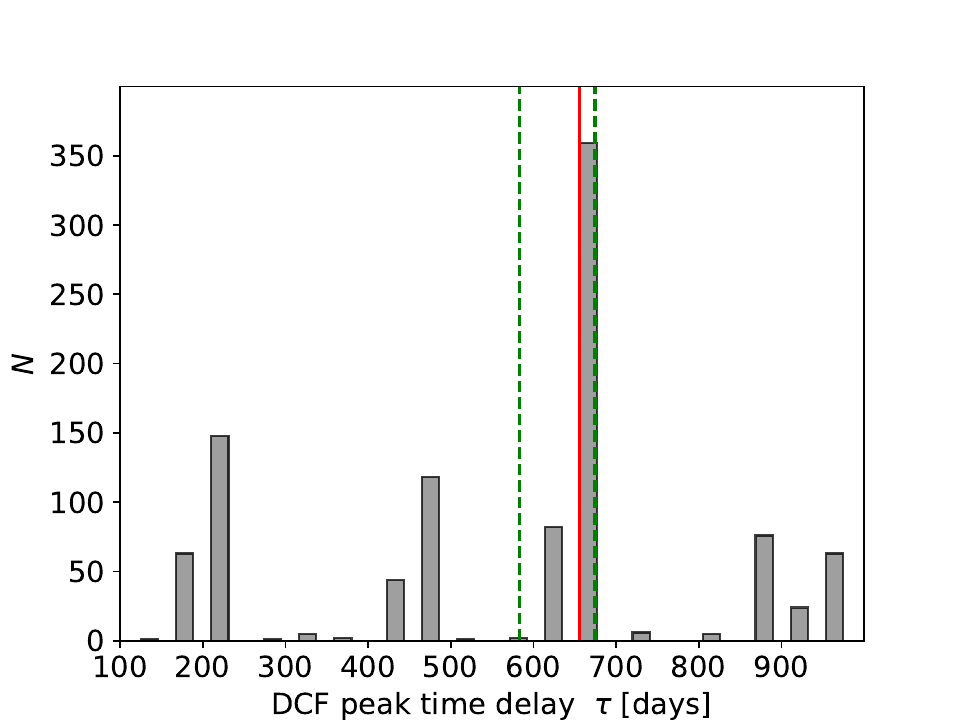}
    \caption{Discrete Correlation Function (DCF) vs. a time-delay in the observer's frame using d11$_{\rm mod}$ template. {\bf Left panel:} A DCF vs. time-delay in the observer's frame. A red vertical line denotes the peak DCF. {\bf Right panel:} Histogram of peak DCF time-delays constructed from 1000 bootstrap realizations. The red and dashed green vertical lines denote the histogram peak and its left and right standard deviations within 30\%-surrounding of the main peak, respectively.}
    \label{fig_DCF}
\end{figure*}

The DCF is then calculated as a mean over $M$ light curve points that are located in a given bin,
\begin{equation}
    \mathrm{DCF}(\tau)=\frac{1}{M}\sum_{ij} \mathrm{UDCF}_{ij}\,.
    \label{eq_DCF}
\end{equation}
The uncertainty can be estimated using the relation,
\begin{equation}
    \sigma_{\rm DCF}=\frac{1}{M-1}\sqrt{\sum[\mathrm{UDCF_{ij}}-\mathrm{DCF}(\tau)]^2}\,.
    \label{eq_error_DCF}
\end{equation}
We made use of the python script pyDCF \citep{2015MNRAS.453.3455R}, where the general procedure described using Eqs.~\ref{eq_udcf}, \ref{eq_DCF}, and \ref{eq_error_DCF} is implemented. We searched for the DCF peak in the time interval from 100 to 1000 days, with the time-step of 50 days. Using the Gaussian weighting scheme, we obtained a better defined DCF peak than using the slot weighting. In Fig.~\ref{fig_DCF} (left panel), we show the DCF vs. time-delay in the observer's frame. The time-delay with the highest DCF value of $0.41 \pm 0.16$ is at 675 days. To determine the uncertainty of the peak, we ran 1000 bootstrap simulations, where at each step we constructed a pair of new light curves by randomly selecting a light curve subset. The histogram of peak DCF time delays constructed from 1000 bootstrap realizations is shown in Fig.~\ref{fig_DCF} (right panel). The peak of the distribution is at $\tau_{\rm DCF}=656^{+18}_{-73}$ days in the observer's frame. The left and right uncertainties represent standard deviations, where we included the surrounding of the main peak within its 30\%.

\subsection{$Z$-transformed Dicrete Correlation Function (zDCF)}
\label{subsec_zDCF}

\begin{figure}
    \centering
    \includegraphics[width=0.5\columnwidth]{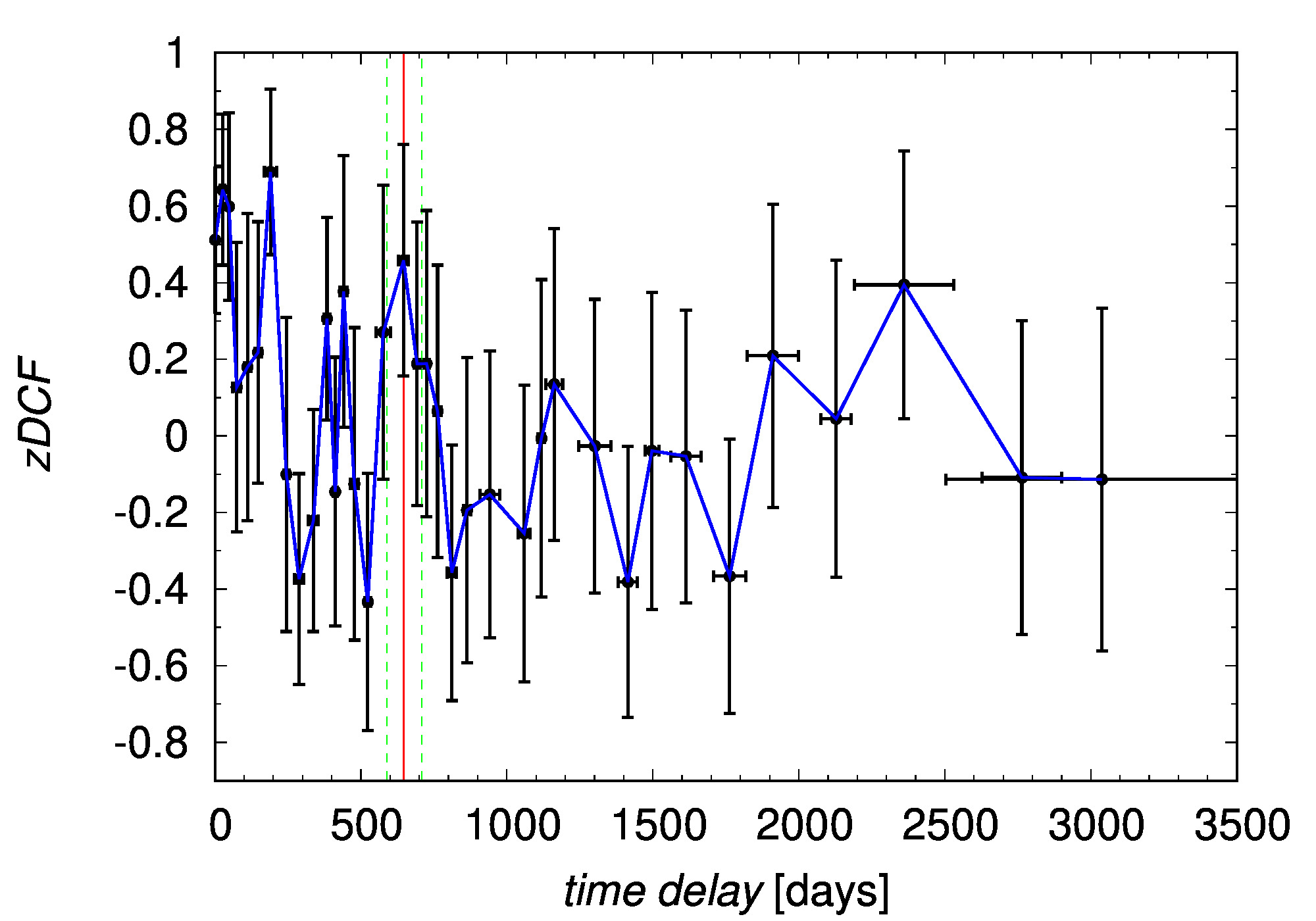}
    \caption{The zDCF as a function of the time delay in the observer's frame. The uncertainties in each population bin were inferred from 5000 Monte-Carlo generated pairs of light curves (using the observed measurement errors). The red vertical line marks the most prominent, maximum-likelihood peak of $646^{+63}_{-57}$ days, while the green vertical dashed lines mark the uncertainties.}
    \label{fig_zDCF}
\end{figure}

In the following, we applied the $z$-transformed Dicrete Correlation Function \citep[zDCF; ][]{1997ASSL..218..163A}, which improves the classical DCF by replacing equal time binning by the equal population binning. This is achieved by applying Fisher's $z$-transform. With this property, the zDCF processes satisfactorily especially undersampled, sparse, and heterogenous light curves, which is also to some extent our case, since the continuum light curve originates from three different instruments and the MgII light curve is relatively sparse with respect to the continuum (24 vs. 79 points). In our analysis, we applied the zDCF method several times, changing the minimum number of light-curve pairs per a population bin. Finally, we set the minimum number of light-curve pairs to eight and using 5000 Monte-Carlo generated pairs of light curves, we determined the errors in each bin. In Fig.~\ref{fig_zDCF}, we show the zDCF values as a function of a time delay in the observer's frame. There are some peaks with a large DCF value within the first 200 days in the rest frame, e.g. $190.3$ days (DCF$=0.69$). These are, however, too short to correspond to the realistic rest-frame time-delay for our highly luminous quasar. The most prominent peak at larger time-delays is at 646 days (DCF$=0.46$).

Next we calculate the maximum likelihood (ML) using the zDCF values and we focus on the prominent peak in the interval between 500 and 800 days. From the ML analysis, we obtain the ML peak of $\tau_{\rm zDCF}=646^{+63}_{-57}$ days in the observer's frame. This peak and the corresponding uncertainties are also highlighted in Fig.~\ref{fig_zDCF} by vertical lines. 

\subsection{The JAVELIN package}
\label{subsec_javelin}

Another technique to evaluate the time-delay distribution is to model the AGN continuum variability as a stochastic process using the damped random walk process
\citep[DRW;][]{2009ApJ...698..895K,2010ApJ...721.1014M,2010ApJ...708..927K,2016ApJ...826..118K}. Subsequently, the line-emission is assumed to be time-delayed, smoothed, and a scaled version of the continuum emission. This method is implemented in the JAVELIN package \citep[Just Another Vehicle for Estimating Lags In Nuclei;][]{2011ApJ...735...80Z,2013ApJ...765..106Z,2016ApJ...819..122Z}\footnote{\url{https://bitbucket.org/nye17/javelin/src/develop/}}. The package uses Markov Chain Monte Carlo (MCMC) to first determine the posterior probabilities for the continuum variability timescale and the amplitude. Based on that, the line emission variability is modelled to obtain the posterior probabilities for the time lag, smoothing width of the top hat function, and the scaling ratio (the ratio between the line and the continuum amplitudes, $A_{\rm l}/A_{\rm c}$). 

\begin{figure*}
    \centering
    \includegraphics[width=0.49\columnwidth]{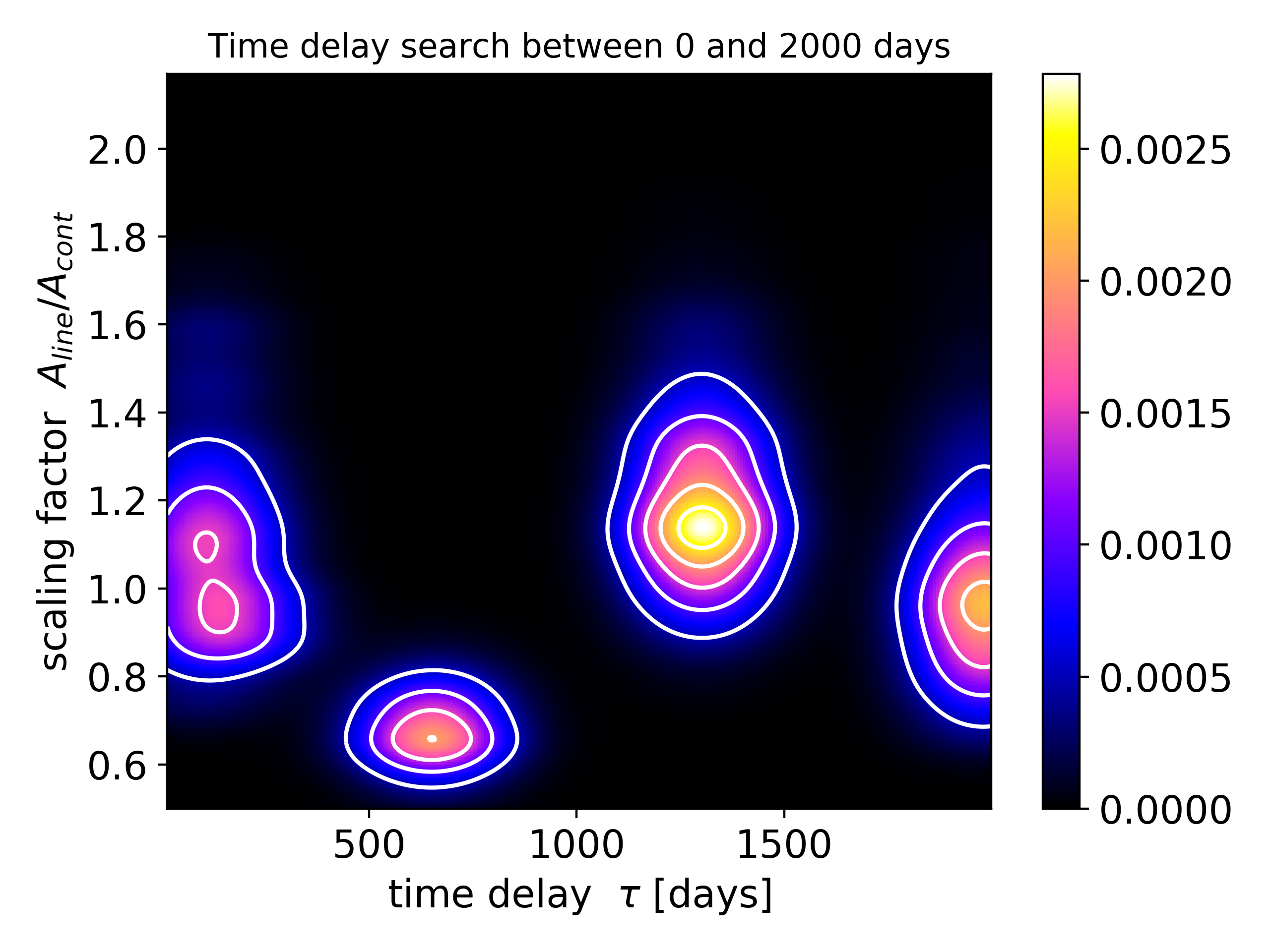}
     \includegraphics[width=0.49\columnwidth]{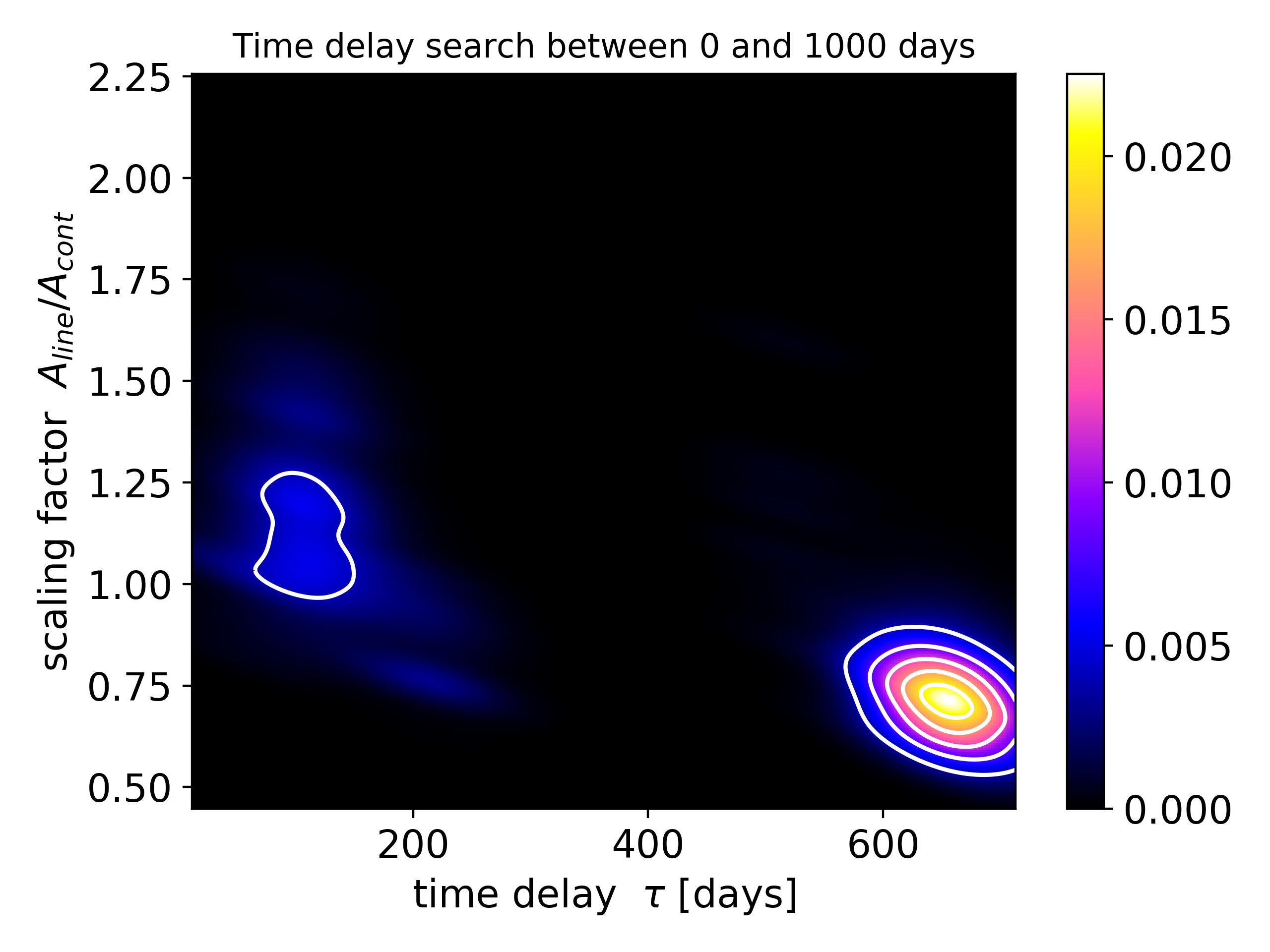}
    \caption{The time-delay determination for HE 0435-4312 using the JAVELIN package. {\bf Left panel:} The time-delay search in the interval between 0 and 2000 days in the observer's frame. There are four distinct peaks at $\sim 100$, $\sim 650$, $\sim1300$, and $\sim 2000$ days. {\bf Right panel:} The time-delay search in the narrower interval between 0 and 1000 days yields a clear peak at $\sim 652$ days.}
    \label{fig_javelin}
\end{figure*}

First, we searched for the time-delay in the longer interval between 0 and 2000 days in the observer's frame. There are four distinct peaks, see Fig.~\ref{fig_javelin} (left panel). at $\sim 100$, $\sim 650$, $\sim 1300$, and $\sim 2000$ days. The first peak is too short, while the last two appear too long for our dataset. In the zDCF analysis in Subsection~\ref{subsec_zDCF}, the time-delay peaks at 1000 days and more also have large uncertainties. Therefore, in the next search, we perform a time-delay search in the narrower interval between 0 and 1000 days to focus on the intermediate peak at $\sim 650$ days. We obtain a prominent peak at $\sim 652$ days, see Fig.~\ref{fig_javelin} (right panel). 

\begin{figure*}
    \centering
    \includegraphics[width=0.47\textwidth]{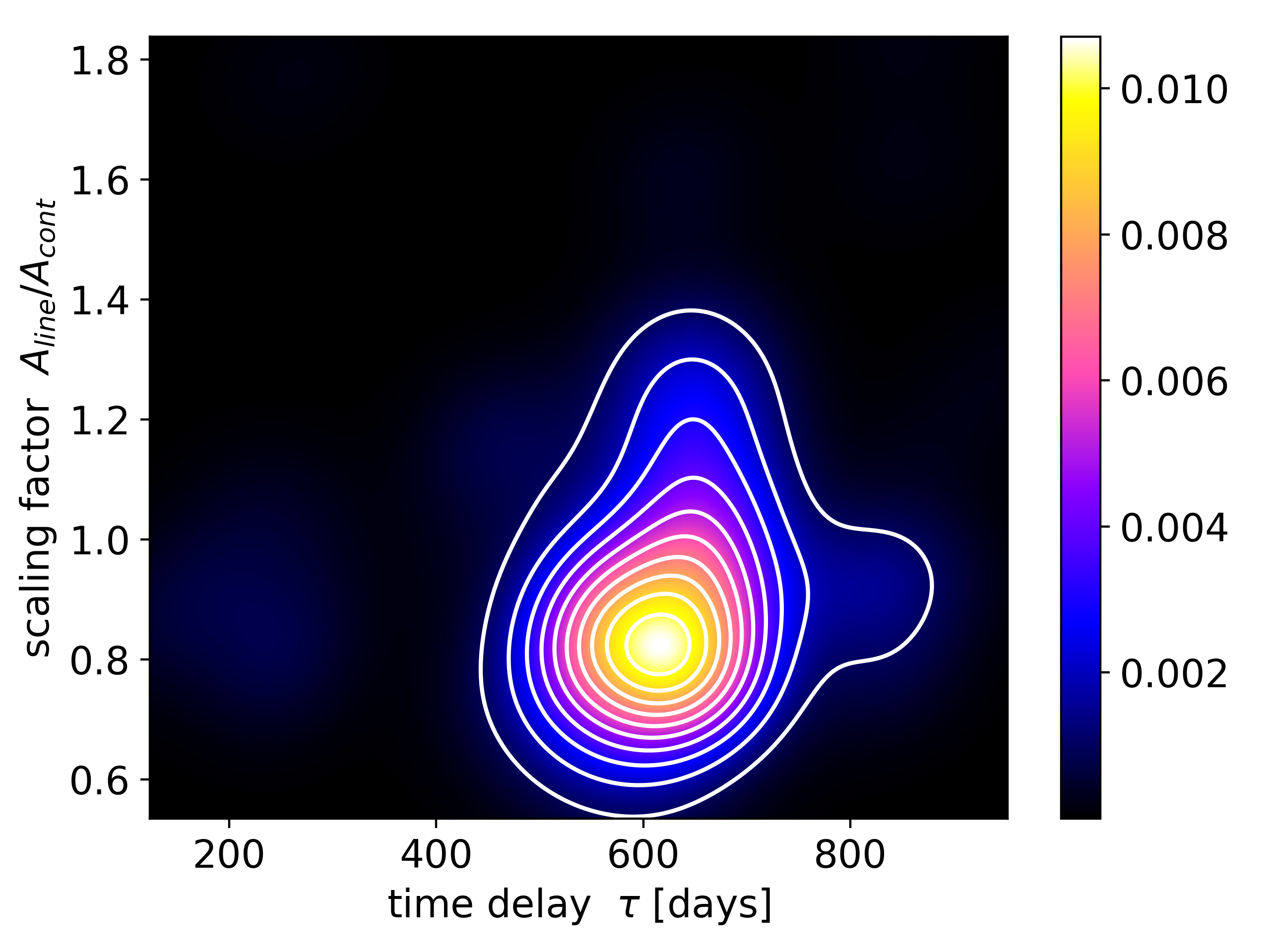}
     \includegraphics[width=0.47\textwidth]{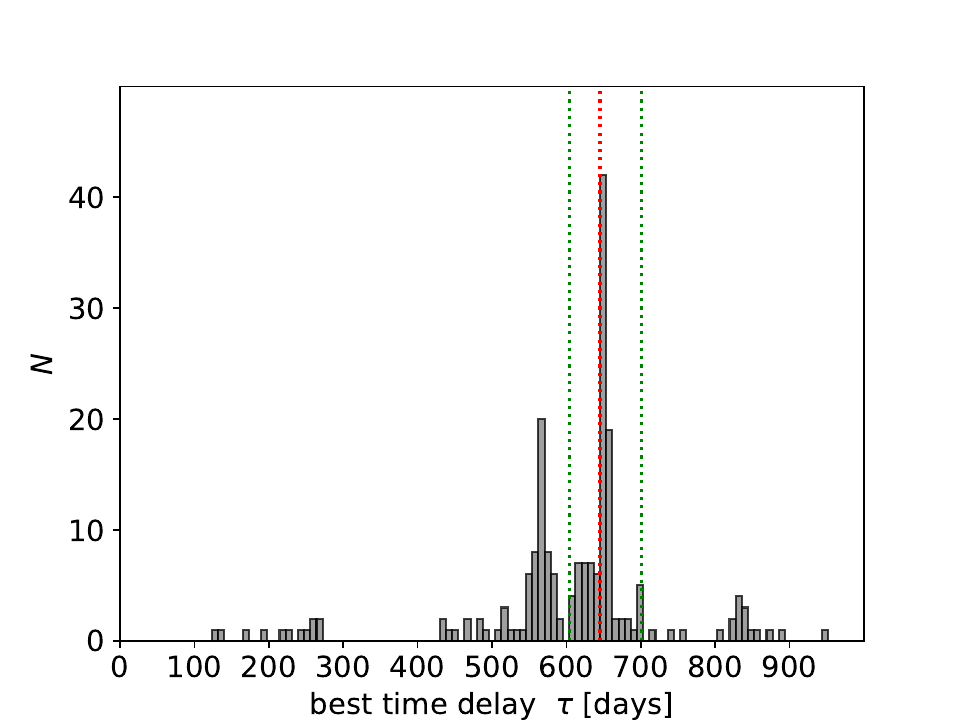}
    \caption{Bootstrap analysis of the BLR time delay in HE 0435-4312 using the JAVELIN package. {\bf Left panel:} The distribution of 200 best time delays in the plane of the time delay (in the observer's frame) and the scaling factor $A_{\rm line}/A_{\rm cont}$. There is a prominent peak in the distribution close to 600 days. {\bf Right panel:} The histogram of 200 best time delays. The peak value is at $\tau_{\rm peak}=645^{+55}_{-41}$ days, which is denoted by the vertical red and green dotted lines. The uncertainty is determined as the standard deviation within 30\%-surrounding of the main peak.}
    \label{fig_javelin_bootstrap}
\end{figure*}

To determine the precise position of the time-delay peak, we performed 200 bootstrap simulations - i.e. generating 200 subsets of the original continuum and MgII light curves. Then we applied the JAVELIN to determine the peak time delay for each individual pair of light curves. From 200 best time delays, we first construct a density plot in the plane of the time-delay and the scaling factor, see Fig.~\ref{fig_javelin_bootstrap} (left panel). We see that there is a peak in the distribution close to 600 days in the observer's frame. In Fig.~\ref{fig_javelin_bootstrap} (right panel), we show the histogram of the best time delays with the peak value of $\tau_{\rm peak}=645^{+55}_{-41}$ days, where the uncertainties were determined within 30\%-surrounding of the peak value.

\begin{figure*}
    \centering
    \includegraphics[width=0.49\columnwidth]{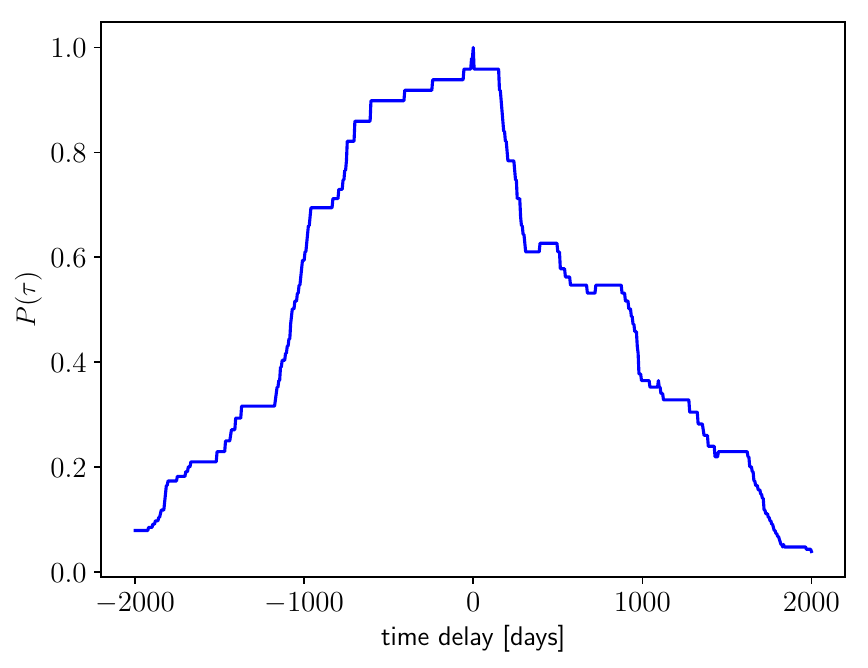}
    \includegraphics[width=0.49\columnwidth]{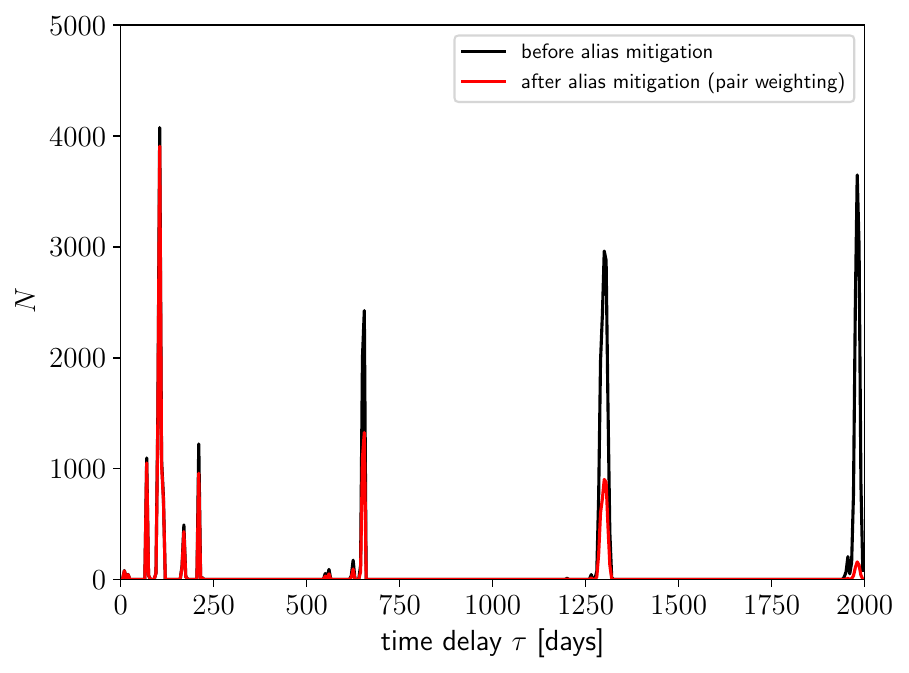}
    \caption{Alias mitigation for the JAVELIN time-delay distribution. {\bf Left panel:} Prior probability distribution $P(\tau)=[N(\tau)/N(0)]^2$ based on the number of overlapping data points for a given time delay. {\bf Right panel:} Posterior time-delay distributions before alias mitigation (black line) and after alias mitigation by using $P(\tau)$ as an effective weight for individual time-delay peaks.}
    \label{fig_javelin_alias_mitigation}
\end{figure*}

For the JAVELIN time-delay determination, we typically notice several narrow comparable peaks in the histogram of time delays for a single MCMC run as can be seen in Fig.~\ref{fig_javelin}. These secondary peaks typically arise due to a limited duration of observational runs and a sparse and/or non-uniform sampling of continuum and line-emission light curves. However, they might also reflect the complex BLR geometry seen in advanced data analysis or models \citep[e.g.][]{grier2017,huchen2020,horne2021,naddaf2021}. As mentioned in \citet{grier2017}, these alias peaks can be of a comparable significance as the true peak in the time-delay distribution, which is also our case, see Fig.~\ref{fig_javelin_alias_mitigation} (right panel, black line), where time delays at $\sim 105$, $\sim 1300$, and $\sim 1980$ days appear to have a larger significance than the fiducial peak of $\sim 650$ days in the observer's frame.

Following \citet{grier2017}, we weight the posterior probability distribution of time delays by the number of overlapping light curve points, where the prior probability distribution is given by $P(\tau)=[N(\tau)/N(0)]^2$, with $N(\tau)$ denoting the number of overlapping light curve points for the time delay $\tau$, and $N(0)$ is the number of data points in the overlap region for the zero time lag. Hence, the weight is $P(\tau)=1$ by definition for the zero time lag and it will decrease for larger positive and negative time lags, being zero when there is no overlap. For our light curves, we construct the prior probability distribution $P(\tau)$, which is shown in Fig.~\ref{fig_javelin_alias_mitigation} (left panel).  

Using $P(\tau)$, we weight the posterior time-delay distribution based on the number of overlapping light curve points. This is demonstrated in Fig.~\ref{fig_javelin_alias_mitigation} (right panel, red line). The alias mitigation based on $P(\tau)$ helps to effectively suppress longer time-delays, which drop in significance below the fiducial time lag at $\sim 655$ days. However, this technique is not efficient to suppress aliases at shorter time delays of $\sim 100$ days, where the decrease is negligible due to the definition of $P(\tau)$. This peak at smaller time-delays is effectively mitigated by the bootstrap or random subset selection technique. As we showed in Fig.~\ref{fig_javelin_bootstrap}, this peak is not further represented in the posterior time-delay distribution, which leaves us with the fiducial peak close to $\sim 650$ days. In addition, in Section~\ref{appendix_alias_mitigation} we show that the artefact peaks at $\lesssim 200$ days arise in the probability density distributions (see Figs.~\ref{fig_tk_iccf} and \ref{fig_tk}) constructed from mock light curves that are sampled with the same cadence as the original continuum and line-emission light curves.

\subsection{Measures of data regularity/randomness -- von Neumann and Bartels estimators}
\label{subsec_vonneumann}

We also analyzed the time-delay between the continuum and MgII light curves using a novel technique of the measures of the data regularity or randomness \citep{2017ApJ...844..146C}, which were previously applied extensively in cryptography and data compression. The advantage of these regularity measures is that they do not require neither the interpolation as for ICCF or $\chi^2$ technique nor do they require binning in the correlation space as the DCF and zDCF methods. In addition, no AGN variability model is needed for the continuum light curve as for the JAVELIN.

One of the suitable estimators to analyze the time delay between two light curves is the optimized von Neumann scheme, which works with the combined light curve $F(t,\tau)={(t_i,f_i)}_{i=1}^N=F_1 \cup F_2^{\tau}$, where $F_1$ is the continuum light curve and $F_2^{\tau}$ is the time-shifted line-emission light curve. The optimized von Neumann estimator for a given time-delay $\tau$ is defined as the mean of the successive differences of $F(t,\tau)$,
\begin{equation}
    E(\tau)=\frac{1}{N-1}\sum_{i=1}^{N-1}[F(t_i)-F(t_{i+1})]^2\,.
    \label{eq_von_Neumann}
\end{equation}
The aim of the von Neumann scheme is to find the minimum $E(\tau')$, where $\tau'\sim \tau_0$ is supposed to correspond to the actual time delay. 

\begin{figure*}
    \centering
    \includegraphics[width=0.47\columnwidth]{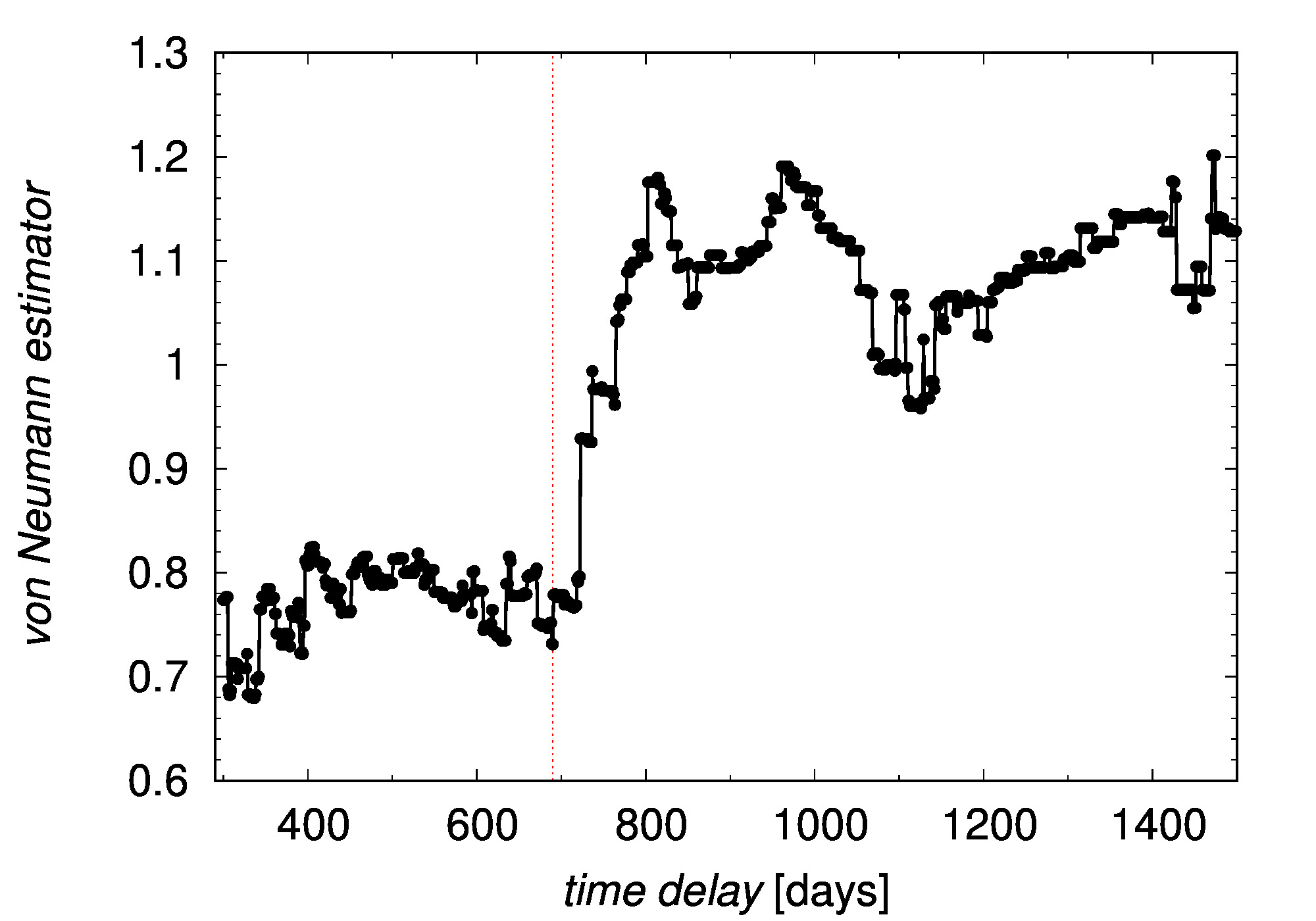}
    \includegraphics[width=0.47\columnwidth]{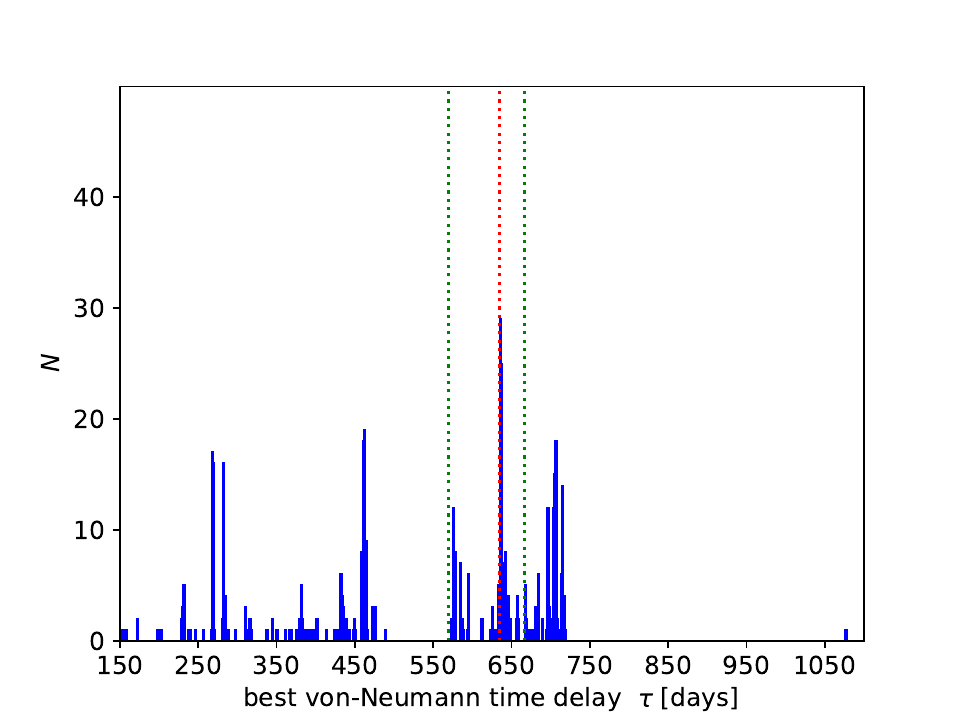}
    \caption{The results of the time-delay analysis using the von Neumann estimator. {\bf Left panel:} The von Neumann estimator value as a function of the time-delay in the observer's frame. The red vertical value denotes the minimum at 690 days. {\bf Right panel:} The histogram of best von-Neumann time delays constructed from 10000 bootstrap simulations of random subset selections of both continuum and line-emission light curves. The red and green dotted vertical lines mark the peak value of $635^{+32}_{-66}$ days and the corresponding uncertainties.}
    \label{fig_vonneumann}
\end{figure*}

In Fig.~\ref{fig_vonneumann} (left panel), we calculate $E(\tau)$ for HE 0435-4312 using the python script based on \citet{2017ApJ...844..146C}\footnote{For the script and the corresponding documentation, see \url{www.pozonunez.de/astro_codes/python/vnrm.py}.}. The minimum is reached for the time-delay between 300 and 400 days in the observer's frame (333-334 days and 391-392 days), and then for 690 days, which is consistent with the results of previous methods. To determine the peak value and its uncertainty close to this minimum, we performed $\sim 10000$ bootstrap realizations, from which we constructed the histogram of the best von Neumann time delays, see Fig.~\ref{fig_vonneumann} (right panel). In this histogram, we detect two prominent peaks at $\sim 27$ days and $\sim 1200$ days and a smaller peak at $\sim 635$ days. After focusing on the interval between 150 and 1100 days, the best peak is at $635^{+32}_{-66}$ days in the observer's frame, where the uncertainties correspond to standard deviations within 30\% surroundings of the best peak.  

\begin{figure*}
    \centering
    \includegraphics[width=0.47\columnwidth]{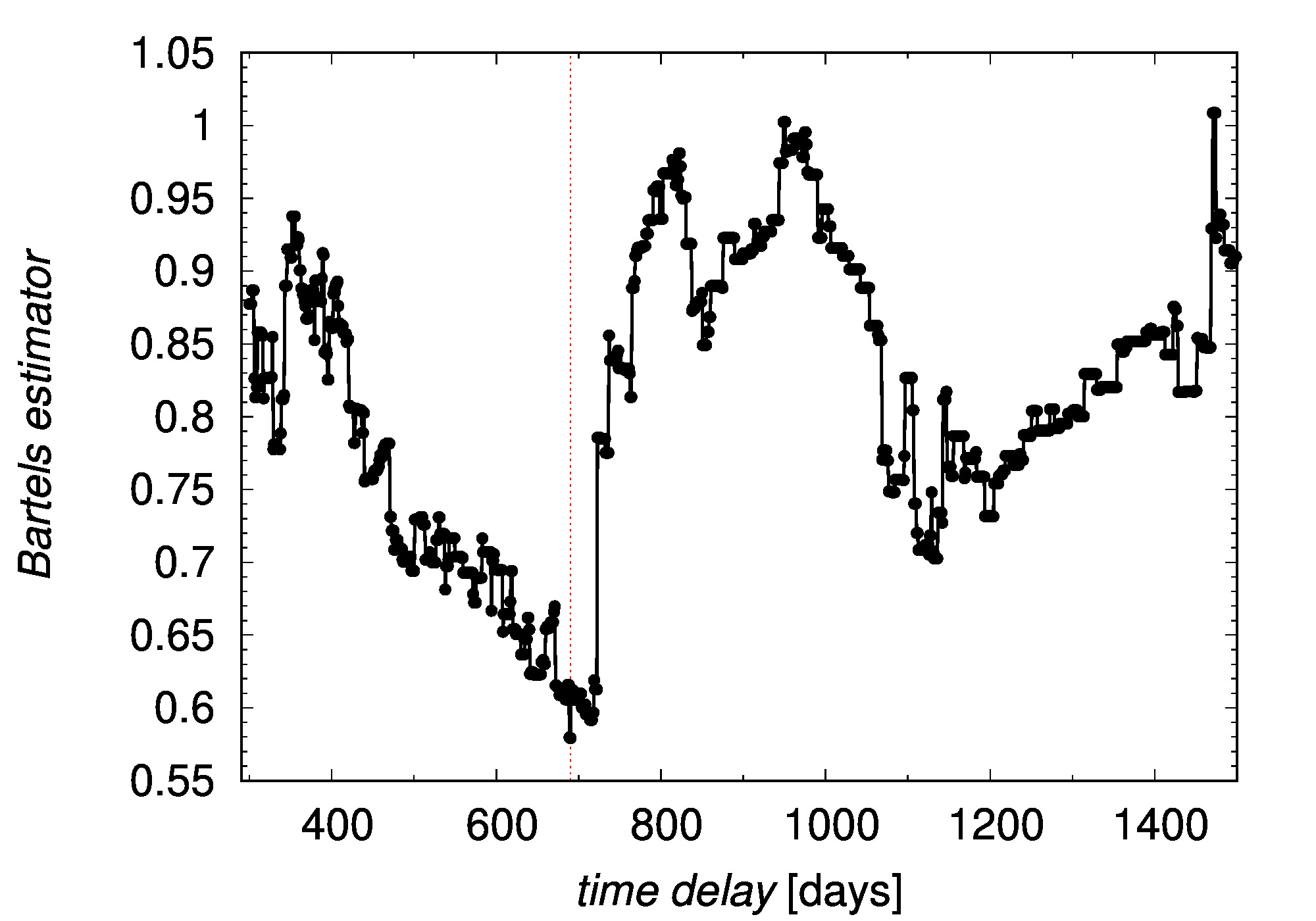}
    \includegraphics[width=0.47\columnwidth]{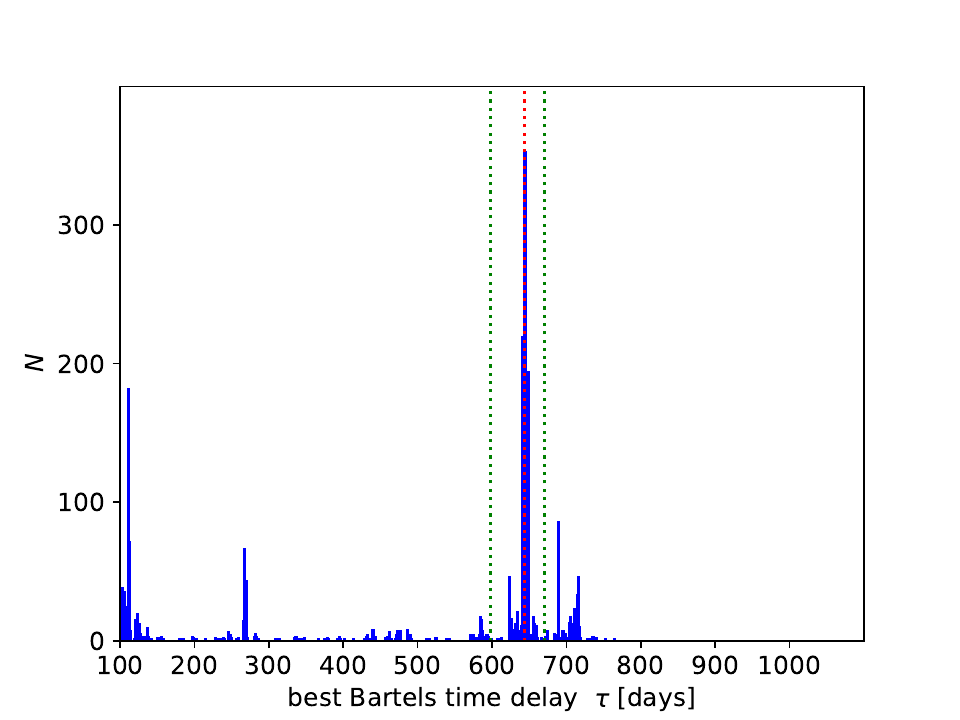}
    \caption{The results of the time-delay analysis using the Bartels estimator. {\bf Left panel:} The Bartels estimator value as a function of the time-delay in the observer's frame. The minimum estimator value is reached for the time-delay of 690 days. {\bf Right panel:} The histogram of best Bartels time delays constructed from 10000 bootstrap simulations of random subset selections of both continuum and line-emission light curves. The red and green dotted vertical lines mark the peak value of $644^{+27}_{-45}$ days and the corresponding uncertainties.}
    \label{fig_bartels}
\end{figure*}

For comparison, we also applied the Bartels estimator to both light curves, which uses the ranked version of the combined light curve $F_{\rm R}(t,\tau)$. In Fig.~\ref{fig_bartels} (left panel), we show the Bartels estimator as a function of the time delay in the observer's frame. The minimum is clearer than for the von Neumann estimator and is located at 690 days. After running 10\,000 bootstrap realizations, we detect three peaks as for von Neumann estimator. Between 100 and 1100 days, there is a peak of $644^{+27}_{-45}$ days, see Fig.~\ref{fig_bartels} (right panel), where the uncertainties were calculated within 30\% surroundings of this time-delay peak.

\subsection{The $\chi^2$ method}
\label{subsec_chi2}

Inspired by the time-delay analysis in the quasar lensing studies, we applied the $\chi^2$ method to our set of light curves. The $\chi^2$ method performed well and consistently in comparison with other time-delay analysis techniques for the previous two SALT quasars -- CTS C30.10 \citep{czerny2019} and HE 0413-4031 \citep{2020ApJ...896..146Z}. It also performs better than the classical ICCF for the case when the AGN variability can be interpreted as a red-noise process \citep{2013A&A...556A..97C}. We subtracted the mean from both the continuum and the line-emission light curve and subsequently, they were normalized by their corresponding variances. The similarity between the continuum and the time-shifted line-emission light curve is evaluated using the $\chi^2$. We make use of the symmetric interpolation.

\begin{figure*}
    \centering
    \includegraphics[width=0.49\textwidth]{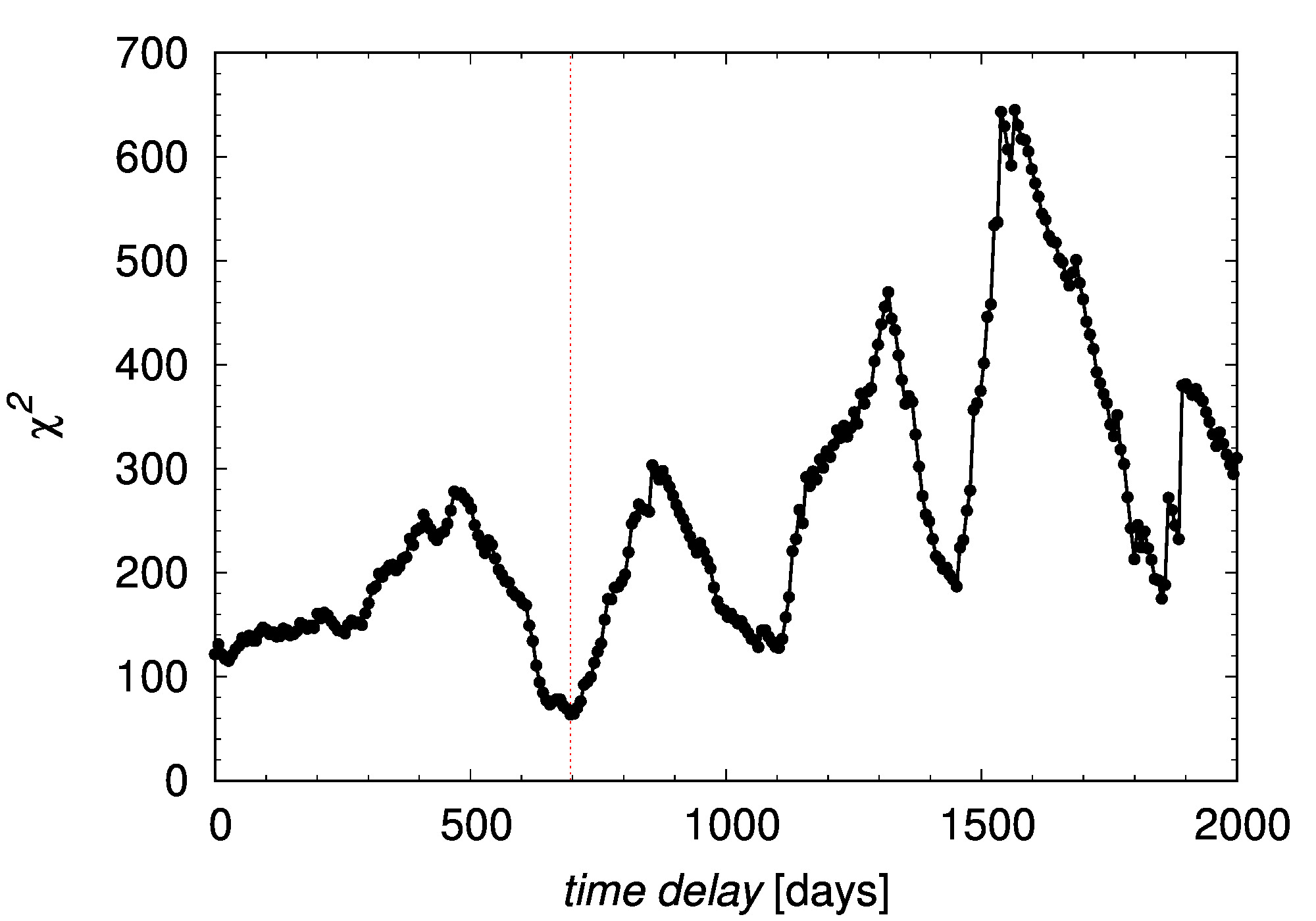}
    \includegraphics[width=0.49\textwidth]{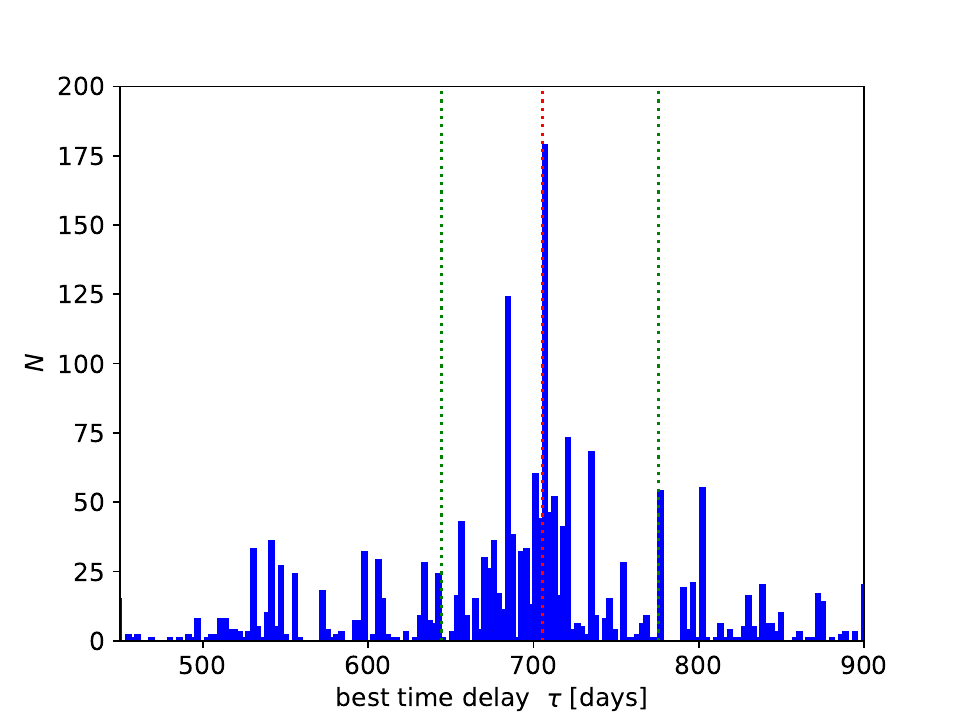}
    \caption{Time-delay determination using $\chi^2$ method. {\bf Left panel:} The $\chi^2$ value as a function of the time-delay in the observer's frame. The global minimum is at $\sim 696$ days, which is denoted by a red vertical line. {\bf Right panel:} The histogram of best time delays constructed from $10\,000$ bootstrap simulations around the global $\chi^2$ minimum of $\sim 700$ days. The peak value is $706^{+70}_{-61}$ days, which is shown using vertical red and green dotted lines.}
    \label{fig_chi2}
\end{figure*}

In Fig.~\ref{fig_chi2} (left panel), we show the $\chi^2$ as a function of the time-delay in the observer's frame. There is a global minimum at $696$ days. The exact position and the uncertainty of this peak is inferred from the histogram of the best time-delays constructed from 10\,000 bootstrap realizations. We obtain $706^{+70}_{-61}$ days, where the uncertainties were determined as the left and the right standard deviations from the time-delays that are positioned within 30\% of the main peak, see Fig.~\ref{fig_chi2} (right panel).

\section{Alias mitigation using generated light curves}
\label{appendix_alias_mitigation}

In this section, we investigate the alias problem, i.e. the occurrence of secondary peaks in the time-delay distribution from a different perspective than presented in Subsection~\ref{subsec_javelin}, where we applied the down-weighting technique on the posterior time-delay distribution based on the number of data points in the overlap region. Here we instead generate a large set of synthetic light curves similar in basic temporal properties to those of our studied quasar, while the line-emission light curve has a known prior time-delay with respect to the continuum light curve. Then we apply several time-delay determination methods to see how well we can recover the true time delay and what the distribution width is close to the assumed time-delay.

\begin{figure*}
    \centering
    \includegraphics[width=0.49\columnwidth]{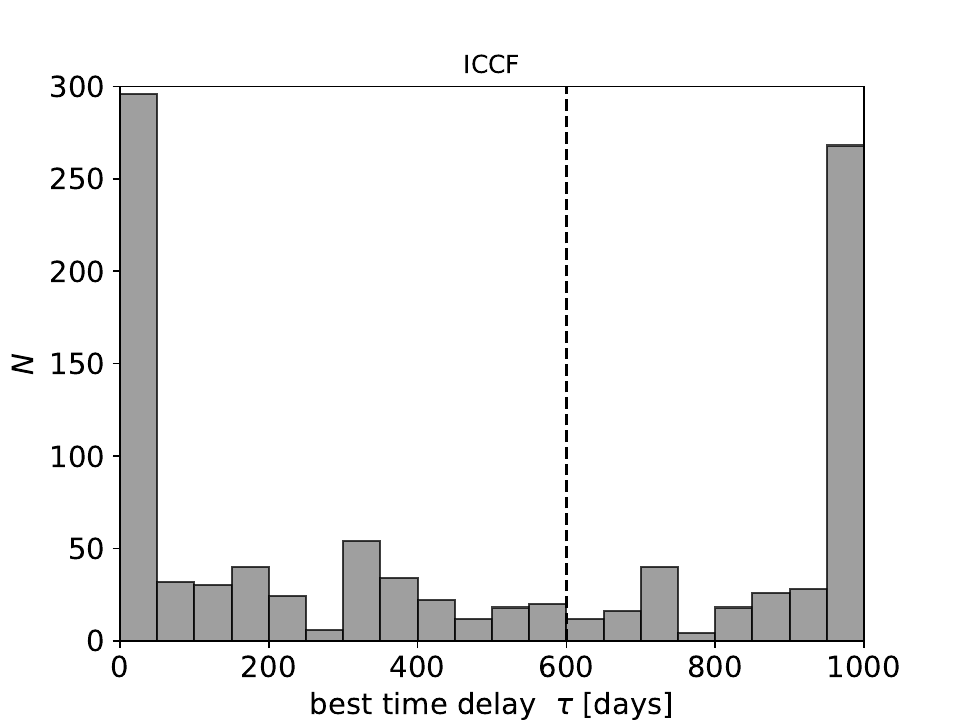}
    \caption{Best time-delay distribution constructed from the analysis of 1000 pairs of light curves generated using the Timmer-Koenig algorithm \citep{1995A&A...300..707T} and subsequently analyzed by the interpolated cross-correlation function (ICCF). The histogram of the best time delays has a bin size of 50 days. The dashed vertical line marks the time delay of 600 days in the observer's frame assumed for the time-shifted line-emission light curve.}
    \label{fig_tk_iccf}
\end{figure*}

\begin{figure*}
    \centering
    \includegraphics[width=0.49\columnwidth]{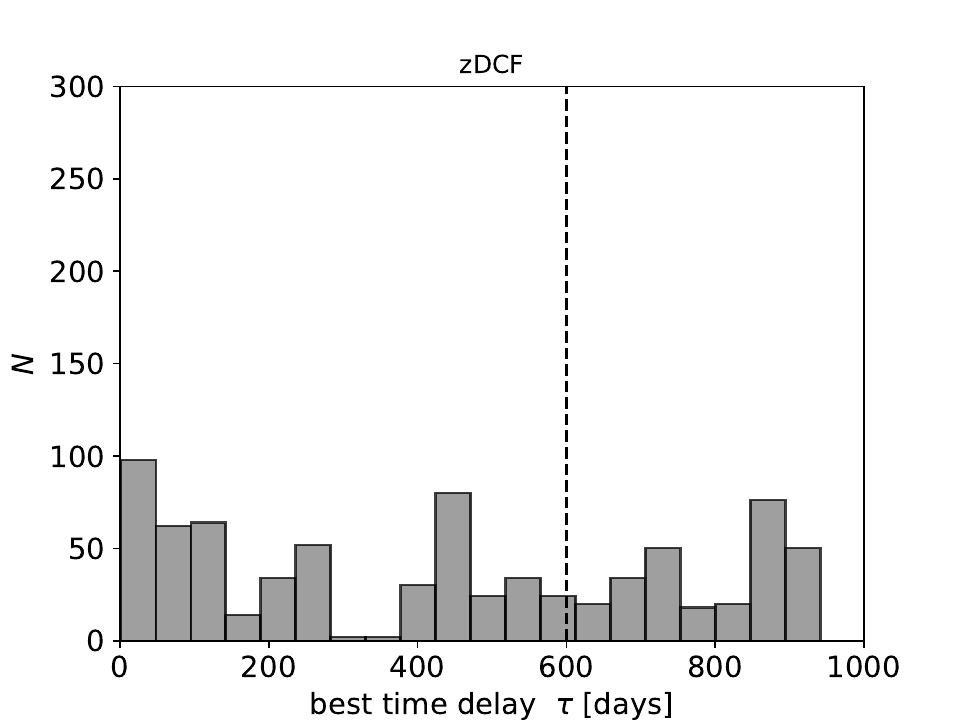}
    \includegraphics[width=0.49\columnwidth]{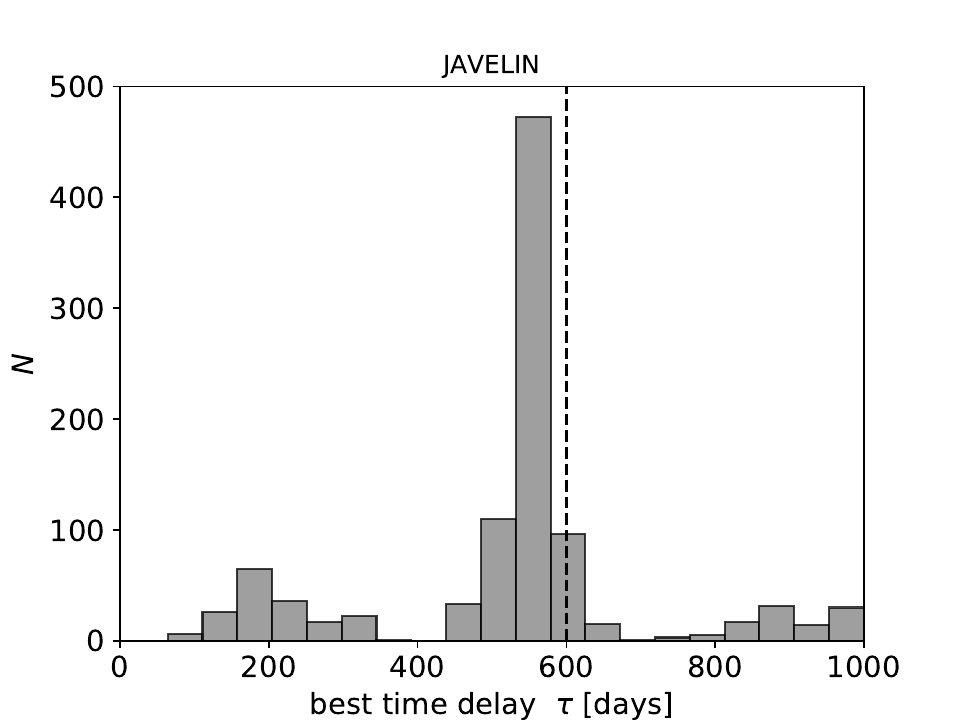}
    \includegraphics[width=0.49\columnwidth]{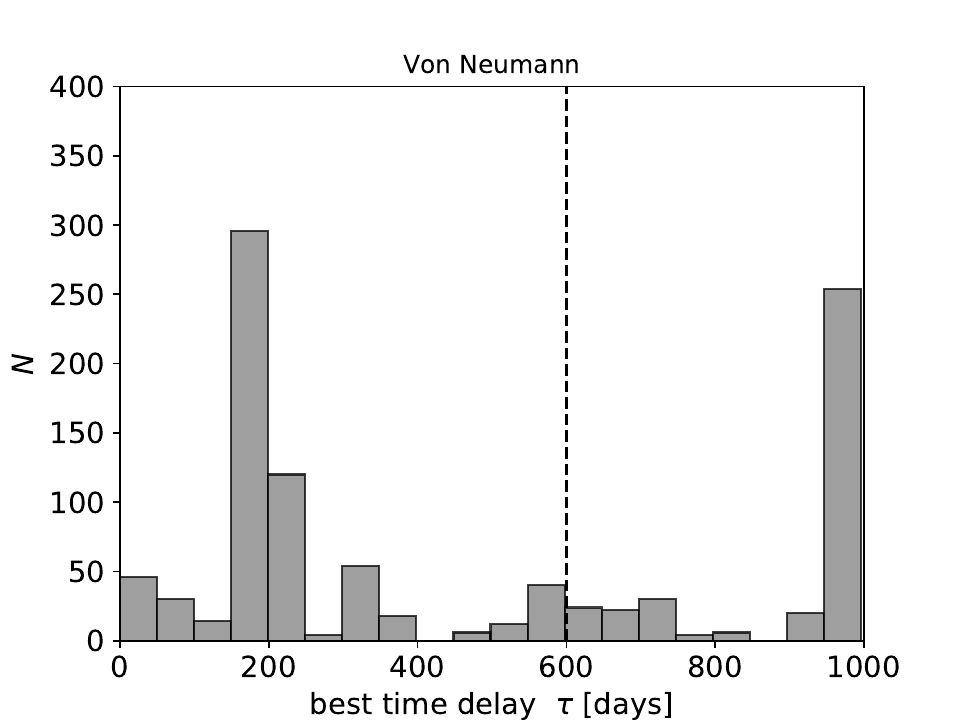}
    \includegraphics[width=0.49\columnwidth]{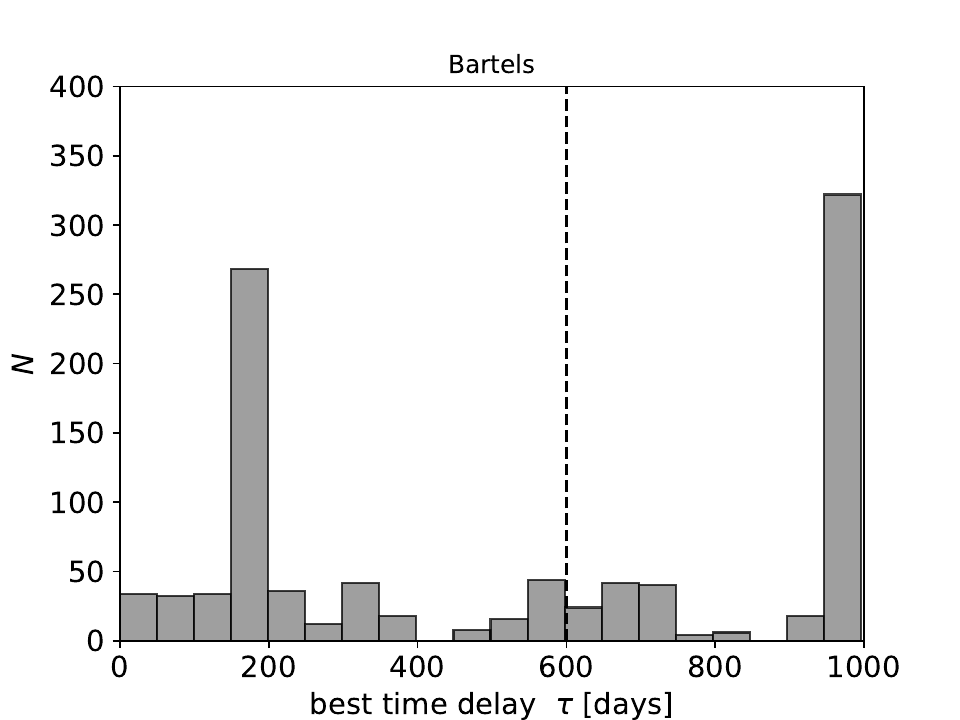}
     \includegraphics[width=0.49\columnwidth]{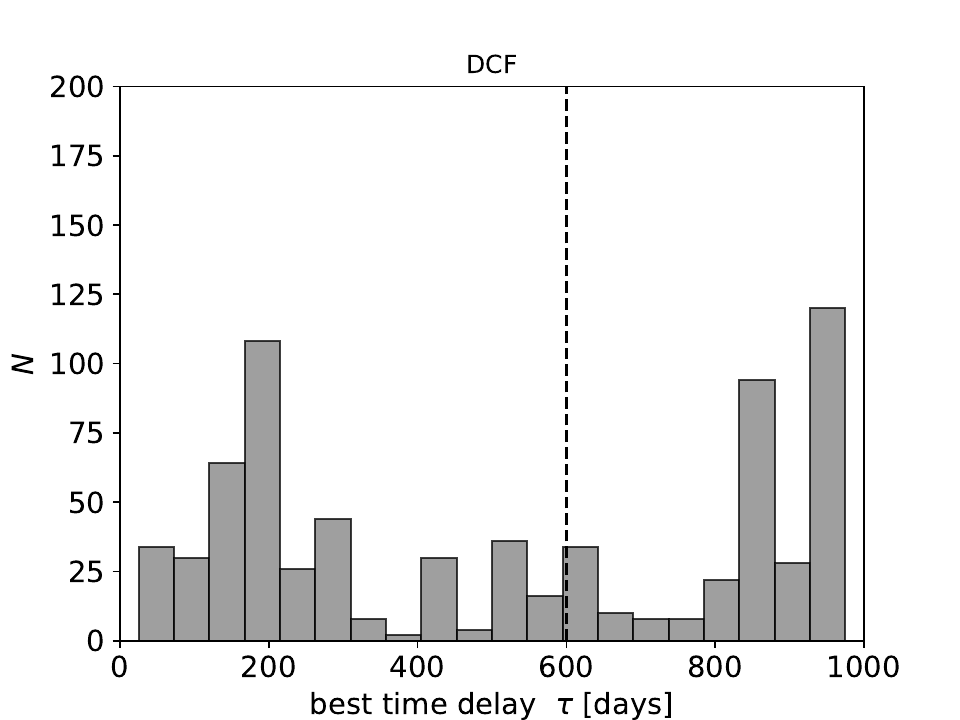}
     \includegraphics[width=0.49\columnwidth]{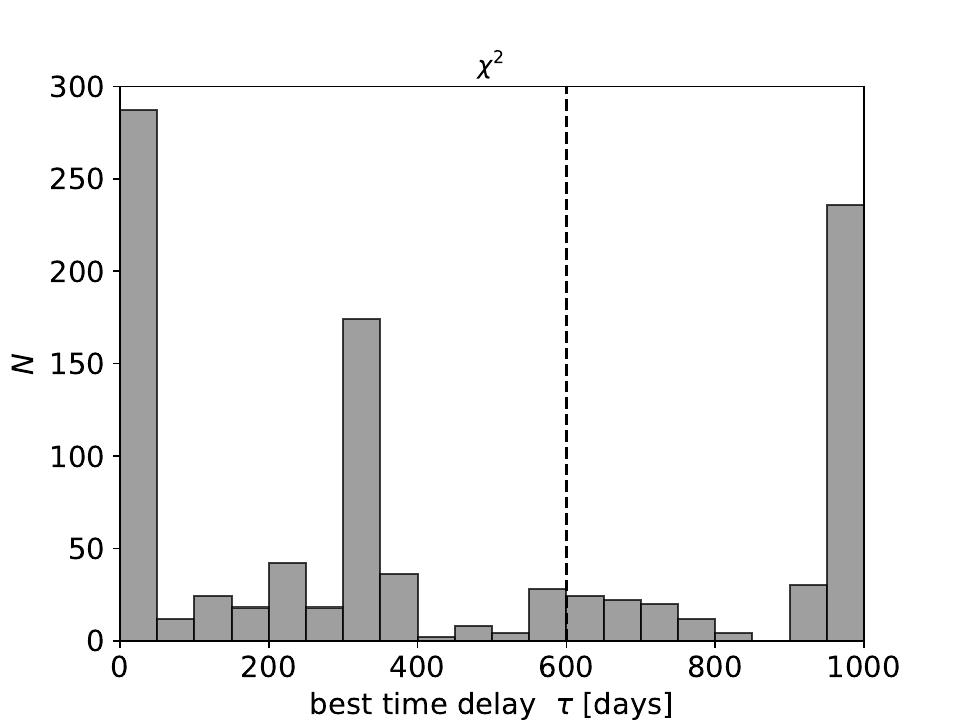}
    \caption{Best time-delay distributions constructed from 1000 pairs of light curves generated using the Timmer-Koenig algorithm \citep{1995A&A...300..707T}. {\bf Top left:} The histogram of the best time delays constructed for the zDCF method. The bin size is 50 days. The dashed vertical line marks the time delay of 600 days in the observer's frame assumed for the time-shifted line-emission light curve. {\bf Top right:} The analogous histogram of the best time delays as in the top left panel constructed for the JAVELIN method. {\bf Middle left:} The analogous histogram of the best time delays as in the top left panel constructed for the von-Neumann estimator. {\bf Middle right:} The analogous histogram of the best time delays as in the top left panel constructed for the Bartels estimator. {\bf Bottom left:} The analogous histogram of the best time delays as in the top left panel constructed for the DCF method. {\bf Bottom right:} The analogous histogram of the best time delays as in the top left panel constructed for the $\chi^2$ method.}
    \label{fig_tk}
\end{figure*}

\begin{table*}[]
    \centering
    \caption{Peaks close to $\tau=600$ days (in the range between 500 and 800 days in the observer's frame) in the time-delay distributions (see Fig.~\ref{fig_tk}) for individual methods.}
    \begin{tabular}{c|c}
    \hline
    \hline
    Method & Time-delay peak (in days) close to $\tau=600$ days\\
    \hline
    ICCF & $720^{+79}_{-218}$\\
    zDCF & $724^{+79}_{-221}$ \\
    JAVELIN     &  $564^{+131}_{-165}$ \\
    Von Neumann  & $594^{+143}_{-115}$  \\
    Bartels      & $707^{+78}_{-199}$   \\
    DCF          & $525^{+127}_{-109}$  \\
    $\chi^2$     & $632^{+128}_{-178}$ \\
    \hline
    \end{tabular}
    \label{table_tk}
\end{table*}

To generate mock light curves, we use the method of \citet{1995A&A...300..707T} (hereafter TK) to produce synthetic light curves from the assumed shape and the normalization of the power density spectrum (PDS). The TK method can generate mock light curves based on any shape of the PDS unlike the damped random walk approach \citep[DRW;][]{2009ApJ...698..895K,2010ApJ...721.1014M,2011ApJ...735...80Z,2013ApJ...765..106Z,2016ApJ...819..122Z}, which generates light curves with the PDS slope close to $-2$. The assumed shape of the PDS is a broken power-law function with two break frequencies corresponding to $\sim 10\,000$ days at lower frequencies and to $\sim 700$ days at higher frequencies. The slopes from lower to higher frequencies vary consequently from $0.0$, through $-1.2$, up to $-2.5$. The line emission light curve is first time-shifted by the assumed time delay, here fixed to 600 days in the observer's frame. Then the line emission is smeared by the timescale equal to $10\%$ of the time delay, where we adopt the Gaussian shape of the BLR transfer function. Finally, dense continuum and line emission light curves are interpolated to the actual observational epochs of our observed light curves. We generate in total 1000 mock light curves with the actual observational cadence, which has a potential to reveal artefact peaks in the time-delay probability distributions due to the particular sampling pattern of our observations \citep[see also][for a similar analysis of cross-correlation artefacts]{2014MNRAS.445..437M}. These corresponding continuum-line pairs are further analyzed by seven different time-delay methods. Subsequently, we construct histograms of best time delays, which are utilized as probability density distributions to determine the peak time delays and the corresponding 1$\sigma$ uncertainties, see Fig.~\ref{fig_tk_iccf} for the ICCF method and \ref{fig_tk} for the zDCF, JAVELIN, von Neumann, Bartels, DCF, and $\chi^2$ methods. In particular, we focus on how well we can determine the prior (true) time delay and what its uncertainty is. In Table~\ref{table_tk}, we list for individual methods the most prominent peaks and their uncertainties in the interval between 500 and 800 days in the observer's frame.

In general, the cadence of our observed light curves appears to produce two-three secondary peaks besides the prior peak close to 600 days, which are often more prominent than the assumed peak. Two peaks close to $\lesssim 200$ days and $\sim 900-1000$ days are especially apparent. For ICCF, DCF, and zDCF methods, the prior time-delay peak is not well defined, i.e. the distribution is highly smeared and broad in the interval between 500 and 800 days. For other methods, the peak close to 600 days is better defined, however, with a large standard deviation of the order of $100$ days. The true peak at $\sim 600$ days is best recovered using the JAVELIN method, which also suppresses effectively the secondary peaks detected in other methods. This result is consistent with the study of \citet{2020MNRAS.491.6045Y}, who report that the JAVELIN outperforms the ICCF in terms of time-delay error estimation. Furthermore, \citet{2019ApJ...884..119L} used mock light curves to compare the JAVELIN, ICCF, and zDCF time-lag recovery and error estimation and concluded that the JAVELIN produces higher quality time delay estimations over both the ICCF and the zDCF. In Table~\ref{table_tk}, we list the inferred peaks in the range between 500 and 800 days, i.e. close to the true peak of 600 days in the observer's frame. We see that due to the sparse cadence of our observations, the detected peak can be shifted by as much as $\sim 100$ days shortwards or longwards of the assumed peak. 

The time-delay analysis of the generated light curves thus shows that the secondary peaks can be created in the time-delay distribution due to the specific cadence and the duration of the observed light curves. The secondary peaks longwards of the true peak at 600 days can be mitigated using the down-weighting technique as we showed in Subsection~\ref{subsec_javelin} for the JAVELIN method; see also \citet{2017ApJ...851...21G}. Here we show that the prominent peaks between 0 and $\sim 200$ days, which are not mitigated by the down-weighting due to a large number of overlapping light curve pairs, can arise due to a lower cadence of our data light curves, especially the line-emission light curve with only 24 points. Hence, they do not reflect the physical response of the BLR transfer function. In this sense, the reported mean peak time-delay of $658^{+29}_{-31}$ days in the observer's frame ($296^{+13}_{-14}$ days in the rest frame for $z=1.2231$) for the quasar HE 0435-4312 can be considered as the best candidate of the true time delay, albeit with the large uncertainty up to $\sim 100$ days. The other peaks appear to be aliases due to the specific cadence and the duration of our light curves.

\end{document}